\documentclass[reqno,11pt]{amsart}
\usepackage[utf8]{inputenc}
\usepackage{graphicx}
\usepackage{slashed}
\usepackage{amscd}
\usepackage{amssymb}
\usepackage{esint}
\usepackage{enumitem}
\usepackage{comment}
\usepackage{amsmath}
\usepackage{dsfont}
\usepackage[mathscr]{eucal}
\usepackage[linktocpage]{hyperref}
\usepackage{pstricks}
\usepackage{pst-all}

\numberwithin{equation}{section}
\allowdisplaybreaks[1]
\definecolor{labelkey}{gray}{.65}

\title[Holographic Mixing and Fock Space Dynamics of Causal Fermion Systems]{Holographic Mixing and Fock Space Dynamics \\ of Causal Fermion Systems}

\author[C.\ Dappiaggi]{Claudio Dappiaggi}
\address{Dipartimento di Fisica \\ Universit{\`a} degli Studi di Pavia and INFN \& INDAM, Sezione di Pavia \\ Via Bassi, 6 --  I-27100 Pavia \\ Italy}
\email{claudio.dappiaggi@unipv.it}

\author[F.\ Finster]{Felix Finster}
\address{Fakult\"at f\"ur Mathematik \\ Universit\"at Regensburg \\ D-93040 Regensburg \\ Germany}
\email{finster@ur.de}

\author[N.\ Kamran]{Niky Kamran}
\address{Department of Mathematics and Statistics \\ McGill University \\ Montr{\'e}al, QC \\ H3A 0B9\\ Canada}
\email{niky.kamran@mcgill.ca}

\author[M.\ Reintjes]{Moritz Reintjes \\ \\ October 2024 / May 2025}
\address{Department of Mathematics \\ City University of Hong Kong \\ SAR Hong Kong}
\email{moritzreintjes@gmail.com}

\newtheorem{Def}{Definition}[section]
\newtheorem{Thm}[Def]{Theorem}
\newtheorem{Prp}[Def]{Proposition}
\newtheorem{Lemma}[Def]{Lemma}
\newtheorem{Remark}[Def]{Remark}

\newcommand{\Thanks}{\vspace*{.5em} \noindent \thanks}
\newcommand{\beq}{\begin{equation}}
\newcommand{\eeq}{\end{equation}}
\newcommand{\Proof}{\begin{proof}}
\newcommand{\QED}{\end{proof} \noindent}
\newcommand{\QEDrem}{\ \hfill $\Diamond$}

\newcommand{\la}{\langle}
\newcommand{\ra}{\rangle}
\newcommand{\bra}{\mathopen{<}}
\newcommand{\ket}{\mathclose{>}}
\newcommand{\bbra}{\mathopen{\ll}}
\newcommand{\kket}{\mathclose{\gg}}
\newcommand{\Sl}{\mathopen{\prec}}
\newcommand{\Sr}{\mathclose{\succ}}

\newcommand{\C}{\mathbb{C}}
\newcommand{\R}{\mathbb{R}}
\newcommand{\1}{\mathds{1}}

\newcommand{\N}{\mathbb{N}}
\newcommand{\Pdd}{\mbox{$\partial$ \hspace{-1.2 em} $/$}}

\renewcommand{\H}{\mathscr{H}}
\newcommand{\U}{{\rm{U}}}

\newcommand{\G}{\mathscr{G}}

\newcommand{\bep}{\begin{pmatrix}}
\newcommand{\enp}{\end{pmatrix}}

\renewcommand{\O}{\mathscr{O}}

\newcommand{\F}{{\mathscr{F}}}
\newcommand{\Dir}{{\mathcal{D}}}

\newcommand{\K}{{\mathcal{K}}}

\newcommand{\B}{{\mathscr{B}}}
\renewcommand{\O}{{\mathscr{O}}}
\renewcommand{\L}{{\mathcal{L}}}
\newcommand{\Sact}{{\mathcal{S}}}
\newcommand{\s}{{\mathfrak{s}}}

\newcommand{\Lin}{\text{\rm{L}}}

\newcommand{\T}{{\mathscr{T}}}

\newcommand{\Fock}{{\mathcal{F}}}
\newcommand{\fermi}{{\mathrm{\tiny{f}}}}
\newcommand{\bose}{{\mathrm{\tiny{b}}}}
\newcommand{\macro}{{\mathrm{\tiny{macro}}}}

\newcommand{\scrU}{{\mathscr{U}}}
\newcommand{\scrA}{{\mathscr{A}}}

\newcommand{\A}{\myscr A}
\newcommand{\dyn}{{\text{\rm{\tiny{dyn}}}}}

\newcommand{\hol}{\text{\rm{hol}}}

\newcommand{\starD}{\triangleright}

\DeclareFontFamily{OT1}{rsfso}{}
\DeclareFontShape{OT1}{rsfso}{m}{n}{ <-7> rsfso5 <7-10> rsfso7 <10-> rsfso10}{}
\DeclareMathAlphabet{\myscr}{OT1}{rsfso}{m}{n}

\newcommand\Felix[1]{}

\DeclareMathOperator{\re}{Re}
\DeclareMathOperator{\im}{Im}

\renewcommand{\Tr}{\text{\rm{Tr}}}
\DeclareMathOperator{\tr}{tr}

\DeclareMathOperator{\supp}{supp}

\newcommand{\bu}{{\mathbf{u}}}

\newcommand{\bitem}{\begin{itemize}[leftmargin=2em]}
\newcommand{\eitem}{\end{itemize}}

\newcommand{\ml}{\text{\rm{\tiny{ml}}}}

\begin{document}

\maketitle

\begin{abstract}
A limiting case is considered in which the causal action principle for causal fermion systems describing Minkowski space gives rise to the linear Fock space dynamics of quantum electrodynamics. The quantum nature of the bosonic field is a consequence of the stochastic description of a multitude of fluctuating fields coupled to non-commuting operators, taking into account dephasing effects. The scaling of all error terms is specified. Our analysis leads to the concept of holographic mixing, which is introduced and explained in detail.
\end{abstract}

\tableofcontents

\section{Introduction}
This paper is part of the research program aimed at getting a precise connection between
causal fermion systems and quantum field theory (QFT).
This research program was initiated in~\cite{fockbosonic, fockfermionic} by the abstract construction
of a quantum state of a causal fermion system for any fixed time. In~\cite{fockentangle}, this construction was 
extended, allowing for the description of general entangled states, but the question remained of how to derive dynamical equations for the time evolution of this quantum state.
In the present paper we resolve this question by deriving these dynamical equations and formulating
them in terms of a unitary time evolution on bosonic and fermionic Fock spaces.
More precisely, we obtain standard quantum electrodynamics (QED) in Minkowski space
with an ultraviolet cutoff in the limiting case
where different types of error terms can be neglected. We specify the scaling behavior of all these
error terms. This analysis opens the door for going beyond standard QED
by working out correction terms (as will be discussed in the outlook in Section~\ref{secoutlook}).
We also plan to write a detailed survey article on the connection between causal fermion systems
and QFT~\cite{qftlimit}.

Compared to the previous papers cited above,
where the goal was to relate the mathematical structures of a causal fermion system to corresponding
objects of QFT, we now need to delve deeper into the dynamical equations
of causal fermion systems. More precisely, the analytic backbone of the theory of causal fermion systems is the
{\em{causal action principle}}, a variational principle for the measure
describing the causal fermion system.
The dynamics of the causal fermion system are then governed by the corresponding {\em{Euler-Lagrange}} (EL)
{\em{equations}},
being nonlinear equations formulated in spacetime. By linearizing these equations, one obtains the
so-called {\em{linearized field equations}}, which are the starting point for the detailed
analysis of the causal action principle, including the resulting nonlinear dynamics.
The present work is based on and makes essential use of results in~\cite{nonlocal},
where the linearized field equations were analyzed in detail for causal fermion systems describing
Minkowski space.
Our main objective is to show that the dynamics of these linearized fields together with their 
coupling to matter can be described in a well-defined limiting case as a unitary time evolution of a quantum state
on fermionic and bosonic Fock spaces.

We now give a brief outline of our methods and results. We take the following results from~\cite{nonlocal}
as the basic input:
\bitem
\item[(i)] The dynamics of the wave functions in a causal fermion system describing Min\-kowski space
can be described by a Dirac equation of the form
\beq \label{dirnonloc}
\big( i \Pdd + \B - m \big) \psi = 0 \:,
\eeq
where~$\B$ is a {\em{nonlocal}} potential, i.e.\ an integral operator of the form
\beq \label{Bnonlocal}
\big( \B \psi \big)(x) = \int_M \B(x,y)\, \psi(y)\: d^4y \:.
\eeq
The integral kernel~$\B(x,y)$ is nonlocal on a scale~$\ell_{\min}$ lying between
the length scale~$\varepsilon$ of the ultraviolet regularization (which can be thought of as the Planck scale)
and the length scale~$\ell_\macro$ of macroscopic physics (which can be thought of as the
Compton scale),
\beq \label{ellmindef}
\varepsilon \ll \ell_{\min} \ll \ell_\macro \:.
\eeq
In simple terms, $\ell_{\min}$ can be regarded as the minimal length scale on which the
analysis on the light cone and the resulting formalism of the continuum limit
as developed in~\cite{pfp, cfs} apply.

\item[(ii)] The nonlocal potential~$\B$ is composed of a collection of vector potentials~$A_a$
with~$a  \in \{1,\ldots N\}$. More precisely,
\beq \label{hatBjdefasticintro}
\B(x,y) = \sum_{a=1}^N \slashed{A}_a \Big( \frac{x+y}{2} \Big) \:L_a(y-x) \:,
\eeq
where the~$L_a$ are fixed complex-valued kernels.
The number~$N$ of these fields is very large and scales like
\beq \label{Nscaleintro}
N \simeq \frac{\ell_{\min}}{\varepsilon} \:.
\eeq
\eitem
We note that, in the special case~$L_a(y-x)=\delta^4(y-x)$, the potential~$A_a$
can be regarded as a classical electromagnetic potential. If~$L_a$ is a nonlocal kernel,
the potential~$A_a$ can still be regarded as being classical, but its coupling to the wave functions
is modified by~$L_a$.
We also remark that the ansatz~\eqref{hatBjdefasticintro} poses no restrictions on the form
of the nonlocal potential (for details see the approximation argument in Lemma~\ref{lemmaapprox}),
but the scaling of the number~$N$ of summands~\eqref{Nscaleintro} clearly gives constraints
for the form of the potential.

To summarize, the Dirac equation~\eqref{dirnonloc}
describes the propagation of Dirac waves in the presence of a multitude of classical background fields,
which are nonlocal on a small scale~$\ell_{\min}$.
Apart from these results from~\cite{nonlocal}, our analysis is based on two additional assumptions.
First, in order for the potentials~$A_a$ to have a notable effect, they must clearly be nonzero.
It seems easiest and physically sensible to describe them stochastically  by Gaussian fields:
\begin{itemize}[leftmargin=3em]
\item[(iii)] The potentials~$A_a$ are non-vanishing and can be described stochastically by Gaussian fields.
\eitem
As we shall see in detail later in this paper, the above assumptions do not yet give rise to
bosonic quantum fields. One important ingredient which is still missing is related to the local gauge
freedom of electrodynamics. Namely, a classical electromagnetic potential~$A_a$ leads to
local phase transformations of the Dirac wave functions (for basics, see the beginning of
Section~\ref{secgaugemom}). Likewise, the potentials in~\eqref{hatBjdefasticintro} also lead to
local phase transformations, but only of those wave functions to which the corresponding potential couples.
This gives rise to a decomposition of the wave functions into many components, each
experiencing different relative phase transformations. We refer to these components as
{\em{holographic components}} and the resulting mechanism as {\em{holographic mixing}}.
It is one of the main tasks of the present paper to model holographic mixing mathematically
and to work out the resulting effects.
Here we do not yet enter the details but merely state in general our second additional assumption:
\begin{itemize}[leftmargin=3em]
\item[(iv)] The potentials~$A_a$ lead to dephasing effects.
\eitem
Here by {\em{dephasing}} we mean the common effect in quantum theory that
a superposition of wave functions becomes small if the summands are ``not in phase''
(for example by involving independently chosen or random phase factors).
This effect is often described by the closely related notions of {\em{destructive interference}}
or {\em{decoherence}}. We here prefer the more specific notion of dephasing.

The main lesson from the present paper is that, implementing the above assumptions~(i)--(iv)
in a coherent mathematical setting gives rise to bosonic quantum fields.
Moreover, in a specific limiting case, the coupling of the bosonic fields to the Dirac wave functions
gives rise to QED. This limiting case involves a specific choice of the covariances of the Gaussian
stochastic fields, in agreement with Lorentz invariance on macroscopic length scales.
In this way, we obtain a derivation of QFT in the time-dependent setting with Lorentzian signature
from the fundamental and broader causal action principle.
Even more generally and independent of causal fermion systems, our
methods and results apply to any physical theory satisfying the above assumptions~(i)--(iv),
thereby giving a connection to quantum field theory.

We proceed by explaining in words how the above assumptions are connected to the appearance of bosonic 
quantum fields. Let us begin by noting that studying similarities and connections between random
processes and QFT has a long history.
The striking similarities between the Wiener process and the Euclidean path integral
(see for example~\cite{glimm+jaffe, kleinert}) inspired attempts to describe the ``quantization'' of a classical system
using stochastic notions. The most prominent approach in this direction is Nelson's mechanics~\cite{nelson}.
From the more mathematical point of view,
there is a close and well-established connection between a suitable class of random
distributions and the $n$-point functions of Euclidean QFTs, 
as highlighted by the
stochastic quantization program~\cite{ParisiWu, namiki}. On the other hand, the dynamics of a random
distribution is often encoded by
nonlinear stochastic partial differential equations with an additive or a multiplicative white noise. When the linear
part of these equations is ruled either by a parabolic or by an elliptic partial differential operator, the solution
theory is well-understood provided that the nonlinearity is sufficiently tame. Both para-controlled
calculus~\cite{Gubinelli} and regularity structures~\cite{Hairer} are efficient frameworks to prove existence
and uniqueness of solutions. For our purposes, it is not necessary to enter the details of these formulations,
but it suffices to highlight that they rely on an algorithmic construction of the solution, based on a graph expansion.
This is necessary in order to cope with ill-defined products of distributions which are a consequence of the singular
structure of the white noise.
We point out that this graph expansion is reminiscent and has close similarities to the
expansion into Feynman diagrams, being at the heart of the textbook approach to perturbative QFT.
These similarities and the presence of the same structural hurdles as in QFT (like for example renormalization) 
is not a mere accident, as has been studied more in depth in~\cite{DDRZ}.
Herein, an algorithm was devised to construct the
solution and the correlation functions of a stochastic, nonlinear partial differential equation adapting the
language and the tools of the algebraic approach to QFT, in particular the Epstein-Glaser
renormalization scheme.

In view of these connections and similarities, the reader may wonder what the difference between a quantum field
and a classical stochastic field actually is. So, what is it that makes a field ``quantum''?
This questions can be answered in various ways.
On the level of Feynman diagrams, one difference is that, for a classical stochastic field,
the bosonic lines in the loops are described by fundamental solutions (meaning by solutions of the
homogeneous equations; for details see again~\cite{loop}),
whereas quantum fields give rise to propagators (i.e.\ Green's operators like the Feynman propagator,
being inhomogeneous solutions).
Another important difference is that, in statistical physics, one takes the statistical average
by integrating over the stochastic fields at the very end when computing the statistical mean.
In quantum physics, however, the path integral, being an integral over field configurations,
describes the time evolution of the state~$|\Psi \ket$,
and probabilities are obtained by first computing the path integral and then taking the expectation value
of observables.
Thus, denoting the path integral or the probability integral by~${\mathcal{D}} \phi$, the
difference can be summarized symbolically as follows,
\begin{align*}
\text{Statistical physics:} &&& \qquad\quad \int \mathcal{D} \phi \: \la \Psi | A | \Psi \ra \\
\text{Quantum physics:} &&& \bigg\la \Big( \int \mathcal{D} \phi \:\Psi \Big) \:\bigg|\: A \:\bigg|
\:\Big( \int \mathcal{D} \phi\: \Psi \Big) \bigg\ra \:,
\end{align*}
where~$A$ denotes the observable.
This means in particular that, in quantum physics, we have separate Feynman diagrams for bra and ket,
but there are no bosonic lines connecting bra and ket.

One way of understanding how this basic difference between quantum fields and classical stochastic fields
comes about is that in QFT one has {\em{non-commuting operators}}.
For example, the Feynman propagator arises from a time-ordering of the bosonic field operators
which satisfy the canonical commutation relations (CCR).
It is important to observe that with the above assumption~(i) non-commutativity comes into play.
Namely, regarding the nonlocal potential as a convolution operator~\eqref{Bnonlocal},
the convolution operators corresponding to different potentials~$A_a$ do in general not commute with
each other. However, our analysis will reveal that the assumption~(i) together with the stochasticity assumption~(iii)
alone are not sufficient for getting QFT. Instead, the fact that we have many stochastic fields~(ii)
which give rise to dephasing effects~(iv) will be essential for obtaining QED.

We next explain in some more detail how bosonic quantum fields arise in our setting.
This connection is easier to make in momentum space. Taking the Fourier transform of the
kernel of the nonlocal potential~\eqref{Bnonlocal} by setting
\beq \label{hatBdef}
\hat{\B}(p,k) := \int_M d^4x \int_M d^4y\: \B(x,y)\: e^{i p x - i k y} \:,
\eeq
in operator products the adjacent momentum variables always coincide
(like for example in products involving Green's operators~\eqref{pertretarded}).
Therefore, the variable~$p-k$ tells us about the momentum change (or momentum transfer)
of the operator~$\B$. The field operator~$\hat{\B}_q$ of momentum~$q$ is obtained by fixing this
momentum change to~$q$, i.e.\
\[ \hat{\B}_q(p,k) := (2 \pi)^4\: \delta^4(p-k-q)\: \hat{\B}(p,k) \]
(for simplicity of presentation, we here disregard the spinorial degrees of freedom;
for more precise formulas for scalar and vector fields see~\eqref{phidefgen}, \eqref{hatAdef}
and~\eqref{hatBjdef}).
By defining the field operators in this way, we have arranged that~$q$ gives the desired
momentum transfer inside the Feynman diagrams.
Note that the field operator is again a nonlocal operator. Moreover, the field operators
for different momenta will in general not commute with each other.
In this way, non-commutativity arises, being a basic feature of field operators.
The question whether this non-commutativity indeed gives rise to the CCR
is more subtle. As already mentioned, it can be answered affirmatively only at the end of our constructions
using the assumptions~(iii) and~(iv) as additional ingredients (see Theorem~\ref{thmccr}).
In this setting, the CCR even arise naturally (see Remark~\ref{remnatural}).

We next specify the structure of the error terms. The causal fermion systems being considered involve several length scales:
\begin{align*}
&\;\;\varepsilon && \text{regularization length (length scale of ultraviolet cutoff)} \\
&\ell_{\min} && \text{length scale of nonlocality of potential~$\B$} \\
&\;\,\ell_\Lambda && \text{length scale of holographic dephasing} \\
&\!\!\ell_\macro && \text{length scale of macroscopic physics} \:.
\end{align*}
The regularization length~$\varepsilon$ can be thought of as a length scale as small as the Planck length.
By the length scale of macroscopic physics we mean the smallest length scale accessible to experiments.
The length scales~$\ell_{\min}$ and~$\ell_\Lambda$ are largely unknown, except that they should lie
between the two other length scales, i.e.\
\[ \varepsilon \ll \ell_{\min}, \ell_\Lambda, \ll \ell_\macro \:. \]
Our results apply to the two limiting cases~$\ell_\Lambda \ll \ell_{\min}$ and~$\ell_{\min} \ll \ell_\Lambda$.
The corresponding relative error terms are
\begin{align}
&\times \Big( 1 + \O \Big( \frac{\ell_\Lambda}{\ell_{\min}} \Big) \Big)\label{err1} \\
\intertext{respectively} 
&\times \Big( 1 + \O \Big( \frac{\ell_\Lambda}{\ell_\macro} \Big) + \O \Big( \frac{\ell_{\min}}{\ell_\Lambda}\Big)
\Big) \label{err2} \:.
\end{align}
The error terms~\eqref{err1} are easier to obtain. They will be derived in Section~\ref{secstationaryphase}
(see Theorem~\ref{thmerr1}).
The error terms~\eqref{err2} are physically more convincing because they allow the dephasing effects
to happen on larger length scales (i.e., lower frequencies and smaller momenta).
They will be derived with a different method
in Section~\ref{secimproved} (see Theorem~\ref{thmerr2}).
We note that the present methods do not cover the case that~$\ell_\Lambda$
and~$\ell_{\min}$ are of the same order of magnitude. This mainly technical issue 
will be excluded from our analysis.

As a final remark, we note that dephasing effects were first considered in~\cite{fockentangle},
and they were shown to be essential for the description of entangled quantum states.
However, from the mathematical point of view, the treatment of dephasing
in~\cite{fockentangle} is quite different from the methods in the present paper.
More precisely, in~\cite{fockentangle} unitary group integrals were considered, and the different
``dephased components'' were recovered as saddle points of these group integrals.
In the present paper, however, the dephasing is analyzed using stationary phase methods in position space.
These methods seem so complementary that it is not yet clear if and how these techniques are related
to each other (see Section~\ref{secstate} for a discussion of this point).

The paper is organized as follows. Section~\ref{secprelim} provides the necessary background
on causal fermion systems and the causal action principle. It also reviews the relevant results on the
linearized field equations obtained in~\cite{nonlocal}.
In Section~\ref{secstoch} we outline our strategy for obtaining bosonic field operators, which
should satisfy the CCR (Section~\ref{secstrategy}).
The general idea is that the nonlocality of the bosonic potentials also introduces
a non-commutativity, giving rise to non-trivial commutation relations.
The basic question is whether it is possible to arrange the CCR.
This question is first analyzed for a single scalar stochastic field (Section~\ref{secscalex}).
This stochastic field can be regarded as a classical bosonic background field. However, it is nonlocal on the
scale~$\ell_{\min}$.  We show that this setting allows to arrange that the
CCR are satisfied in the stochastic average, but not as operator equations.
We proceed by analyzing if the situation improves if, instead of a single stochastic field, one considers
a large number~$N$ of fields (thus modeling the findings of~\cite{nonlocal} outlined above).
This gives additional freedom and flexibility, but not quite to the extent that the CCR can be satisfied
as operator equations.
In Section~\ref{secgauge} the holographic phases are introduced. It is shown that, making use of
resulting dephasing effects, it does become possible to realize the CCR. The errors of the dephasing effects
are specified with the help of a stationary phase analysis.
In Section~\ref{secdynholo} the Dirac dynamics in the presence of holographic phases is developed.
It is a subtle question how to incorporate the phases into the Green's operators.
We show that, by doing this properly, one can improve the scaling of the error terms in the stationary
phase analysis.
In Section~\ref{secfock} the dynamics is formulated in the language of Fock spaces.
In Section~\ref{secstate} we explain how
our constructions relate to the quantum state as constructed in~\cite{fockfermionic}.
Section~\ref{secoutlook} concludes the paper with a brief discussion of our results and of open problems.
The appendices provide mathematical tools for a more computational analysis of Dirac waves
in the presence of nonlocal potentials and holographic phases.

\section{Preliminaries} \label{secprelim}
This section provides the necessary background on causal fermion systems
and the linearized field equations.
\subsection{Causal Fermion Systems and the Causal Action Principle}
We begin with the abstract setting.
\begin{Def} \label{defcfs} (causal fermion systems) {\em{ 
Given a separable complex Hilbert space~$\H$ with scalar product~$\la .|. \ra_\H$
and a parameter~$n \in \N$ (the {\em{``spin dimension''}}), we let~$\F \subset \Lin(\H)$ be the set of all
symmetric operators on~$\H$ of finite rank, which (counting multiplicities) have
at most~$n$ positive and at most~$n$ negative eigenvalues. On~$\F$ we are given
a positive measure~$\rho$ (defined on a $\sigma$-algebra of subsets of~$\F$).
We refer to~$(\H, \F, \rho)$ as a {\em{causal fermion system}}.
}}
\end{Def} \noindent

A causal fermion system describes a spacetime together
with all structures and objects therein.
In order to single out the physically admissible
causal fermion systems, one must formulate physical equations. To this end, we impose that
the measure~$\rho$ should be a minimizer of the causal action principle.
which we now introduce. For brevity of the presentation, we only consider the
{\em{reduced causal action principle}} where the so-called boundedness constraint has been built
incorporated by a Lagrange multiplier term. This simplification is no loss of generality, because
the resulting EL equations are the same as for the non-reduced action principle
as introduced for example in~\cite[Section~\S1.1.1]{cfs}.

For any~$x, y \in \F$, the product~$x y$ is an operator of rank at most~$2n$. 
However, in general it is no longer a symmetric operator because~$(xy)^* = yx$,
and this is different from~$xy$ unless~$x$ and~$y$ commute.
As a consequence, the eigenvalues of the operator~$xy$ are in general complex.
We denote the rank of~$xy$ by~$k \leq 2n$. Counting algebraic multiplicities, we choose~$\lambda^{xy}_1, \ldots, \lambda^{xy}_{k} \in \C$ as all the nonzero eigenvalues and set~$\lambda^{xy}_{k+1}, \ldots, \lambda^{xy}_{2n}=0$.
We refer to the resulting collection of complex numbers~$\lambda^{xy}_1, \ldots, \lambda^{xy}_{2n}$
as the {\em{non-trivial eigenvalues}} of~$xy$.
Given a parameter~$\kappa>0$ (which will be kept fixed throughout this paper),
we introduce the $\kappa$-Lagrangian and the causal action by
\begin{align}
\text{\em{$\kappa$-Lagrangian:}} && \L(x,y) &= 
\frac{1}{4n} \sum_{i,j=1}^{2n} \Big( \big|\lambda^{xy}_i \big|
- \big|\lambda^{xy}_j \big| \Big)^2 + \kappa\: \bigg( \sum_{j=1}^{2n} \big|\lambda^{xy}_j \big| \bigg)^2 \label{Lagrange} \\
\text{\em{causal action:}} && \Sact(\rho) &= \iint_{\F \times \F} \L(x,y)\: d\rho(x)\, d\rho(y) \:. \label{Sdef}
\end{align}
The {\em{reduced causal action principle}} is to minimize~$\Sact$ by varying the measure~$\rho$
under the following constraints,
\begin{align}
\text{\em{volume constraint:}} && \rho(\F) = 1 \quad\;\; \label{volconstraint} \\
\text{\em{trace constraint:}} && \int_\F \tr(x)\: d\rho(x) = 1 \:. \label{trconstraint}
\end{align}
This variational principle is mathematically well-posed if~$\H$ is finite-dimensional.
For the existence theory and the analysis of general properties of minimizing measures
we refer to~\cite{continuum, lagrange} or~\cite[Chapter~12]{intro}.
In the existence theory one varies in the class of regular Borel measures
(with respect to the topology on~$\Lin(\H)$ induced by the operator norm),
and the minimizing measure is again in this class. With this in mind, we always assume
that~$\rho$ is a {\em{regular Borel measure}}.

We finally point out that the causal action principle is invariant under unitary transformations, i.e.\
to joint transformations~$x \mapsto \scrU x \scrU^{-1}$ of all spacetime point operators
with~$\scrU$ a unitary linear operator on~$\H$. The conservation laws corresponding to this
symmetry will be considered in Section~\ref{secconserve}.

\subsection{The Physical Wave Functions and the Wave Evaluation Operator} \label{secweo}
Let~$\rho$ be a {\em{minimizing}} measure. Defining {\em{spacetime}}~$M$ as the support
of this measure,
\[ 
M := \supp \rho \subset \F \:. \]
the spacetimes points are symmetric linear operators on~$\H$.
These operators contain a lot of information which, if interpreted correctly,
gives rise to spacetime structures like causal and metric structures, spinors
and interacting fields (for details see~\cite[Chapter~1]{cfs}).
Here we restrict attention to those structures needed in what follows.
We begin with a basic notion of causality.

\begin{Def} (causal structure) \label{def2}
{\em{The points~$x$ and~$y$ are
called {\em{spacelike}} separated if all the non-trivial eigenvalues~$\lambda^{xy}_1, \ldots, \lambda^{xy}_{2n}$
have the same absolute value.
They are said to be {\em{timelike}} separated if the non-trivial eigenvalues are all real and do not all 
have the same absolute value.
In all other cases (i.e.\ if the~$\lambda^{xy}_j$ are not all real and do not all 
have the same absolute value),
the points~$x$ and~$y$ are said to be {\em{lightlike}} separated. }}
\end{Def} \noindent
Restricting the causal structure of~$\F$ to~$M$, we get corresponding causal relations in spacetime.
Before going on, we point out that this ``spectral definition'' of causal structures is closely related but
not equivalent to the standard notions of causality in classical spacetimes.
The correspondence is obtained by considering so-called regularized Dirac sea configurations
in Minkowski space, in which case the above notions agree with the corresponding causal notions
in Minkowski space in the limiting case when the regularization length~$\varepsilon$ tends to zero
(this is worked out in detail in~\cite[\S1.2.5]{cfs}).
This analysis applies similarly in Lorentzian spacetimes (see~\cite{nrstg, lqg}).
On the fundamental level, the above definition can be supplemented by
a functional distinguishing a time direction (for details see~\cite[eq.~(1.1.11) and~\S1.2.5]{cfs}).
Other constructions of causal cone structures were given in~\cite[Section~4.1]{linhyp} and~\cite[Section~5]{localize}.
In simple terms, this analysis shows that, on large scales, a causal fermion system
has transitive causal order relations similar to a causal set (for the general context see the
review~\cite{surya}). On small scales and on the fundamental level, one still has
causal relations, but they are not necessarily transitive.

Next, for every~$x \in \F$ we define the {\em{spin space}}~$S_xM$ by~$S_xM = x(\H)$;
it is a subspace of~$\H$ of dimension at most~$2n$.
It is endowed with the {\em{spin inner product}} $\Sl .|. \Sr_x$ defined by
\beq \label{ssp}
\Sl u | v \Sr_x = -\la u | x v \ra_\H \qquad \text{(for all $u,v \in S_xM$)}\:.
\eeq
A {\em{wave function}}~$\psi$ is defined as a function
which to every~$x \in M$ associates a vector of the corresponding spin space,
\beq \label{psirep}
\psi \::\: M \rightarrow \H \qquad \text{with} \qquad \psi(x) \in S_xM \quad \text{for all~$x \in M$}\:.
\eeq
We remark that a wave function~$\psi$ is said to be {\em{continuous}} if
for every~$x \in M$ and~$\varepsilon>0$ there is~$\delta>0$ such that
\beq \label{wavecontinuous}
\big\| \sqrt{|y|} \,\psi(y) -  \sqrt{|x|}\, \psi(x) \big\|_\H < \varepsilon
\qquad \text{for all~$y \in M$ with~$\|y-x\| \leq \delta$}
\eeq
(where~$|x|$ is the absolute value of the symmetric operator~$x$ on~$\H$, and~$\sqrt{|x|}$
is the square root thereof).
We denote the set of continuous wave functions by~$C^0(M, SM)$.

It is an important observation that every vector~$u \in \H$ of the Hilbert space gives rise to a distinguished
wave function. In order to obtain this wave function, denoted by~$\psi^u$, we simply project the vector~$u$
to the corresponding spin spaces,
\beq \label{psiudef}
\psi^u \::\: M \rightarrow \H\:,\qquad \psi^u(x) = \pi_x u \in S_xM \:,
\eeq
where~$\pi_x : \H \rightarrow S_xM$ denotes the orthogonal projection operator to~$S_xM \subset \H$.
We refer to~$\psi^u$ as the {\em{physical wave function}} of~$u \in \H$.
A direct computation shows that the physical wave functions are continuous
(in the sense~\eqref{wavecontinuous}). Associating to every vector~$u \in \H$
the corresponding physical wave function gives rise to the {\em{wave evaluation operator}}
\beq \label{weo}
\Psi \::\: \H \rightarrow C^0(M, SM)\:, \qquad u \mapsto \psi^u \:.
\eeq
The wave evaluation operator describes the whole family of physical wave functions.
All the structures in spacetime are encoded in~$\Psi$ and therefore in the family of physical
wave functions. This can be see for example by expressing the spacetime point operators~$x \in M$
as (for the derivation see~\cite[Lemma~1.1.3]{cfs})
\beq
x = - \Psi(x)^* \,\Psi(x) \label{Fid} \:.
\eeq
Here~$\Psi(x) : \H \rightarrow S_x$, and its adjoint~$\Psi(x)^* : S_x \rightarrow \H$ is
taken with respect to the corresponding inner products, meaning that the relation
\[ \Sl \Psi(x)\, u \:|\: \phi \Sr_x = \la u \:|\: \Psi(x)^*\, \phi\ra_\H \qquad \text{holds for all~$u \in \H$ and~$\phi \in S_x$} \:. \]

\subsection{The Restricted Euler-Lagrange Equations}
We now state the Euler-Lagrange equations.
\begin{Prp} Let~$\rho$ be a minimizer of the reduced causal action principle.
Then the local trace is constant in spacetime, meaning that
\[ 
\tr(x) = 1 \qquad \text{for all~$x \in M$} \:. \]
Moreover, there are parameters~$\mathfrak{r}, \s>0$ such that
the function~$\ell$ defined by
\beq \label{elldef}
\ell \::\: \F \rightarrow \R\:,\qquad \ell(x) := \int_M \L(x,y)\: d\rho(y) - \mathfrak{r}\, \big( \tr(x) -1 \big) - \s
\eeq
is minimal and vanishes in spacetime, i.e.\
\beq \label{EL}
\ell|_M \equiv \inf_\F \ell = 0 \:.
\eeq
\end{Prp} \noindent
For the proof of the EL equations and more details we refer for example to~\cite{nonlocal}.
The parameter~$\mathfrak{r}$ can be viewed as the Lagrange parameter corresponding to the
trace constraint. Likewise, $\s$ is the Lagrange parameter of
the volume constraint.

We now work out what the EL equations mean for first variations of the spacetime points.
The starting point of our consideration is the formula~\eqref{Fid}, which expresses the spacetime point operator
in terms of the wave evaluation operator.
Using this formula, first variations of the wave evaluation operator~$\Psi(x)$ 
(see~\eqref{weo}) at a given spacetime point~$x \in M$ give rise to corresponding variations of the spacetime point operator, i.e.\
\beq \label{ufermi}
\bu := \delta x = -\delta \Psi(x)^*\, \Psi(x) - \Psi(x)^*\, \delta \Psi(x) \:.
\eeq
The minimality of~$\ell$ on~$M$ as expressed by~\eqref{EL} implies that the
derivative of~$\ell$ in the direction of~$\bu$ vanishes, i.e.\
\beq \label{Dul}
D_\bu \ell(x) = 0
\eeq
for all variations of the form~\eqref{ufermi} for which the directional derivative
in~\eqref{Dul} exists. Here the derivative~$D_\bu$ can be understood geometrically as follows.
The linear operators in~$\F$ of maximal rank (the so-called {\em{regular points}}) form a
manifold (for details see~\cite{gaugefix, banach} or~\cite[Section~3.1]{intro}).
The vector~$\bu$ in~\eqref{ufermi} is a tangent vector of this manifold at~$x$, and the left side in~\eqref{Dul}
is the derivative of the function~$\ell$ in the direction of this tangent vector.

For the computations, it is more convenient to reformulate
the restricted EL equations in terms of variations of the kernel of the fermionic projector, as we now explain.
In preparation, we use~\eqref{elldef} in order to write~\eqref{Dul} as
\beq \label{resELL}
\int_M D_{1,\bu} \L(x,y)\: d\rho(y) = \mathfrak{r}\, D_\bu \tr(x) \:,
\eeq
where the meaning of the index is that the directional derivative acts on the first argument of the Lagrangian.
For the computation of the first variation of the Lagrangian, one can make use of the fact
that for any $p \times q$-matrix~$A$ and any~$q \times p$-matrix~$B$,
the matrix products~$AB$ and~$BA$ have the same nonzero eigenvalues, with the same
algebraic multiplicities (as is explained in detail in~\cite[\S1.1.2 after eq.~(1.1.8)]{cfs}).
As a consequence, applying again~\eqref{Fid},
\beq
x y 
= \Psi(x)^* \,\big( \Psi(x)\, \Psi(y)^* \Psi(y) \big)
\simeq \big( \Psi(x)\, \Psi(y)^* \Psi(y) \big)\,\Psi(x)^* \:, \label{isospectral}
\eeq
where $\simeq$ means that the operators are isospectral (in the sense that they
have the same non-trivial eigenvalues with the same algebraic multiplicities). Thus, introducing
the {\em{kernel of the fermionic projector}} $P(x,y)$ by
\beq \label{Pxydef}
P(x,y) := -\Psi(x)\, \Psi(y)^* \::\: S_yM \rightarrow S_xM \:,
\eeq
we can write~\eqref{isospectral} as
\[ x y \simeq P(x,y)\, P(y,x) \::\: S_xM \rightarrow S_xM \:. \]
In this way, the eigenvalues of the operator product~$xy$ as needed for the computation of
the Lagrangian~\eqref{Lagrange} are recovered as
the eigenvalues of a $2n \times 2n$-matrix. Since~$P(y,x) = P(x,y)^*$,
the Lagrangian~$\L(x,y)$ in~\eqref{Lagrange} can be expressed in terms of the kernel~$P(x,y)$.
Consequently, the first variation of the Lagrangian can be expressed in terms
of the first variation of this kernel. Being real-valued and real-linear in~$\delta P(x,y)$,
it can be written as
\beq \label{delLdef}
\delta \L(x,y) = 2 \re \Tr_{S_xM} \!\big( Q(x,y)\, \delta P(x,y)^* \big) \:,
\eeq
where~$Q(x,y)$ is a kernel which is again symmetric (with respect to the spin inner product), i.e.
\beq \label{Qxydef}
Q(x,y) \::\: S_yM \rightarrow S_xM \qquad \text{and} \qquad Q(x,y)^* = Q(y,x) \:.
\eeq
Here~$\Tr_{S_xM}$ denotes the trace on the spin space~$S_xM$, and~$\delta P(x,y)$ is the
first variation of~$P(x,y)$ as a linear operator from~$S_yM$ to~$S_xM$.
More details on this method and many computations can be found in~\cite[Sections~1.4 and~2.6
as well as Chapters~3-5]{cfs}.

Expressing the variation of~$P(x,y)$ in terms of~$\delta \Psi$,
the first variations of the Lagrangian can be written as
\begin{align*}
D_{1,\bu} \L(x,y) = -2\, \re \tr \big( \delta \Psi(x)^* \, Q(x,y)\, \Psi(y) \big) \\
D_{2,\bu} \L(x,y) = -2\, \re \tr \big( \Psi(x)^* \, Q(x,y)\, \delta \Psi(y) \big)
\end{align*}
(where~$\tr$ denotes the trace of a finite-rank operator on~$\H$).
Using these formulas, the restricted EL equation~\eqref{resELL} becomes
\[ \re \int_M \tr \big( \delta \Psi(x)^* \, Q(x,y)\, \Psi(y) \big)\: d\rho(y) = \mathfrak{r}\,\re \tr \big( \delta \Psi(x)^* \,\Psi(x) \big)  \:. \]
Using that the variation can be arbitrary at every spacetime point, we obtain
\[ \int_M Q(x,y)\, \Psi(y) \:d\rho(y) = \mathfrak{r}\, \Psi(x) \qquad \text{for all~$x \in M$}\:, \]
where~$\mathfrak{r} \in \R$ is the Lagrange parameter of the trace constraint.

\subsection{The Conserved Commutator Inner Product} \label{secconserve}
The connection between symmetries and conservation laws made by Noether's theorem
extends to causal fermion systems~\cite{noether}. However, the conserved quantities of a causal fermion system
have a rather different structure, being formulated in terms of so-called surface layer integrals.
A {\em{surface layer integral}} is a double integral of the form
\[ 
\int_\Omega \bigg( \int_{M \setminus \Omega} (\cdots)\: \L(x,y)\: d\rho(y) \bigg)\, d\rho(x) \]
where the two variables~$x$ and~$y$ are integrated over~$\Omega$ and its complement,
and~$(\cdots)$ stands for variational derivatives acting on the Lagrangian.
Since in typical applications,
the Lagrangian is small if~$x$ and~$y$ are far apart, the main contribution to the
surface layer integral is obtained when both~$x$ and~$y$ are near the boundary~$\partial \Omega$.
With this in mind, a surface layer integral can be thought of as a ``thickened'' surface integral,
where we integrate over a spacetime strip of a certain width. For systems in Minkowski space as considered
here, the length scale of this strip is the Compton scale~$m^{-1}$.
For more details on the concept of a surface layer integral we refer to~\cite[Section~9.1]{intro}.

There are various {\em{Noether-like theorems}} for causal fermion systems, which relate
symmetries to conservation laws (for an overview 
see~\cite{osi} or~\cite[Chapter~9]{intro}). The conserved quantity of relevance here is the
commutator inner product (for more details see~\cite[Section~9.4]{intro}
or~\cite[Section~5]{noether} and~\cite[Section~3]{dirac}):
The causal action principle is invariant under unitary transformations of the measure~$\rho$, i.e.\ under
transformations
\[ \rho \rightarrow \scrU \rho \qquad \text{with} \qquad
(\scrU \rho)(\Omega) := \rho \big( \scrU^{-1} \,\Omega\, \scrU \big) \:, \]
where~$\scrU$ is a unitary operator on~$\H$ and~$\Omega \subset \F$ is any measurable subset.
The conserved quantity corresponding to this symmetry is the so-called {\em{commutator inner product}}     
\beq \label{cip}
\la \psi | \phi \ra^\Omega := -2i \,\bigg( \int_{\Omega} \!d\rho(x) \int_{M \setminus \Omega} \!\!\!\!\!\!\!\!d\rho(y) 
- \int_{M \setminus \Omega} \!\!\!\!\!\!\!\!d\rho(x) \int_{\Omega} \!d\rho(y) \bigg)\:
\Sl \psi(x) \:|\: Q(x,y)\, \phi(y) \Sr_x \:,
\eeq
where~$\psi, \phi$ are wave functions~\eqref{psirep} and~$\Omega \subset M$ describes a
spacetime region ($Q(x,y)$ is again the kernel in~\eqref{Qxydef}).
Here {\em{conservation}} means that the commutator inner product of any two physical wave
functions~$\psi$ and~$\phi$ (as defined by~\eqref{psiudef}) vanishes for any compact~$\Omega \subset M$.
If~$\Omega$ is chosen to be non-compact, the commutator inner product is in general nonzero.
But, taking exhaustions, the conservation law can be stated that the commutator inner product~\eqref{cip}
does not depend on the choice of~$\Omega$ within a certain class of sets.
This ``class of sets'' can be specified systematically by working with equivalence classes
(for details see~\cite[Section~3.2]{dirac}).
For our purposes, it suffices to restrict attention to {\em{Minkowski-type spacetimes}}
as introduced in~\cite[Section~2.8]{fockfermionic}. In this case, there is a global
time function~$T : M \rightarrow \R$, and the commutator inner product is well-defined
if~$\Omega$ is chosen as the past of any time~$t$,
\[ \Omega = \Omega_t := T^{-1}\big( (-\infty, t] \big) \:. \]
The set~$\Omega_t$ can be thought of as the past of the Cauchy surface
at time~$t$, so that the surface layer integral describes a ``thickened'' integral over the Cauchy surface.
The conservation law states that the commutator product~$\la \psi | \phi \ra^{\Omega_t}$ does not depend
on~$t \in \R$.

We finally remark that the commutator inner product is also independent of the choice of the time function,
within a large class of time functions, making it possible to describe more general Cauchy surfaces.
As this generalization will not be needed here, we refer for the details to~\cite{dirac}.

\subsection{The Linearized Field Equations in Minkowski Space} \label{secdir}
The linearized field equations describe variations of the measure~$\rho$ which preserve the EL equations.
The linearized field equations play a central role in the analysis of causal fermion systems,
both conceptually and computationally.
From the conceptual point of view, the analysis of the linearized field equations reveals
the causal nature of the dynamics and thereby clarifies the causal structure of spacetime itself.
From the computational point of view, being a linear equation, it becomes possible to analyze
the equations explicitly using methods of functional analysis and Fourier analysis.
Moreover, the linearized field equations are an important first step toward the analysis of
the nonlinear dynamics as described by the EL equations (for example perturbatively using the methods developed in~\cite{perturb}). The linearized field equations were derived and formulated in~\cite{jet}.
They were first analyzed in~\cite{linhyp} using energy methods.
In~\cite{nonlocal} the EL equations and linearizations thereof (the so-called linearized field equations)
were studied in detail for causal fermion systems describing Minkowski space.
In this setting, the abstract structures of a causal fermion system become more concrete,
opening the door for a detailed analysis.

We now explain what these findings mean for the structure of the dynamics in the presence
of linearized fields for causal fermion systems describing Minkowski space.
In this setting, the physical wave functions~\eqref{psiudef} can be represented
by usual spinorial wave functions in Minkowski space. Moreover, the spin inner product~\eqref{ssp}
goes over to the usual pointwise inner product on Dirac spinors, i.e.\
\beq \label{sip}
\Sl \psi | \phi \Sr(x) = \overline{\psi(x)} \phi(x) = \psi(x)^\dagger \gamma^0 \phi(x) \:,
\eeq
where~$\overline{\psi(x)}$ is sometimes referred to as the adjoint spinor
(for more details on the correspondence of the abstract objects with objects in Minkowski space
see~\cite[Section~1.2]{cfs}).
Moreover, in the Minkowski vacuum, the linearized field equations can be described by
a Dirac equation in Minkowski space
\[ \big( i \Pdd - m \big) \psi(x) = 0 \:, \]
where, for ease of notation, the superscript~$u$ of the wave function was omitted.
In the interacting situation, one must insert potentials into the Dirac equation. It turns out that
the linearized field equations do not allow only for homogeneous classical fields
(like plane electromagnetic waves), but instead for a plethora of fields coupling to different wave
packets propagating in different directions. The reason for this surprisingly large space
of linearized solutions can be understood in non-technical terms from the fact that the causal 
Lagrangian~\eqref{Lagrange} is invariant under phase transformations of the
kernel of the fermionic projector~\eqref{Pxydef} of the form
\beq \label{dirlocintro}
P(x,y) \rightarrow e^{i \lambda(x,y)}\: P(x,y)
\eeq
with a real-valued function~$\lambda(x,y)$ which is anti-symmetric (i.e.~$\lambda(x,y) = -\lambda(y,x)$
for all~$x,y \in M$). This invariance generalizes the local gauge invariance of electrodynamics,
which is recovered by choosing~$\lambda(x,y) = \Lambda(x) - \Lambda(y)$ with~$\Lambda$ a real-valued
function  (for basics see again the beginning of Section~\ref{secgaugemom}).
This more general invariance, which can be understood as a direction-dependent local gauge freedom,
is the underlying reason for the appearance of many additional linearized fields.
More details on this point and its mathematical underpinning can be found in~\cite{nonlocal}.

We here proceed by stating how the dynamics in the presence of this multitude of linearized fields
is described mathematically. As already mentioned in the introduction, the resulting Dirac equation
involves a {\em{nonlocal potential}}~$\B$ with integral kernel~$\B(x,y)$ (see~\eqref{dirnonloc}
and~\eqref{Bnonlocal}). This integral kernel is of the form
\beq \label{hatBjdefastic}
\B(x,y) = \sum_{a=1}^N B_a \Big( \frac{x+y}{2} \Big) \:L_a(y-x) \:,
\eeq
where~$B_a(x)$ are multiplication operators acting on the spinors and the~$L_a$ are 
smooth complex-valued functions. The factors in the nonlocal potential are symmetric in the sense that
\beq \label{symmpot}
B_a(x)^* = B_a(x) \qquad \text{and} \qquad \overline{L_a(\xi)} = L_a(-\xi)
\eeq
(where the star is the adjoint with respect to the spin inner product).
The number~$N$ of these potentials is very large and scales like~\eqref{Nscaleintro}
with~$\ell_{\min}$ in the range~\eqref{ellmindef}.
The scale~$\ell_{\min}$ also determines the scale of the nonlocality of the potential.
It is an important consequence of~\eqref{symmpot} that the nonlocal potential
is {\em{symmetric}}, meaning that
\beq \label{Bsymm}
\B(x,y)^* = \B(x,y) \:.
\eeq

The conserved commutator inner product~\eqref{cip} takes the following form
\begin{align}
\la \psi | \phi \ra_t &:= \int \Sl \psi \,|\, \gamma^0\, \phi \Sr_{(t,\vec{x})} \: d^3x \label{c11} \\
&\quad\;\; -i \int_{x^0<t} d^4x \int_{y^0>t} d^4y\;
\Sl \psi(x) \,|\, \B(x,y)\, \phi(y) \Sr_x \label{c2} \\
&\quad\;\; +i \int_{x^0>t} d^4x \int_{y^0<t} d^4y\;
\Sl \psi(x) \,|\, \B(x,y)\, \phi(y) \Sr_x \:. \label{c3}
\end{align}
Note that~\eqref{c11} is the usual scalar product on Dirac wave functions.
The additional summands~\eqref{c2} and~\eqref{c3} can be understood as correction terms
which take into account the nonlocality of the potential~$\B$ in~\eqref{sip}.
The mathematical structure of these additional terms is again that of a surface layer integral.
Making use of the symmetry of the nonlocal potential~\eqref{Bsymm}, a straightforward computation
shows that {\em{current conservation}} holds in the sense that the quantity
is independent of the time~$t$ (for the derivation see~\cite[Proposition~B.1]{baryogenesis}).
We remark that this conservation law holds more generally for arbitrary Cauchy surfaces
(for more details see~\cite[Appendix~B]{baryogenesis}).

In~\cite{nonlocal} the structure of the potentials in~\eqref{hatBjdefastic} is specified in some
more detail, as we now outline.
Considering the kernel~$L_a(y-x)$ as a convolution operator, its Fourier transform~$\hat{L}_a$
defined by
\beq \label{Lahat}
\hat{L}_a(k) := \int_M L_a(\xi)\: e^{-i \xi k}\: d^4 \xi
\eeq
is a multiplication operator in momentum space. The symmetry property in~\eqref{symmpot} means
that the function~$\hat{L}_a$ is real-valued. In Figure~\ref{fighom} the functions~$L_a$ and~$\hat{L}_a$
are depicted in a typical example.
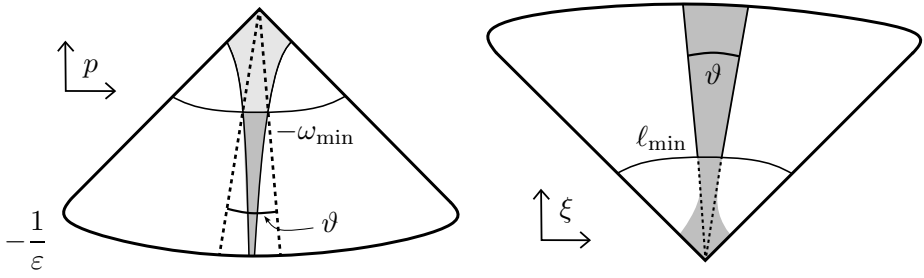
\begin{figure}
\psset{xunit=.5pt,yunit=.5pt,runit=.5pt}
\begin{pspicture}(654.39143624,196.84522734)
{
\newrgbcolor{curcolor}{0.74901962 0.74901962 0.74901962}
\pscustom[linestyle=none,fillstyle=solid,fillcolor=curcolor]
{
\newpath
\moveto(484.9008189,79.87636687)
\curveto(484.9008189,79.87636687)(486.38605984,70.55285065)(486.97048819,56.03074167)
\curveto(487.55491654,41.50862891)(471.66034016,21.46118702)(471.66034016,21.46118702)
\lineto(489.96292913,2.4394749)
\lineto(509.0304189,20.71977222)
\curveto(509.0304189,20.71977222)(498.10365354,31.97315301)(498.81669921,49.4397823)
\curveto(499.52974488,66.90641159)(502.60170709,79.42466419)(502.60170709,79.42466419)
\closepath
}
}
{
\newrgbcolor{curcolor}{0.74901962 0.74901962 0.74901962}
\pscustom[linewidth=1.20188972,linecolor=curcolor]
{
\newpath
\moveto(484.9008189,79.87636687)
\curveto(484.9008189,79.87636687)(486.38605984,70.55285065)(486.97048819,56.03074167)
\curveto(487.55491654,41.50862891)(471.66034016,21.46118702)(471.66034016,21.46118702)
\lineto(489.96292913,2.4394749)
\lineto(509.0304189,20.71977222)
\curveto(509.0304189,20.71977222)(498.10365354,31.97315301)(498.81669921,49.4397823)
\curveto(499.52974488,66.90641159)(502.60170709,79.42466419)(502.60170709,79.42466419)
\closepath
}
}
{
\newrgbcolor{curcolor}{0.90196079 0.90196079 0.90196079}
\pscustom[linestyle=none,fillstyle=solid,fillcolor=curcolor]
{
\newpath
\moveto(159.31893543,114.03735836)
\lineto(153.27110173,113.2111612)
\lineto(140.69603906,113.45769978)
\lineto(139.67797039,124.07638828)
\lineto(138.32090835,134.82361096)
\lineto(136.62054803,143.91922923)
\lineto(135.23355591,150.17405632)
\lineto(133.03922646,156.89477821)
\lineto(130.66084157,162.6976645)
\lineto(128.58723024,166.81147994)
\lineto(152.86016126,191.86696214)
\lineto(177.39190299,166.97203805)
\lineto(173.56362331,162.60630198)
\lineto(170.18796472,156.24387931)
\lineto(166.74466394,146.9633212)
\lineto(165.11775874,141.62339301)
\lineto(163.0972422,132.71962734)
\lineto(161.61350551,126.21741128)
\lineto(160.80185575,121.14235238)
\closepath
}
}
{
\newrgbcolor{curcolor}{0 0 0}
\pscustom[linewidth=0,linecolor=curcolor]
{
\newpath
\moveto(159.31893543,114.03735836)
\lineto(153.27110173,113.2111612)
\lineto(140.69603906,113.45769978)
\lineto(139.67797039,124.07638828)
\lineto(138.32090835,134.82361096)
\lineto(136.62054803,143.91922923)
\lineto(135.23355591,150.17405632)
\lineto(133.03922646,156.89477821)
\lineto(130.66084157,162.6976645)
\lineto(128.58723024,166.81147994)
\lineto(152.86016126,191.86696214)
\lineto(177.39190299,166.97203805)
\lineto(173.56362331,162.60630198)
\lineto(170.18796472,156.24387931)
\lineto(166.74466394,146.9633212)
\lineto(165.11775874,141.62339301)
\lineto(163.0972422,132.71962734)
\lineto(161.61350551,126.21741128)
\lineto(160.80185575,121.14235238)
\closepath
}
}
{
\newrgbcolor{curcolor}{0.74901962 0.74901962 0.74901962}
\pscustom[linestyle=none,fillstyle=solid,fillcolor=curcolor]
{
\newpath
\moveto(140.67810217,113.66265751)
\lineto(149.29360176,113.66265751)
\lineto(158.56370494,113.14926917)
\lineto(153.27341254,62.30635238)
\lineto(149.0039554,4.79805884)
\lineto(145.09012611,4.88377852)
\closepath
}
}
{
\newrgbcolor{curcolor}{0.74901962 0.74901962 0.74901962}
\pscustom[linewidth=0,linecolor=curcolor]
{
\newpath
\moveto(140.67810217,113.66265751)
\lineto(149.29360176,113.66265751)
\lineto(158.56370494,113.14926917)
\lineto(153.27341254,62.30635238)
\lineto(149.0039554,4.79805884)
\lineto(145.09012611,4.88377852)
\closepath
}
}
{
\newrgbcolor{curcolor}{0.74901962 0.74901962 0.74901962}
\pscustom[linestyle=none,fillstyle=solid,fillcolor=curcolor]
{
\newpath
\moveto(502.47403465,80.07801222)
\lineto(493.85852598,80.07801222)
\lineto(484.6872,80.07801222)
\lineto(473.45380157,196.33007017)
\lineto(522.83408504,193.87349821)
\closepath
}
}
{
\newrgbcolor{curcolor}{0.74901962 0.74901962 0.74901962}
\pscustom[linewidth=0,linecolor=curcolor]
{
\newpath
\moveto(502.47403465,80.07801222)
\lineto(493.85852598,80.07801222)
\lineto(484.6872,80.07801222)
\lineto(473.45380157,196.33007017)
\lineto(522.83408504,193.87349821)
\closepath
}
}
{
\newrgbcolor{curcolor}{0 0 0}
\pscustom[linewidth=2.26771663,linecolor=curcolor]
{
\newpath
\moveto(10.34325165,47.94309317)
\lineto(153.1038652,191.91919899)
\lineto(297.71344252,45.49890293)
}
}
{
\newrgbcolor{curcolor}{0 0 0}
\pscustom[linewidth=2.26771663,linecolor=curcolor]
{
\newpath
\moveto(331.9672252,159.04237758)
\lineto(490.33555276,1.60155364)
\lineto(647.6992063,159.11102513)
}
}
{
\newrgbcolor{curcolor}{0 0 0}
\pscustom[linewidth=1.19999845,linecolor=curcolor]
{
\newpath
\moveto(86.90170356,125.60923767)
\curveto(94.96008416,121.46687698)(103.01686375,117.32534173)(120.92631836,115.25100699)
\curveto(138.83577298,113.17667225)(166.59617461,113.16896202)(184.4675452,115.31554621)
\curveto(202.33891124,117.46213039)(210.32011691,121.76262803)(218.30102324,126.06296239)
}
}
{
\newrgbcolor{curcolor}{0.74901962 0.74901962 0.74901962}
\pscustom[linestyle=none,fillstyle=solid,fillcolor=curcolor]
{
\newpath
\moveto(523.00435276,194.6981949)
\lineto(502.72586457,79.60664466)
}
}
{
\newrgbcolor{curcolor}{0 0 0}
\pscustom[linewidth=1.36062998,linecolor=curcolor]
{
\newpath
\moveto(523.00435276,194.6981949)
\lineto(502.72586457,79.60664466)
}
}
{
\newrgbcolor{curcolor}{0 0 0}
\pscustom[linewidth=2.04094496,linecolor=curcolor,linestyle=dashed,dash=3.24000001 3.24000001]
{
\newpath
\moveto(152.98944378,191.67302702)
\lineto(122.76131906,5.61345411)
}
}
{
\newrgbcolor{curcolor}{0 0 0}
\pscustom[linewidth=2.04094496,linecolor=curcolor,linestyle=dashed,dash=3.24000001 3.24000001]
{
\newpath
\moveto(153.00550677,191.56175773)
\lineto(168.95405102,5.52815773)
}
}
{
\newrgbcolor{curcolor}{0 0 0}
\pscustom[linewidth=1.13385827,linecolor=curcolor]
{
\newpath
\moveto(145.10768277,5.35355868)
\curveto(144.49650142,31.91405254)(143.88522028,58.47876435)(142.7336829,81.85747049)
\curveto(141.58215005,105.23613128)(139.89120907,125.41557921)(137.47319282,138.95628246)
\curveto(135.05517657,152.4969857)(131.91245254,159.38719295)(128.76746532,166.28235742)
}
}
{
\newrgbcolor{curcolor}{0 0 0}
\pscustom[linewidth=1.13385827,linecolor=curcolor]
{
\newpath
\moveto(149.28273487,5.02610041)
\curveto(151.14253002,34.02751899)(153.0027515,63.03560466)(155.87324523,87.89789632)
\curveto(158.74373896,112.76016983)(162.62190161,133.45837606)(166.3419908,146.01600151)
\curveto(170.06208454,158.57362243)(173.6214788,162.98048958)(177.18165317,167.38832731)
}
}
{
\newrgbcolor{curcolor}{0 0 0}
\pscustom[linewidth=1.19999845,linecolor=curcolor]
{
\newpath
\moveto(424.36583433,67.92078387)
\curveto(432.42416504,72.06313096)(440.4809537,76.20466167)(458.39042646,78.27894198)
\curveto(476.29989921,80.3532223)(504.06028724,80.36075112)(521.93166236,78.21408529)
\curveto(539.80303748,76.06741947)(547.78422047,71.76692183)(555.76513134,67.46660561)
}
}
{
\newrgbcolor{curcolor}{0 0 0}
\pscustom[linewidth=1.58740153,linecolor=curcolor,linestyle=dashed,dash=2.51999998 2.51999998]
{
\newpath
\moveto(502.70065512,79.59489789)
\lineto(490.46545512,1.43038639)
}
}
{
\newrgbcolor{curcolor}{0 0 0}
\pscustom[linewidth=1.58740153,linecolor=curcolor,linestyle=dashed,dash=2.51999998 2.51999998]
{
\newpath
\moveto(484.76203465,80.24292057)
\lineto(490.49482205,2.00343836)
}
}
{
\newrgbcolor{curcolor}{0 0 0}
\pscustom[linewidth=1.36062998,linecolor=curcolor]
{
\newpath
\moveto(473.38996535,195.84198576)
\lineto(484.5672,79.752096)
}
}
{
\newrgbcolor{curcolor}{0 0 0}
\pscustom[linewidth=1.36062998,linecolor=curcolor]
{
\newpath
\moveto(6.16537323,166.96453191)
\lineto(6.51195591,129.89537537)
\lineto(42.71458394,129.8087108)
}
}
{
\newrgbcolor{curcolor}{0 0 0}
\pscustom[linewidth=1.36062998,linecolor=curcolor]
{
\newpath
\moveto(366.11990551,54.04755175)
\lineto(366.46648819,16.97841411)
\lineto(402.66914646,16.89178734)
}
}
{
\newrgbcolor{curcolor}{0 0 0}
\pscustom[linewidth=1.36062998,linecolor=curcolor]
{
\newpath
\moveto(0.52695307,159.9660601)
\lineto(6.24604346,166.97014072)
\lineto(12.48505323,159.9660601)
}
}
{
\newrgbcolor{curcolor}{0 0 0}
\pscustom[linewidth=1.36062998,linecolor=curcolor]
{
\newpath
\moveto(360.32301732,47.86997443)
\lineto(366.04208504,54.87404372)
\lineto(372.28114016,47.86997443)
}
}
{
\newrgbcolor{curcolor}{0 0 0}
\pscustom[linewidth=1.36062998,linecolor=curcolor]
{
\newpath
\moveto(36.48791055,136.13650923)
\lineto(43.66922457,129.8108538)
\lineto(36.57456378,123.91844561)
}
}
{
\newrgbcolor{curcolor}{0 0 0}
\pscustom[linewidth=1.36062998,linecolor=curcolor]
{
\newpath
\moveto(395.67456378,23.21980498)
\lineto(402.85589291,16.89414576)
\lineto(395.76122835,11.00171112)
}
}
{
\newrgbcolor{curcolor}{0 0 0}
\pscustom[linewidth=2.26771654,linecolor=curcolor]
{
\newpath
\moveto(10.59259767,48.19371742)
\curveto(6.02296471,42.59200403)(1.45496451,36.99228624)(8.78267414,31.15822261)
\curveto(16.1103874,25.32415899)(35.333952,19.25701947)(59.02701354,14.31276246)
\curveto(82.72007509,9.36850545)(110.88159043,5.54735773)(141.66337058,5.00133695)
\curveto(172.44515074,4.45531616)(205.84694173,7.18451301)(234.74607723,11.80103962)
\curveto(263.64520819,16.41761159)(288.04117569,22.92151332)(297.74259855,28.90710576)
\curveto(307.44401688,34.89274356)(302.45112189,40.35916498)(297.45691162,45.82703773)
}
}
{
\newrgbcolor{curcolor}{0 0 0}
\pscustom[linewidth=2.26771654,linecolor=curcolor]
{
\newpath
\moveto(332.44140472,158.44238589)
\curveto(327.35646312,164.91079332)(322.27255106,171.37789455)(329.13500145,176.98484183)
\curveto(335.99745638,182.59179364)(354.80612409,187.33758463)(385.07642079,190.83749669)
\curveto(415.34671748,194.33740422)(457.07737323,196.59124233)(496.28315339,195.38116158)
\curveto(535.48893354,194.17108535)(572.16938457,189.49708573)(599.25957165,185.27706028)
\curveto(626.34980409,181.05703484)(643.8492737,177.29106973)(650.20477606,172.86595654)
\curveto(656.56027843,168.44084787)(651.77126929,163.35654576)(646.98262299,158.27263824)
}
}
{
\newrgbcolor{curcolor}{0 0 0}
\pscustom[linewidth=1.65543311,linecolor=curcolor]
{
\newpath
\moveto(128.56686614,40.77888529)
\curveto(134.5310332,39.26790576)(140.49505512,37.75697159)(146.88093128,37.39078072)
\curveto(153.26681197,37.02477128)(160.07400718,37.80305159)(166.88133392,38.5817401)
}
}
{
\newrgbcolor{curcolor}{0 0 0}
\pscustom[linewidth=0.74834649,linecolor=curcolor]
{
\newpath
\moveto(159.44250557,31.56406828)
\curveto(161.69267906,28.55825537)(163.94283439,25.55244246)(169.68108699,24.66431395)
\curveto(175.41933959,23.77618545)(184.64524951,25.00583206)(193.87132724,26.23543332)
}
}
{
\newrgbcolor{curcolor}{0 0 0}
\pscustom[linestyle=none,fillstyle=solid,fillcolor=curcolor]
{
\newpath
\moveto(161.74778562,31.98108634)
\lineto(156.61712103,35.33825611)
\lineto(158.392952,29.46963341)
\curveto(158.74218406,31.03980248)(160.13863355,32.08519477)(161.74778562,31.98108634)
\closepath
}
}
{
\newrgbcolor{curcolor}{0 0 0}
\pscustom[linewidth=1.65543311,linecolor=curcolor]
{
\newpath
\moveto(476.82078614,156.59390891)
\curveto(483.59405178,157.7677697)(490.36714961,158.94158513)(496.88788687,158.8695171)
\curveto(503.40862413,158.79740372)(509.67642935,157.47949758)(515.9443525,156.16150072)
}
\rput[bl](20,140){$p$}
\rput[bl](380,30){$\xi$}
\rput[bl](165,90){$-\omega_{\min}$}
\rput[bl](438,85){$\ell_{\min}$}
\rput[bl](200,23){$\vartheta$}
\rput[bl](490,135){$\vartheta$}
\rput[bl](-40,-5){$\displaystyle -\frac{1}{\varepsilon}$}
}
\end{pspicture}
\caption{A homogeneous solution in momentum and position space.}
\label{fighom} 
\end{figure}
On the left side, the support of~$\hat{L}_a$ is shown on the lower mass shell.
Given a parameter~$\omega_{\min}$ is in the range
\beq \label{omegaminrange}
\frac{1}{\ell_{\min}} \lesssim \omega_{\min} \leq \frac{1}{\varepsilon} \:,
\eeq
for frequencies smaller than~$-\omega_{\min}$, the function~$\hat{L}_a$ is supported inside
a cone of opening angle
\[ 
\vartheta = \frac{1}{\sqrt{\ell_{\min} \,\omega_{\min}}} \:. \]
This function is smooth in the sense that its derivatives have the scaling behavior
\beq \label{DhatL}
\big| D^s \hat{L}(p) \big| \lesssim \frac{1}{\big( \vartheta\,(\omega_{\min} + |\omega|) \big)^s}
\eeq
(for any~$s \in \N$). 
For frequencies larger than~$-\omega_{\min}$, the support of the function~$\hat{L}_a$ is a bit more
spread out, as indicated by the light gray region on the left of Figure~\ref{fighom}.
Taking the Fourier transform, the corresponding function~$L_a$
can be thought of as a wave packet localized in space on the scale~$\ell_{\min}$, as shown on the
right of Figure~\ref{fighom}.

Next, the potential~$B_a$ in~\eqref{hatBjdefastic} is vectorial. Thus it can be written as
\beq \label{Bavec}
B_a = (A_a)_j\: \gamma^j
\eeq
with potentials~$A_a$ which can be thought of as nonlocal generalizations of an electromagnetic potential.
In a suitable gauge, they satisfy the homogeneous wave equation with an error term,
\beq \label{Aerr}
\Box A_a = \O \big( \ell_{\min} \:D^3 A_a \big) \:,
\eeq
where~$D^3A_a$ denotes the third derivatives of~$A_a$. Thus the error term is of the 
multiplicative order~$\ell_{\min}/\ell_\macro$, where~$\ell_\macro$ is the length scale
on which the potential~$A_a$ varies. These error terms and their scaling behavior are worked out
in detail in~\cite{nonlocal}. In this paper, it is also shown that the functions~$L_a$ and~$A_a$ can
be computed iteratively in an expansion in powers of~$\ell_{\min}/\ell_\macro$.
Here we do not need the details, but it suffices to work with the error term~\eqref{Aerr}.

The symmetry of the nonlocal potential~\eqref{Bsymm} can be expressed more conveniently
by introducing the indefinite inner product
\beq \label{stip}
\begin{split}
\bra .|. \ket &\::\: C^\infty(M, SM) \times C^\infty_0(M, SM) \rightarrow \C \:, \\
&\bra \psi|\phi \ket = \int_M \Sl \psi(x) \,|\, \phi(x) \Sr \: d^4x
\end{split}
\eeq
(where~$C^\infty(M, SM)$ denotes the smooth wave functions in Minkowski space,
and the subscript zero indicates compact support).
This indefinite inner product endows the wave functions with a Krein structure;
we denote the corresponding Krein space by~$(\K, \bra .|. \ket)$.
The symmetry of the nonlocal potential implies that the Dirac operator in~\eqref{dirnonloc}
is symmetric with respect to the Krein inner product.
Taking the Fourier transform and using Plancherel's theorem, the Krein inner product can
also be expressed in momentum space. In what follows, it will sometimes be convenient to consider
wave functions in spacetime as vectors in the Krein space~$(\K, \bra .|. \ket)$.

\section{Dynamics in the Presence of Stochastic Nonlocal Potentials} \label{secstoch}
\subsection{Strategy for Obtaining Bosonic Field Operators} \label{secstrategy}
We consider the Dirac equation~\eqref{dirnonloc} in the presence of a nonlocal potential
of the form~\eqref{Bnonlocal} and~\eqref{hatBjdefastic}.
We again assume that the potential is symmetric (see~\eqref{Bsymm} and~\eqref{symmpot}).
Solutions of the Dirac equation can be constructed with the help of the retarded perturbation series
\beq \label{pertser}
\tilde{\psi} = \sum_{n=0}^{\infty} \big( -s_m^{\wedge}\, \mathscr{B} \big)^n \,\psi \:,
\eeq
where~$s_m^\wedge$ is the retarded Dirac Green's operator defined as follows.
As is common in QFT, we usually prefer to work in momentum space,
denoting the four-momenta by~$p$, $k$ and~$q$. Then the retarded Dirac Green's operator
is the operator of multiplication by the distribution
\beq \label{smwedge}
s_m^\wedge(p) = \lim_{\varepsilon \searrow 0} \frac{\slashed{p} + m}{p^{2}-m^{2}+i \varepsilon p^{0}} \:.
\eeq
The Fourier transform of~$\B$ given by~\eqref{hatBdef} has the corresponding symmetry property
\[ 
\hat{\B}(p,k)^* = \hat{\B}(k,p) \:. \]
Now the operator products in the perturbation series~\eqref{pertser} can be built up
of integral expressions of the form
\beq \label{pertretarded}
\int \frac{d^4 p_2}{(2 \pi)^4}  \:\cdots\:
s_m^{\wedge}(p_1)\: \hat{\B}(p_1, p_2)\: s_m^{\wedge}(p_2)\: \hat{\B}(p_2, p_3)\: \cdots \:.
\eeq
We remark that this perturbation expansion is well-defined to every order, provided that the dynamical potentials
contained in~$\B$ are smooth and fall off sufficiently fast at infinity
(this can be proved similar to~\cite[Lemma~2.1.2]{cfs}, noting that the nonlocal kernels give rise to
additional convolutions). 
Following the usual procedure in
perturbative QFT, we shall not enter the question of convergence of this perturbation expansion.
Instead, we consider the perturbation series as an asymptotic series.
In simple situations (for example, if~$\B$ is a smooth multiplication operator), the left side of~\eqref{pertser}
can be defined even non-perturbatively (for example using the theory of linear hyperbolic systems;
see for example~\cite[Chapter~13]{intro}). But, even in this case, it is not obvious whether the
right side of~\eqref{pertser} converges and coincides with the left side.
 In~\cite{cauchynonloc} it was shown that the series converges
in suitable function spaces, provided that the nonlocal potential~$\B$ is sufficiently small.
Even if this result applies, the subtle issue remains that it is not clear whether the stochastic averages
may be taken order by order in perturbation theory. As already mentioned, here we shall disregard these issues,
taking the naive point of view that the perturbative description should make sense provided that the
dynamical potentials are sufficiently small.

In order to get a connection to bosonic {\em{quantum}} fields, the factors~$\hat{\B}$ in the above perturbation
expansion would have to be replaced by field operators acting on the bosonic Fock space.
Our general strategy is to show that, making use of the nonlocality of the potential~$\hat{B }$,
these factors can indeed be interpreted as such field operators, without the need to ``quantize'' them.
The crucial step for making this strategy work is to show that these operators satisfy the
CCR, as we now make precise.
Decomposing~$\B$ into the Dirac covariants,
\beq \label{vectorscalar}
\B(x,y) = \B_j(x,y)\: \gamma^j + \Phi(x,y)\: \1 \:,
\eeq
we can speak of the {\em{vector}} and {\em{scalar components}} of these fields (the 
pseudo-scalar, axial and bilinear contributions will not be considered here).

Having a Maxwell field in mind, we will be mainly concerned with the vector component.
Nevertheless, for ease of presentation, we begin with the scalar component.
We are aiming at getting a connection to a {\em{massless real scalar field}}.
The corresponding field operators, denoted by~$\phi(x)$, satisfy the CCR
\beq \label{ccr}
\big[ \phi(x), \phi(y) \big] = K(x,y) \:,
\eeq
where~$K(x,y)$ is the {\em{causal fundamental solution}}
\beq \label{Kdef}
K(x,y) := \int \frac{d^4k}{(2 \pi)^4}\: \delta(k^2)\: \epsilon(k^0)\: e^{-ik(x-y)}
= \frac{i}{4 \pi^2}  \:\delta\big( \xi^2 \big)\:\epsilon(\xi^0)
\eeq
(where we set~$\xi := y-x$, and~$\epsilon(t)$ is the usual sign function). Taking the Fourier transform,
\[ 
\phi(x) = \int \frac{d^4q}{(2 \pi)^4} \:\hat{\phi}_q\: e^{-i q x} \qquad \text{with} \qquad
\hat{\phi}_q^* = \hat{\phi}_{-q} \:, \]
the CCR take the form
\beq \label{hatccr}
\big[ \hat{\phi}_q, \hat{\phi}_{q'} \big] = (2 \pi)^4\: \delta^4\big( q + q' \big)\: \delta(q^2)\: \epsilon(q^0) \:,
\eeq
because then
\begin{align*}
\big[ \phi(x), \phi(y) \big] &= 
\int \frac{d^4q}{(2 \pi)^4} \int \frac{d^4q'}{(2 \pi)^4} \big[ \hat{\phi}_q, \hat{\phi}_{q'} \big]\: e^{-i q x - i q' y} \\
&= \int \frac{d^4q}{(2 \pi)^4}\: \delta(q^2)\: \epsilon(q^0) \: e^{-i q x +i q y} = K(x,y) \:,
\end{align*}
as desired.
The coupling of the scalar quantum field to the Dirac field is described again by~\eqref{pertser}, but
with~$\B$ replaced by factors~$\phi$. Likewise, in momentum space,
the operator products in~\eqref{pertretarded} become
\[ s_m^{\wedge}(p_1)\: \hat{\phi}_{p_1-p_2}\: s_m^{\wedge}(p_2)\: \hat{\phi}_{p_2-p_3}\: \cdots \:. \]
This motivates us to {\em{define}} the field operators by
\beq \label{phidefgen}
\hat{\phi}_q(k_L, k_R) := (2 \pi)^4 \,\delta^4(k_L - k_R - q)\: \hat{\Phi}(k_L, k_R) \:.
\eeq
Our goal is to show that, under suitable assumptions on~$\B$, the so-defined field operators
really satisfy the desired commutation relations~\eqref{hatccr}.
To this end, we proceed in several steps. We first try to arrange the commutation relations
by choosing~$\Phi$ as a nonlocal Gaussian field (Section~\ref{secscalex}). Our analysis will
show that~\eqref{hatccr} can be arranged in the statistical mean, but not as an operator equation.
Next, following~\eqref{hatBjdefastic}, we consider instead of one scalar field a multitude of
stochastic vector potentials (Section~\ref{secmultinonloc}).
This gives us more freedom, but, as we shall see, it will not be sufficient for satisfying the
CCR as operator equations. Nevertheless, this analysis will be a preparation for 
the constructions in Section~\ref{secgauge}, where it will be shown that the CCR
hold naturally once holographic mixing is taken into account.

\subsection{Example of a Nonlocal Stochastic Scalar Field} \label{secscalex}
We consider a nonlocal scalar potential. Thus in~\eqref{vectorscalar} we 
set~$\B_j$ to zero and choose~$\Phi$ as
\beq \label{Phiansatz}
\Phi(x,y) = W\Big( \frac{x+y}{2} \Big)\: L(y-x) \:,
\eeq
where the kernel~$L$ is complex-valued and symmetric in the sense that
\[ \overline{L(\xi)} = L(-\xi) \:. \]
Moreover, we choose~$W$ as a real-valued Gaussian stochastic field with mean zero,
\beq \label{WGauss}
\bbra W(x) \kket = 0 \:,\qquad \bbra W(x) \, W(y) \kket = h(y-x) \:,
\eeq
with a covariance~$h(y-x)$ which is real and symmetric, i.e.\
\beq \label{scalsymm}
\overline{h(\xi)} = h(\xi) = h(-\xi) \qquad \text{for all~$\xi \in M$} \:.
\eeq
Transforming to momentum space, we obtain
\beq \label{scalcov}
\bbra \hat{W}(q) \kket = 0 \:,\qquad \bbra \hat{W}(q) \, \hat{W}(q') \kket = 
(2 \pi)^4\: \delta^4(q+q')\: \hat{h}(q) \:,
\eeq
and~\eqref{scalsymm} translates into
\beq \label{scalhprop}
\overline{\hat{h}(q)} = \hat{h}(-q) = \hat{h}(q) \qquad \text{for all~$q \in \hat{M}$} \:.
\eeq
Moreover, being defined as the covariance of a real-valued field, the function~$h$
must have the positivity property that for any test function~$f \in C^\infty_0(M, \R)$,
\beq \label{posarg}
0 \leq \bbra \bigg( \int_M f(x)\: W(x)\: d^4x \bigg)^2 \kket
= \int_M d^4x \int_M d^4y\; h(y-x)\: f(x)\: f(y) \:.
\eeq
Using that convolution in position space corresponds to multiplication in momentum space,
this inequality is equivalent to~$\hat{h}$ being positive,
\beq \label{hpos}
\hat{h}(q) \geq 0 \quad \text{for all~$q \in \hat{M}$} \:.
\eeq

We next transform the nonlocal potential to momentum space by taking the
inverse transformation to~\eqref{hatBdef},
\begin{align}
\Phi(x,y) &= \int \frac{d^4k_L}{(2 \pi)^4} \int \frac{d^4k_R}{(2 \pi)^4}\: \hat{\Phi}\big( k_L, k_r\big)\:
e^{-i k_L x + i k_R y} \notag \\
&= \int \frac{d^4p}{(2 \pi)^4} \int \frac{d^4q}{(2 \pi)^4}\: \hat{\Phi}\Big(p + \frac{q}{2} , p - \frac{q}{2} \Big)\:
e^{i p \xi + i q \zeta } \:, \label{Phirep}
\end{align}
where we set
\[ 
\xi := y-x\:,\qquad \zeta := \frac{y+x}{2}\:. \]
Using~\eqref{Phiansatz}, we obtain
\beq \label{WL}
\hat{\Phi}\Big(p + \frac{q}{2} , p - \frac{q}{2} \Big) = \hat{W}(q)\: \hat{L}(p) \:.
\eeq

\subsubsection{The Statistical Mean of the Commutator}
Taking~\eqref{phidefgen} as the definition of the field operator~$\hat{\phi}_q$,
we would like to satisfy the CCR~\eqref{hatccr}.
The question is whether these commutation relations can be obtained by a suitable
choice of the covariance~$h$.
As an intermediate step toward answering this question, we now compute
the statistical mean of the commutator. First, using~\eqref{phidefgen},
\begin{align*}
\big( \hat{\phi}_{q'} \,\hat{\phi}_{q} \,\hat{\psi} \big)(k+q+q') &=
\int \frac{d^4p}{(2 \pi)^4} \int \frac{d^4p'}{(2 \pi)^4}\:
\hat{\phi}_{q'}(k+q'+q, p) \,\hat{\phi}_{q}(p,p') \,\hat{\psi}(p') \\
&= \hat{\Phi}(k+q+q',k+q)\, \hat{\Phi}(k+q,k) \,\hat{\psi}(k) \:.
\end{align*}
For the operator product on the left side, we also use the short notation
\beq \label{shortnot}
\hat{\phi}_{q'} \,\hat{\phi}_{q} \big|_k\,\hat{\psi}(k) \:.
\eeq
We thus obtain
\begin{align}
& \big[ \hat{\phi}_{q'}, \hat{\phi}_{q}  \big] \big|_k
= \hat{\Phi}(k+q+q',k+q)\, \hat{\Phi}(k+q,k) - \hat{\Phi}(k+q+q',k+q')\, \hat{\Phi}(k+q',k) \notag \\
&\overset{\eqref{WL}}{=}
\hat{W}(q)\, \hat{W}(q') \: \bigg(  \hat{L} \Big( k + q + \frac{q'}{2} \Big) \:
\hat{L} \Big( k + \frac{q}{2} \Big) - \hat{L} \Big( k + q' + \frac{q}{2} \Big) \:
\hat{L} \Big( k + \frac{q'}{2} \Big) \bigg) \:. \label{commphi}
\end{align}
Taking the statistical mean with the help of~\eqref{scalcov} gives
\begin{align*}
& \bbra \big[ \hat{\phi}_{q'}, \hat{\phi}_{q}  \big] \big|_k \kket \\
&= (2 \pi)^4\: \delta^4(q+q')\: \hat{h}(q)
\: \bigg(  \hat{L} \Big( k + \frac{q}{2} \Big) \:
\hat{L} \Big( k + \frac{q}{2} \Big) - \hat{L} \Big( k - \frac{q}{2} \Big) \:
\hat{L} \Big( k - \frac{q}{2} \Big) \bigg) \:.
\end{align*}
In order to get agreement with~\eqref{hatccr}, we need to arrange that
\beq \label{hLcond}
\hat{h}(q) \: \bigg(  \hat{L} \Big( k + \frac{q}{2} \Big)^2 - \hat{L} \Big( k - \frac{q}{2} \Big)^2 \bigg) 
= \delta(q^2)\: \epsilon(q^0) \:.
\eeq
These equations can be solved explicitly, as illustrated by the following simple example.
\begin{Prp} \label{prpccrmean}
Choosing
\beq \label{Lhchoice}
\hat{L}(p) = \Big( C + \frac{p^0}{C} \Big) \qquad \text{and} \qquad
\hat{h}(q) = \frac{1}{2 \,|q^0|} \:\delta(q^2) \:,
\eeq
the field operators~$\hat{\phi}_q$ defined by~\eqref{phidefgen} and~\eqref{WL} for~$W$
the Gaussian field with covariance~\eqref{scalcov} satisfy the CCR
asymptotically for large~$C$; more precisely,
\beq \label{CCRmean}
\bbra \big[ \hat{\phi}_q, \hat{\phi}_{q'} \big] \kket = (2 \pi)^4\: \delta^4\big( q + q' \big)\: \delta(q^2)\: \epsilon(q^0)
\: + \O \Big( \frac{1}{C^2} \Big) \:.
\eeq
\end{Prp}
\Proof By direct computation,
\begin{align*}
&\hat{L}(p)^2 = \Big( C + \frac{p^0}{C} \Big)^2 = C^2 + 2 p^0 + \Big(\frac{p^0}{C} \Big)^2 \\
&\hat{L} \Big( k + \frac{q}{2} \Big)^2 - \hat{L} \Big( k - \frac{q}{2} \Big)^2 = 2 q^0 + \frac{2 kq}{C^2} \:.
\end{align*}
Using this result together with the formula for~$\hat{h}(q)$ in~\eqref{Lhchoice} in~\eqref{hLcond}
gives the result.
\QED
We note that the function~$\hat{h}$ has the desired symmetry properties~\eqref{scalhprop}
and is positive~\eqref{hpos}.
We also remark that the error term in~\eqref{CCRmean} could be avoided by choosing~$\hat{L}(p)$
as a function involving a square root. This has the disadvantage, however, that the computation of the
Fourier transform becomes more difficult. This is why we here prefer to work with~\eqref{Lhchoice}.

\subsubsection{The Commutator in Position Space} \label{secposex}
In order to complete the picture, we proceed by rewriting the above findings to position space.
We first compute the covariance in position space.
\begin{Lemma} The Fourier transform of the distribution~$\hat{h}(q)$ in~\eqref{Lhchoice} is given by
\[ 
h(x) = \frac{1}{16 \pi^2}\:\frac{1}{|\vec{x}|}\: \Theta \big( |\vec{x}| - |x^0| \big) \:. \]
\end{Lemma}
\Proof The following method was already used in~\cite[Section~5]{action}. We first note that,
for any~$\varepsilon>0$,
\begin{align*}
&\int_{-\infty}^\infty \epsilon(\tau)\: e^{-\varepsilon\, |\tau|} \: e^{-i \omega (t-\tau)} d\tau \\
&= \int_{-\infty}^\infty \epsilon(\tau)\: \frac{1}{i \omega - \varepsilon\: \epsilon(\tau)}\:
\frac{d}{d\tau} e^{-\varepsilon\, |\tau|} \: e^{-i \omega (t-\tau)} d\tau
= -\Big( \frac{1}{i \omega - \varepsilon} + \frac{1}{i \omega + \varepsilon} \Big)\: e^{-i \omega t}
\end{align*}
and thus
\[ \lim_{\varepsilon \searrow 0} \int_{-\infty}^\infty \epsilon(\tau)\: e^{-\varepsilon\, |\tau|} \: e^{-i \omega (t-\tau)} d\tau
= 2i\: \frac{\text{PP}}{\omega}\: e^{-i \omega t} \]
(where ``PP'' denotes the principal part). Hence
\begin{align*}
\frac{1}{2 q^0} \:\delta(q^2)\: \epsilon(q^0)\: e^{-i q x} \:
= -\frac{i}{4} \lim_{\varepsilon \searrow 0} \int_{-\infty}^\infty \epsilon(\tau)\: e^{-\varepsilon\, |\tau|}\:
\delta(q^2)\: \epsilon(q^0)\: e^{-i q x + i q^0 \tau}\: d\tau \:.
\end{align*}
Integrating over~$q$ gives
\[ h(x) = \frac{i}{4} \lim_{\varepsilon \searrow 0} \int_{-\infty}^\infty \epsilon(\tau)\: e^{-\varepsilon\, |\tau|}\:
K_0(x^0 - \tau, \vec{x}) \:, \]
where~$K_0$ is again the causal fundamental solution~\eqref{Kdef}, because
\begin{align*}
K_0(x) &= -\int \frac{d^4q}{(2 \pi)^4}\: \delta(q^2)\: \epsilon(q^0)\: e^{-i q x} \\
&= \int \frac{d^4q}{(2 \pi)^4}\: \delta(q^2)\: \epsilon(q^0)\: e^{i q x}
= K(x) = \frac{i}{4 \pi^2}  \:\delta\big( x^2 \big)\:\epsilon(x^0) \:.
\end{align*}
Thus
\[ h(x) = -\frac{1}{16 \pi^2} \lim_{\varepsilon \searrow 0} \int_{-\infty}^\infty \epsilon(\tau)\: e^{-\varepsilon\, |\tau|}\:
\:\delta\Big( (x^0 - \tau)^2 - |\vec{x}|^2 \Big)\:\epsilon(x^0 - \tau) \:, \]
and carrying out the $\tau$-integral gives
\[ h(x) = -\frac{1}{16 \pi^2} \:
\: \frac{1}{2\, |\vec{x}|}\: (-2)\: \Theta \big( |\vec{x}| - |x^0| \big) \:. \]
This concludes the proof.
\QED
Note that the function~$h$ has the desired symmetry properties~\eqref{scalsymm}.

Finally, it is instructive to see how the CCR arise in position space.
To this end, we introduce the field operator~$\phi(x)$ as the Fourier transform of~$\hat{\phi}_q$
in the variable~$q$,
\[ \phi_x(k_L, k_R) = \int \frac{d^4q}{(2 \pi)^4}\: \hat{\phi}_q(k_L, k_R)\: e^{-i q x}\:. \]
Using~\eqref{phidefgen}, \eqref{WL} and~\eqref{Lhchoice}, we obtain
\begin{align*}
\phi_x(k_L, k_R) &= \hat{\Phi}(k_L, k_R)\,e^{-i (k_L-k_R) x} =
\hat{W}(k_L-k_R)\: \hat{L}\Big( \frac{k_L+k_R}{2} \Big) \,e^{-i (k_L-k_R) x} \\
&= \hat{W}(k_L-k_R)\:  \Big( C + \frac{k_L^0+k_R^0}{2C} \Big) \,e^{-i (k_L-k_R) x}\:.
\end{align*}
Next, we also transform the variables~$k_L$ and~$k_R$ to position space by setting
\begin{align*}
&\phi_x(y_L, y_R) 
= \int \frac{d^4k_L}{(2 \pi)^4} \int \frac{d^4k_R}{(2 \pi)^4}\: \phi_x(k_L, k_R)\:
e^{-i k_L y_L + i k_R y_R} \\
&= \int \frac{d^4k_L}{(2 \pi)^4} \int \frac{d^4k_R}{(2 \pi)^4}\: \hat{W}(k_L-k_R)\:
\Big( C + \frac{k_L^0+k_R^0}{2C} \Big)\: e^{-i k_L y_L + i k_R y_R}\, e^{-i (k_L-k_R) x} \:.
\end{align*}
The term linear in~$C$ depends only on the difference of momenta~$k_L-k_R$ and
thus depends to the operator of multiplication by~$W(x)$.
Using that linear factors in momentum space correspond to partial derivatives in position space,
we can carry out the Fourier integrals to obtain the simple formula
\beq \label{phix}
\phi(x) \equiv \phi_x = C \,W(x) + \frac{1}{2C} \: \big\{ i D_t, W(x) \big\} \:,
\eeq
where~$D_t= \partial_t$ is a differential operator.
Now the commutation relations can be verified by direct computation.
\begin{Prp} The field operators~$\phi(x)$ in~\eqref{phix} satisfy the canonical commutation
relations in the statistical mean up to errors of the order~$\O(C^{-2})$, i.e.\
\[ \bbra \big[ \phi(x), \phi(y) \big] \kket = K(x,y) + \O \Big( \frac{1}{C^2} \Big)\:. \]
\end{Prp}
\Proof Using~\eqref{phix}, we obtain
\begin{align}
\big[ \phi(x), \phi(y) \big] &= C^2\, \big[ W(x), W(y) \big]
+\frac{1}{2} \,\big[ W(x), \{i D_t, W(y)\} \big] +\frac{1}{2} \,\big[ \{i D_t, W(x)\} , W(y) \big] \notag \\
&\quad\, +\frac{1}{4C^2}\:\big[ \{i D_t, W(x)\}, \{i D_t, W(y)\} \big]  + \O \Big( \frac{1}{C^2} \Big) \notag \\
&= i \,\big( W(x) \,\dot{W}(y) - W(y)\, \dot{W}(x) \big) + \O \Big( \frac{1}{C^2} \Big)\:. \label{phic}
\end{align}
Taking the statistical mean gives
\begin{align*}
&\bbra \big[ \phi(x), \phi(y) \big] \kket
= i D_{y^0} h(y-x) - i \,D_{x^0} h(y-x) + \O \Big( \frac{1}{C^2} \Big) \\
&= 2i \,\frac{\partial}{\partial \xi^0} h(\xi) + \O \Big( \frac{1}{C^2} \Big)
= 2 i\: \frac{1}{16 \pi^2}\: \frac{1}{|\vec{\xi}|}\: \frac{\partial}{\partial \xi^0}
\Theta \big( |\vec{\xi}| - \xi^0 \big) + \O \Big( \frac{1}{C^2} \Big) \\
&= \frac{i}{8 \pi^2}\: \frac{1}{|\vec{\xi}|}\: \frac{\partial}{\partial \xi^0} \Theta \big( |\vec{\xi}| - |\xi^0| \big)
+ \O \Big( \frac{1}{C^2} \Big)
= \frac{i}{8 \pi^2}\: \frac{1}{|\vec{\xi}|}\: \delta \big( |\vec{\xi}| - \xi^0) \big) \: \epsilon(\xi^0)
+ \O \Big( \frac{1}{C^2} \Big)\\
&= \frac{i}{4 \pi^2}\: \delta \big(\xi^2 \big)\ \epsilon(\xi^0)  + \O \Big( \frac{1}{C^2} \Big)
\overset{\eqref{Kdef}}{=} K(x,y) + \O \Big( \frac{1}{C^2} \Big)\:,
\end{align*}
as desired.
\QED

\subsubsection{Going Beyond the Statistical Mean}
With the constructions so far, we have arranged that the CCR
are satisfied for the {\em{statistical mean}} of the commutator~\eqref{CCRmean}.
However, in order to get a connection to QFT, we need to make sure that
the CCR hold without taking the statistical mean (because the statistical mean is taken only when computing
the expectation value of an observable in a quantum measurement).
To state it differently, we need to make sure that the CCR also hold when taking the statistical mean
of composite expressions, i.e.\
\beq \label{CCRfull}
\begin{split}
\bbra &\phi(x_1) \cdots \phi(x_p) \;\big[ \phi(x), \phi(y) \big] \;
\phi(x_{p+1}) \cdots \phi(x_q) \kket \\
&= K(x,y)\; \bbra \phi(x_1) \cdots \phi(x_p) \:
\phi(x_{p+1}) \cdots \phi(x_q) \kket \:,
\end{split}
\eeq
where the dots stand for any combination of field operators.
Let us verify whether this relation is satisfied.
Applying the Wick rules gives pairings of the field operators.
If the two field operators inside the commutator are paired with each other,
we get the statistical mean as computed in Proposition~\ref{prpccrmean}.
But we also need to take into account the contributions when the operators inside the commutator
are paired with operators outside. In order to analyze these contributions in a clear setting, it is
convenient to consider the statistical mean of the combination
\[ \big[ \phi(x), \phi(y) \big] \otimes \phi(x_1) \otimes \phi(x_2) \:, \]
where the tensor product means that we do not specify what these operators act on or are multiplied by.
In this formulation, our task is to show that
\beq \label{CCRtensor}
\bbra \big[ \phi(x), \phi(y) \big] \otimes \phi(x_1) \otimes \phi(x_2) \kket 
= K(x,y) \; \1 \otimes \bbra  \phi(x_1) \otimes \phi(x_2) \kket
\eeq
(possibly up to certain error terms).
This amounts to showing that the contributions by pairings of operators inside the commutator
with operators outside are negligible.
We refer to relations of the form~\eqref{CCRtensor} that we the CCR are
satisfied {\em{in the operator sense}}. 

Let us evaluate the condition~\eqref{CCRtensor} in the concrete example of Proposition~\ref{prpccrmean}.
Using~\eqref{phix} and~\eqref{phic},
\begin{align*}
&\bbra \big[ \phi(x), \phi(y) \big] \otimes \phi(x_1) \otimes \phi(x_2) \kket \\
&= - i \,C^2\, \bbra \big( W(x) \,\dot{W}(y) - W(y)\, \dot{W}(x) \big) \;W(x_1)\: W(x_2)\;
\1 \otimes \1 \otimes \1 \kket + \O (C) \:,
\end{align*}
and applying the Wick rules with covariance~\eqref{WGauss} gives
\begin{align}
&\bbra \big[ \phi(x), \phi(y) \big] \otimes \phi(x_1) \otimes \phi(x_2) \kket = C^2\, \Big( h\big( x_1 - x_2 \big) \, K(x,y) \Big) \label{good} \\
&\quad\:- i \,C^2\: \Big( - h\big( x_1 - x\big) \: \dot{h}\big( x_2 - y\big) - h\big( x_2 - x\big) \: \dot{h}\big( x_1 - y\big)
\label{bad1} \\
&\qquad \qquad\quad
+ \dot{h}\big( x_1 - x\big) \: h\big( x_2 - y\big) + \dot{h}\big( x_2 - x\big) \: h\big( x_1 - y\big) \Big) + \O (C)\:.
\label{bad2}
\end{align}
The terms in the last two lines violate the CCR.

\subsection{A Multitude of Nonlocal Vector Potentials} \label{secmultinonloc}
In the previous section we saw that with a classical nonlocal potential with Gaussian distribution
we can realize the CCR hold in the statistical mean~\eqref{prpccrmean}.
But it was impossible to arrange the CCR as operator equations
(see the consideration after~\eqref{CCRfull}).
In order to improve the situation, we proceed in several steps. In this section, instead of a single
potential we consider a multitude of potentials labeled by~$a \in \{1, \ldots, N\}$.
Moreover, we shall work with vector potentials denoted by~$A_a^j$
(with~$j$ a tensor index).
This ansatz reflects the structure of the solutions of the linearized fields
in Minkowski space as discovered in~\cite{nonlocal}.
In Section~\ref{secgauge} we will proceed by building in holographic phases.

Our starting point is again the nonlocal Dirac equation introduced in Section~\ref{secdir}.
We thus consider the Dirac equation~\eqref{dirnonloc} with a nonlocal potential~$\B$ of the
form~\eqref{hatBjdefastic}, where the kernels~$L_a$ are complex-valued and symmetric~\eqref{symmpot}.
Before moving on, we point out that this ansatz by itself is no loss of generality, but that all the
constraints on the form of the potential are imposed merely by the 
the scalings~\eqref{Nscaleintro} and~\eqref{DhatL}.
\begin{Lemma} \label{lemmaapprox}
Every nonlocal potential~$\B(x,y)$ can be approximated
by the ansatz~\eqref{hatBjdefasticintro}, with an error going to zero if~$N \rightarrow \infty$.
\end{Lemma}
\Proof We represent the nonlocal potential similar to~\eqref{Phirep} as
\[ \B(x,y) = \int \frac{d^4p}{(2 \pi)^4} \int \frac{d^4q}{(2 \pi)^4}\: \hat{\B}\Big(p + \frac{q}{2} , p - \frac{q}{2} \Big)\:
e^{i p \xi + i q \zeta } \]
(with~$\hat{\B}$ as in~\eqref{hatBdef}). Now we approximate the integrand by a sum
\[ \hat{\B}\Big(p + \frac{q}{2} , p - \frac{q}{2} \Big) = \sum_{a=1}^N
\hat{W}_a(q)\: \hat{L}_a(p) \:. \]
Similar to a Riemann sum, the error of this approximation can be made arbitrarily small by
choosing the functions~$\hat{L}_a$ to be supported in small cubes and letting the length of the
sides of the cubes tend to zero and~$N$ to infinity. Transforming back to position space gives the ansatz~\eqref{hatBjdefasticintro}.
\QED

Next, we specialize the setting by assuming that the potentials~$B_a$ are vectorial~\eqref{Bavec},
and that the corresponding potentials~$A^j_a$ are real-valued.
This ansatz ensures in particular that the nonlocal potential is symmetric~\eqref{Bsymm}.
This guarantees that a nonlocal version of current conservation holds
(for details see~\cite[Proposition~B.1]{baryogenesis}).
We remark that this setup is similar to that in~\cite{collapse}, except that here
we restrict attention to the linear dynamics.

In analogy to~\eqref{WGauss} we treat the potentials~$A_a$ as Gaussian fields, i.e.\
\[ 
\bbra A^j_a(x) \kket = 0 \:,\qquad \bbra A^j_a(x) \, A^k_b(y) \kket = h_{a,b}^{jk}(y-x) \]
with covariance matrices~$h^{jk}_{a,b}(y-x)$ which are real and symmetric, i.e.\
\beq \label{vecsymm}
\overline{h^{jk}_{a,b}(\xi)} = h^{jk}_{a,b}(\xi) = h^{kj}_{b,a}(-\xi)  \quad \text{for all~$\xi \in M$} \:.
\eeq
These covariances will be specified below.
Transforming to momentum space, similar to~\eqref{scalcov} we obtain
\beq \label{covariancevector}
\bbra \hat{A}^j_a(q) \kket = 0 \:,\qquad \bbra \hat{A}^j_a(q) \, \hat{A}^k_b(q') \kket = 
(2 \pi)^4\: \delta^4(q+q')\: \hat{h}^{jk}_{a,b}(q) \:,
\eeq
and~\eqref{vecsymm} translates into
\beq \label{covsymm}
\overline{\hat{h}^{jk}_{a,b}(q)} = \hat{h}^{jk}_{a,b}(-q) = \hat{h}^{kj}_{b,a}(q) \quad \text{for all~$q \in \hat{M}$} \:.
\eeq
Moreover, extending the positivity argument~\eqref{posarg} to the vector-valued case,
we find similarly to~\eqref{hpos} that the matrices~$\hat{h}_{a,b}^{jk}$ must be positive semi-definite, i.e.\
\beq \label{covpos}
\sum_{j,k,a,b} \hat{h}^{jk}_{a,b}(q)\: \overline{u_j^a} \,u_k^b \geq 0 \qquad \text{for all~$q \in \hat{M}$ and~$u \in M \times \C^N \simeq \C^{4N}$} \:.
\eeq

Using~\eqref{hatBjdefastic} and~\eqref{Bavec}, similar to~\eqref{WL},
in momentum space the potential takes the form
\beq \label{Bvec}
\hat{\B}\Big(p + \frac{q}{2} , p - \frac{q}{2} \Big) = \sum_{a=1}^N \,\, \hat{\!\!\slashed{A}}_a(q)\: \hat{L}_a(p) \:.
\eeq
The symmetry of the potential means that
\[ \hat{\B}(k_L, k_R)^* = \hat{\B}(k_R, k_L) \:, \]
and therefore
\[ \hat{A}(q)^* = \hat{A}(-q) \qquad \text{and} \qquad \hat{L}_a(p)^* = \hat{L}_a(p) \:. \]

Keeping in mind that we now have a vector potential, we
define the corresponding field operators~$\hat{\B}^j_q$ in analogy to~\eqref{phidefgen} by
\beq \label{hatAdef}
\hat{\B}^j_q(k_L, k_R) := (2 \pi)^4 \,\delta^4(k_L - k_R - q)\: \hat{\B}^j(k_L, k_R) \:,
\eeq
where~$\hat{\B}^j$ are the vector components of~\eqref{Bvec}, i.e.\
\[ \hat{\B}^j\Big(p + \frac{q}{2} , p - \frac{q}{2} \Big) := \sum_{a=1}^N \,\, \hat{A}^j_a(q)\: \hat{L}_a(p) \:. \]
Using~\eqref{Bvec} we obtain
\beq \label{hatA}
\hat{\B}^j_q(k_L, k_R) = (2 \pi)^4\,\delta^4(k_L - k_R - q) \sum_{a=1}^N 
\hat{A}^j_a(q)\: \hat{L}_a \Big( \frac{k_L+k_R}{2} \Big) \:.
\eeq

\subsubsection{Computation of the Commutator in Momentum Space} \label{secccrmom}
Similar to~\eqref{CCRtensor} our goal is to compute the statistical mean of the following tensor product,
\beq \label{goaltensor}
\bbra \big[ \hat{\B}^i_q, \hat{\B}^j_{q'} \big]
\big|_p \otimes \hat{\B}^k_r \big|_{k} \otimes \hat{\B}^l_{r'} \big|_{k'} \kket \:,
\eeq
where we again used the notation~\eqref{shortnot}.
We begin by computing the commutator in momentum space similar to~\eqref{commphi},
\begin{align*}
\big[ \hat{\B}^k_{q'}, \hat{\B}^j_q \big] \big|_k
&= \hat{\B}^k(k+q+q',k+q)\, \hat{\B}^j(k+q,k) - \hat{\B}^j(k+q+q',k+q')\, \hat{\B}^k(k+q',k) \\
&= \sum_{a,b=1}^N  \hat{A}^j_a(q)\, \hat{A}^k_b(q') \\
&\qquad \times \: \bigg(  \hat{L}_b \Big( k + q + \frac{q'}{2} \Big) \:
\hat{L}_a \Big( k + \frac{q}{2} \Big) - \hat{L}_a \Big( k + q' + \frac{q}{2} \Big) \:
\hat{L}_b \Big( k + \frac{q'}{2} \Big) \bigg) \:.
\end{align*}

We proceed with a Taylor expansion of the functions~$\hat{L}_a$ and~$\hat{L}_b$
in~$q$ and~$q'$ (this expansion is admissible in view of the smoothness assumption~\eqref{DhatL}),
\begin{align*}
&\big[ \hat{\B}^k_{q'}, \hat{\B}^j_q \big] \big|_k + \O \big( q^j q^k \big)
+ \O \big( q'^j q^k \big) + \O \big( q'^j q'^k \big) \\
&= \sum_{a,b=1}^N  \hat{A}^j_a(q)\, \hat{A}^k_b(q')\;
q^l \,\bigg( \big( \partial_l \hat{L}_b \big)\: \hat{L}_a + \frac{1}{2}\: \hat{L}_b\: \big(\partial_l \hat{L}_a\big)
- \frac{1}{2}\: \big(\partial_l \hat{L}_a \big)\: \hat{L}_b \bigg) \bigg|_k\\
&\quad+ \sum_{a,b=1}^N  \hat{A}^j_a(q)\, \hat{A}^k_b(q')\;
q'^l \,\bigg( \frac{1}{2}\:\big( \partial_l \hat{L}_b \big)\: \hat{L}_a - \big(\partial_l \hat{L}_a \big)\: \hat{L}_b
- \frac{1}{2}\: \hat{L}_a\: \big(\partial_l \hat{L}_b \big) \bigg) \bigg|_k \\
&= \sum_{a,b=1}^N  \hat{A}^j_a(q)\, \hat{A}^k_b(q')\:\Big( 
\big( q^l \,\partial_l \hat{L}_b \big)\: \hat{L}_a - \big( q'^l\, \partial_l \hat{L}_a \big)\: \hat{L}_b \Big) \\
&= \sum_{a,b=1}^N  \Big( q^l \hat{A}^j_a(q)\, \hat{A}^k_b(q') - \hat{A}^j_b(q)\, q'^l \hat{A}^k_a(q') \Big)\:
\big( \partial_l \hat{L}_b \big)\: \hat{L}_a \:.
\end{align*}
Noting that each partial derivative of~$\hat{L}_b$ yields in position space a scaling
actor~$(y-x)_l \sim \ell_{\min}$, we can write the error terms in the shorter form
\begin{align}
\big[ \hat{\B}^k_{q'}, \hat{\B}^j_q \big] \big|_k
&= \sum_{a,b=1}^N  \Big( q^l \hat{A}^j_a(q)\, \hat{A}^k_b(q') - \hat{A}^j_b(q)\, q'^l \hat{A}^k_a(q') \Big)\:
\big( \partial_l \hat{L}_b \big)\: \hat{L}_a \label{comm1} \\
&\qquad \times \Big( 1 + \O \big( q \ell_{\min} \big) + \O \big( q' \ell_{\min} \big) \Big) \:. \notag
\end{align}
When substituting this formula into~\eqref{goaltensor} and writing the two last factors in this
equation with the help of~\eqref{hatA}, we can take the statistical mean by forming Gaussian
pairings using~\eqref{covariancevector}. When doing so, we get two different types of contributions:
Those where the two factors~$\hat{A}_a$ and~$\hat{A}_b$ in~\eqref{comm1} are paired with each
other (so-called {\em{inner pairings}}), and those where each of these factors is paired with one of
the additional factors in the tensor product~\eqref{goaltensor} ({\em{outer pairings}}).
For the contributions involving the inner pairings, the statistical mean factorizes as
\[ \bbra \big[ \hat{\B}^i_q, \hat{\B}^j_{q'} \big]
\big|_p \otimes \hat{\B}^l_r \big|_{k} \otimes \hat{\B}^{l'}_{r'} \big|_{k'} \kket
= \bbra \big[ \hat{\B}^i_q, \hat{\B}^j_{q'} \big]\big|_p \kket \otimes
\bbra  \hat{\B}^l_r \big|_{k} \otimes \hat{\B}^{l'}_{r'} \big|_{k'} \kket \:. \]
Therefore, it suffices to arrange that the statistical mean of the commutator is of the desired form
(similar to what was accomplished for the scalar field in Proposition~\ref{prpccrmean} above).
Similar to the contribution~\eqref{good} for the scalar field, the contributions involving inner pairings
are the good terms which realize the CCR.
For the contributions involving outer pairings, however, we are not allowed to take the statistical mean
of the commutator. Similar to the contributions~\eqref{bad1} and~\eqref{bad2} for the scalar field,
these contributions are {\em{not}} of the desired form.

Having a multitude of fields makes it possible to distinguish the inner and outer pairings also in a different way.
For simplicity, we assume that the covariance~$\hat{h}^{j,k}_{ab}(q)$ in~\eqref{covariancevector} has
is diagonal in the indices~$ab$, i.e.\
\beq \label{covariancevector2}
\hat{h}^{j,k}_{ab}(q) = \delta_{ab}\: \hat{h}^{j,k}_a(q)
\eeq
with new functions~$\hat{h}^{j,k}_a(q)$ (this can indeed be arranged by diagonalizing the covariance,
as will be explained in detail in the proof of Theorem~\ref{thmccr}] below).
Then, in view of the factor~$\delta_{ab}$ in~\eqref{covariancevector2}, for the inner pairings the two indices~$a$
and~$b$ in~\eqref{comm1} coincide, whereas for outer parings these two indices will in general be different.
In particular, the contributions with inner pairings become large compared to the contributions with
outer pairings if
\beq \label{Labsmall}
\big( \partial_l \hat{L}_b \big)\: \hat{L}_a  \ll \hat{L}_a \hat{L}_a \qquad \text{for all~$a$ and~$b \neq a$}\:.
\eeq
Here it is a subtle point to give the symbol~$\ll$ a precise meaning. Before explaining this point, let us
clarify the general mechanism by specifying the scalings if for simplicity we assume that the left side
of~\eqref{Labsmall} is zero for all~$a \neq b$. In this case, the contributions with inner pairings scale like
\[ \sum_{a,b} \big( \cdots \big)_a \otimes (\cdots )_b \otimes (\cdots )_b \sim N^2 \:, \]
whereas the contributions with outer pairings scale like
\[ \sum_{a,b} \big( \delta_{ab} \cdots \big) \otimes (\cdots)_a \otimes (\cdots)_b \sim N \:. \]
Therefore, the contributions with outer pairings are smaller by a scaling factor~$1/N$.

In order to specify what we mean by~\eqref{Labsmall}, we first have a closer look at the form
the nonlocal potentials as shown in Figure~\ref{fighom}. As explained in Section~\ref{secdir}, the
function~$\hat{L}_a$ is supported in the gray region on the left of Figure~\ref{fighom}.
Likewise, the product~$\hat{L}_a \hat{L}_b$ is supported in the intersection of the corresponding
gray regions, as shown in Figure~\ref{figLaLb}.
\begin{figure}
\psset{xunit=.5pt,yunit=.5pt,runit=.5pt}
\begin{pspicture}(365.28684601,227.67940972)
{
\newrgbcolor{curcolor}{0.89803922 0.89803922 0.89803922}
\pscustom[linestyle=none,fillstyle=solid,fillcolor=curcolor]
{
\newpath
\moveto(153.9075326,195.84358358)
\lineto(156.81292724,189.4517335)
\lineto(159.42777449,181.31664436)
\lineto(159.42777449,175.07007177)
\lineto(157.82980535,166.93498263)
\lineto(154.77915213,158.21881759)
\lineto(148.38729449,143.69187082)
\lineto(139.67112945,128.43858956)
\lineto(132.69820346,117.10757728)
\lineto(112.36048252,86.7462809)
\lineto(85.04984315,47.66880814)
\lineto(60.93512315,14.69265318)
\lineto(65.87427402,13.82104499)
\lineto(77.93163591,29.65540594)
\lineto(102.19162583,63.06736688)
\lineto(125.72526992,93.42867082)
\lineto(134.15089134,103.40973035)
\lineto(142.7855811,113.44041129)
\lineto(153.80478236,125.5579924)
\lineto(165.61412787,138.18224562)
\lineto(177.15063307,155.31342673)
\lineto(194.14715339,175.21533035)
\closepath
}
}
{
\newrgbcolor{curcolor}{0 0 0}
\pscustom[linewidth=0,linecolor=curcolor]
{
\newpath
\moveto(153.9075326,195.84358358)
\lineto(156.81292724,189.4517335)
\lineto(159.42777449,181.31664436)
\lineto(159.42777449,175.07007177)
\lineto(157.82980535,166.93498263)
\lineto(154.77915213,158.21881759)
\lineto(148.38729449,143.69187082)
\lineto(139.67112945,128.43858956)
\lineto(132.69820346,117.10757728)
\lineto(112.36048252,86.7462809)
\lineto(85.04984315,47.66880814)
\lineto(60.93512315,14.69265318)
\lineto(65.87427402,13.82104499)
\lineto(77.93163591,29.65540594)
\lineto(102.19162583,63.06736688)
\lineto(125.72526992,93.42867082)
\lineto(134.15089134,103.40973035)
\lineto(142.7855811,113.44041129)
\lineto(153.80478236,125.5579924)
\lineto(165.61412787,138.18224562)
\lineto(177.15063307,155.31342673)
\lineto(194.14715339,175.21533035)
\closepath
}
}
{
\newrgbcolor{curcolor}{0.89803922 0.89803922 0.89803922}
\pscustom[linestyle=none,fillstyle=solid,fillcolor=curcolor]
{
\newpath
\moveto(192.85865575,132.28684027)
\lineto(183.92533039,131.29540594)
\lineto(164.54178142,131.96187649)
\lineto(163.82399622,132.41210893)
\lineto(143.46252472,133.90081948)
\lineto(150.26117669,151.25902169)
\lineto(151.79930835,159.05595271)
\lineto(149.74838929,170.61448499)
\lineto(144.51075024,183.12085098)
\lineto(141.20347087,183.67915901)
\lineto(183.43222677,225.68242074)
\lineto(224.72201575,183.95681066)
\lineto(215.87188913,178.41234956)
\lineto(210.92258646,172.58563334)
\lineto(207.55910173,168.14149476)
\lineto(203.47382551,161.73358169)
\lineto(199.83035339,153.79147901)
\lineto(197.28808441,147.36003082)
\lineto(195.24760819,139.8987483)
\closepath
}
}
{
\newrgbcolor{curcolor}{0 0 0}
\pscustom[linewidth=0,linecolor=curcolor]
{
\newpath
\moveto(192.85865575,132.28684027)
\lineto(183.92533039,131.29540594)
\lineto(164.54178142,131.96187649)
\lineto(163.82399622,132.41210893)
\lineto(143.46252472,133.90081948)
\lineto(150.26117669,151.25902169)
\lineto(151.79930835,159.05595271)
\lineto(149.74838929,170.61448499)
\lineto(144.51075024,183.12085098)
\lineto(141.20347087,183.67915901)
\lineto(183.43222677,225.68242074)
\lineto(224.72201575,183.95681066)
\lineto(215.87188913,178.41234956)
\lineto(210.92258646,172.58563334)
\lineto(207.55910173,168.14149476)
\lineto(203.47382551,161.73358169)
\lineto(199.83035339,153.79147901)
\lineto(197.28808441,147.36003082)
\lineto(195.24760819,139.8987483)
\closepath
}
}
{
\newrgbcolor{curcolor}{0.65098041 0.65098041 0.65098041}
\pscustom[linestyle=none,fillstyle=solid,fillcolor=curcolor]
{
\newpath
\moveto(142.68218079,185.5660957)
\lineto(183.35786457,226.0195168)
\lineto(223.36672252,186.01065129)
\lineto(214.92041953,179.34250263)
\lineto(204.25138394,172.00753948)
\lineto(196.02733984,166.2284831)
\lineto(196.02733984,166.2284831)
\lineto(188.39322709,159.27234578)
\lineto(178.39575307,150.74508909)
\lineto(163.20351496,136.23894862)
\lineto(160.65514583,147.31457822)
\lineto(155.95045039,162.89887806)
\lineto(150.95170394,172.40627586)
\closepath
}
}
{
\newrgbcolor{curcolor}{0 0 0}
\pscustom[linewidth=0,linecolor=curcolor]
{
\newpath
\moveto(142.68218079,185.5660957)
\lineto(183.35786457,226.0195168)
\lineto(223.36672252,186.01065129)
\lineto(214.92041953,179.34250263)
\lineto(204.25138394,172.00753948)
\lineto(196.02733984,166.2284831)
\lineto(196.02733984,166.2284831)
\lineto(188.39322709,159.27234578)
\lineto(178.39575307,150.74508909)
\lineto(163.20351496,136.23894862)
\lineto(160.65514583,147.31457822)
\lineto(155.95045039,162.89887806)
\lineto(150.95170394,172.40627586)
\closepath
}
}
{
\newrgbcolor{curcolor}{0.89803922 0.89803922 0.89803922}
\pscustom[linestyle=none,fillstyle=solid,fillcolor=curcolor]
{
\newpath
\moveto(165.55455027,132.17199267)
\lineto(179.15233194,131.83716886)
\lineto(192.53696277,131.22110285)
\lineto(183.92810487,70.20958637)
\lineto(178.80475631,1.19963412)
\lineto(174.20149493,1.23228924)
\lineto(171.77124661,46.12785182)
\curveto(170.91251445,61.99175818)(169.6739007,77.83281844)(168.05647113,93.63740218)
\lineto(164.25641681,130.76934106)
\curveto(164.17832526,131.53245067)(164.78768783,132.19087557)(165.55455027,132.17199267)
\closepath
}
}
{
\newrgbcolor{curcolor}{0.74901962 0.74901962 0.74901962}
\pscustom[linewidth=0,linecolor=curcolor]
{
\newpath
\moveto(165.55455027,132.17199267)
\lineto(179.15233194,131.83716886)
\lineto(192.53696277,131.22110285)
\lineto(183.92810487,70.20958637)
\lineto(178.80475631,1.19963412)
\lineto(174.20149493,1.23228924)
\lineto(171.77124661,46.12785182)
\curveto(170.91251445,61.99175818)(169.6739007,77.83281844)(168.05647113,93.63740218)
\lineto(164.25641681,130.76934106)
\curveto(164.17832526,131.53245067)(164.78768783,132.19087557)(165.55455027,132.17199267)
\closepath
}
}
{
\newrgbcolor{curcolor}{0 0 0}
\pscustom[linewidth=2.72125995,linecolor=curcolor]
{
\newpath
\moveto(12.41191181,52.97372279)
\lineto(183.72464882,225.74505129)
\lineto(357.25615748,50.04069602)
}
}
{
\newrgbcolor{curcolor}{0 0 0}
\pscustom[linewidth=1.43999814,linecolor=curcolor]
{
\newpath
\moveto(104.2820541,146.17309771)
\curveto(113.95211082,141.20226487)(123.62024632,136.23242258)(145.11159186,133.74322089)
\curveto(166.6029374,131.2540192)(199.91541936,131.24476692)(221.36106406,133.82066795)
\curveto(242.80670332,136.39656897)(252.38415012,141.55716614)(261.96123772,146.71756737)
}
}
{
\newrgbcolor{curcolor}{0 0 0}
\pscustom[linewidth=1.36062992,linecolor=curcolor]
{
\newpath
\moveto(174.12922915,1.86628291)
\curveto(172.78204226,31.13593608)(171.43495877,60.40324896)(169.23947354,86.68240084)
\curveto(167.04398287,112.96149828)(164.00116263,136.24119592)(159.51377794,151.99863471)
\curveto(155.02639869,167.75607895)(149.09861306,175.97943774)(143.1665605,184.20871255)
}
}
{
\newrgbcolor{curcolor}{0 0 0}
\pscustom[linewidth=1.36062992,linecolor=curcolor]
{
\newpath
\moveto(179.13929167,1.47333299)
\curveto(181.37104585,36.27503529)(183.60331162,71.0847381)(187.16762865,98.40526169)
\curveto(190.73194568,125.72578528)(195.62505388,145.53801553)(201.93550851,158.71984174)
\curveto(208.24595769,171.90166796)(215.96957344,178.44272759)(223.6948927,184.98523493)
}
}
{
\newrgbcolor{curcolor}{0 0 0}
\pscustom[linewidth=1.63275595,linecolor=curcolor]
{
\newpath
\moveto(7.39845921,195.79945003)
\lineto(7.81435843,151.31646295)
\lineto(51.25751055,151.21245035)
}
}
{
\newrgbcolor{curcolor}{0 0 0}
\pscustom[linewidth=1.63275595,linecolor=curcolor]
{
\newpath
\moveto(0.63235276,187.4012831)
\lineto(7.49526047,195.80618137)
\lineto(14.9820737,187.4012831)
}
}
{
\newrgbcolor{curcolor}{0 0 0}
\pscustom[linewidth=1.63275595,linecolor=curcolor]
{
\newpath
\moveto(43.78550173,158.80582358)
\lineto(52.4030778,151.21503554)
\lineto(43.88948787,144.14414421)
}
}
{
\newrgbcolor{curcolor}{0 0 0}
\pscustom[linewidth=2.72125984,linecolor=curcolor]
{
\newpath
\moveto(12.71112703,53.2744734)
\curveto(7.22756748,46.55241734)(1.74596724,39.83275599)(10.5392188,32.83187964)
\curveto(19.33247471,25.83100329)(42.40075223,18.55043586)(70.83242608,12.61732745)
\curveto(99.26409993,6.68421904)(133.05791834,2.09884178)(169.99605453,1.44361684)
\curveto(206.93419071,0.78839189)(247.01633991,4.06342811)(281.6953025,9.60326005)
\curveto(316.37425965,15.14314641)(345.64942065,22.94782849)(357.29112809,30.13053942)
\curveto(368.93283009,37.31330477)(362.94135609,43.87301047)(356.94830377,50.43445778)
}
}
{
\newrgbcolor{curcolor}{0 0 0}
\pscustom[linewidth=1.36062992,linecolor=curcolor]
{
\newpath
\moveto(65.74257078,13.38836921)
\curveto(93.21858716,50.49231177)(120.70094408,87.60479907)(142.92957687,113.59747848)
\curveto(165.15821511,139.59016877)(182.11737857,154.44628821)(194.69026847,164.44963394)
\curveto(207.26315837,174.45298513)(215.44022309,179.59572538)(223.61908929,184.73959769)
}
}
{
\newrgbcolor{curcolor}{0 0 0}
\pscustom[linewidth=1.36062992,linecolor=curcolor]
{
\newpath
\moveto(61.26888491,14.9271328)
\curveto(82.14154658,43.74478471)(103.01748465,72.56695391)(119.13839123,95.92058817)
\curveto(135.25930326,119.27424965)(146.61677511,137.14680399)(149.89870156,150.84616484)
\curveto(153.18063345,164.5455257)(148.38507152,174.05944741)(143.58605904,183.58021037)
}
}
{
\newrgbcolor{curcolor}{0 0 0}
\pscustom[linewidth=0.99999871,linecolor=curcolor]
{
\newpath
\moveto(191.90036787,202.09582925)
\curveto(196.54713071,207.47788783)(201.19379906,212.85984058)(207.45327496,215.38461035)
\curveto(213.71274709,217.90938011)(221.58452031,217.57676279)(229.45645984,217.24413791)
}
}
{
\newrgbcolor{curcolor}{0 0 0}
\pscustom[linestyle=none,fillstyle=solid,fillcolor=curcolor]
{
\newpath
\moveto(194.93464924,201.32569206)
\lineto(187.78326772,197.32724625)
\lineto(190.6959088,204.98533665)
\curveto(191.01553081,202.85978005)(192.77990652,201.33645299)(194.93464924,201.32569206)
\closepath
}
}
{
\newrgbcolor{curcolor}{0 0 0}
\pscustom[linewidth=0.99999871,linecolor=curcolor]
{
\newpath
\moveto(182.28339402,77.84804846)
\curveto(191.79893291,74.58726862)(201.31428661,71.3265568)(207.54754394,64.84047554)
\curveto(213.78080126,58.35439428)(216.73146709,48.64346893)(219.68219717,38.93233948)
}
}
{
\newrgbcolor{curcolor}{0 0 0}
\pscustom[linestyle=none,fillstyle=solid,fillcolor=curcolor]
{
\newpath
\moveto(182.70010299,74.74541598)
\lineto(176.32361747,79.89034154)
\lineto(184.51547462,80.04299513)
\curveto(182.66235284,78.95391811)(181.9067044,76.74880078)(182.70010299,74.74541598)
\closepath
}
}
{
\newrgbcolor{curcolor}{0 0 0}
\pscustom[linewidth=0.99999871,linecolor=curcolor]
{
\newpath
\moveto(106.18820031,61.31718484)
\curveto(109.34235969,59.63867397)(112.49645858,57.96019712)(114.89069858,54.52845035)
\curveto(117.28494236,51.09670358)(118.91913827,45.91196279)(120.55336441,40.72711617)
}
}
{
\newrgbcolor{curcolor}{0 0 0}
\pscustom[linestyle=none,fillstyle=solid,fillcolor=curcolor]
{
\newpath
\moveto(106.1087127,58.18770303)
\lineto(100.62667191,64.27679618)
\lineto(108.73947834,63.13128383)
\curveto(106.73737264,62.34915733)(105.64231644,60.29139182)(106.1087127,58.18770303)
\closepath
}
\rput[bl](25,165){$p$}
\rput[bl](190,18){$\supp \hat{L}_a$}
\rput[bl](88,20){$\supp \hat{L}_b$}
\rput[bl](235,208){$\supp \hat{L}_a \cap \supp \hat{L}_b$}
\rput[bl](272,140){$-\omega_{\min}$}
\rput[bl](375,20){$\displaystyle -\frac{1}{\varepsilon}$}
}
\end{pspicture}
\caption{Estimate of~$(\partial_l \hat{L}_b) \,\hat{L}_a$.}
\label{figLaLb}
\end{figure}
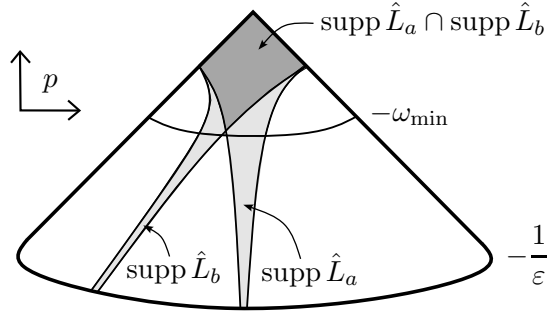%
The derivative in~\eqref{Labsmall} gives a scaling factor of one over the size of this
intersection region, i.e.\
\beq \label{small}
(\partial_l \hat{L}_b)\, \hat{L}_a \simeq \frac{1}{\omega_{\min}}\: \hat{L}_b \hat{L}_a \qquad \text{if~$b \neq a$}\:.
\eeq
In the case~$a=b$, however, the function~$\hat{L}_a^2$ is supported in the whole shaded region
in Figure~\ref{fighom}. Since this shaded region has a spike-like form for large~$|\omega|$,
the derivative only gives a scaling factor~$\ell_{\min}$,
\beq \label{large}
(\partial_l \hat{L}_a)\, \hat{L}_a \simeq \ell_{\min}\: \hat{L}_a \hat{L}_a \:.
\eeq
According to~\eqref{omegaminrange}, the parameter~$\omega_{\min}$ is typically much larger than~$1/\ell_{\min}$,
in which case~\eqref{small} really is much smaller than~\eqref{large}, as desired.

This consideration points to a mechanism to the effect that the contributions with outer pairings
become small. Note that this mechanism is based on two facts: That we have a multitude~$N$ of fields,
and that these fields couple to the sea states as shown in Figure~\ref{fighom}.
Although it might look promising at first sight, this mechanism turns out {\em{not}} to be sufficient
for satisfying the CCR as operator equations. The reason for this shortcoming can be understood
in words as follows.
In order to satisfy the CCR as operator equations, we must make sure
that the equation~\eqref{Labsmall} holds in the operator sense. Having multiplication operators in
momentum space, this means that~\eqref{Labsmall} should hold for any~$p \in \hat{M}$.
However, the scaling behavior in~\eqref{small} and~\eqref{large} does not hold uniformly for all~$p \in \hat{M}$.
In fact, in~\eqref{small} we considered~$p$ with~$|p^0| \lesssim \omega_{\min}$
(otherwise the operator product vanishes), whereas in~\eqref{large} we considered the high-frequency
region~$|p^0| \gg \omega_{\min}$ where the shaded region in Figure~\ref{fighom} has a spike-like form.
With this in mind, the scalings in~\eqref{small} and~\eqref{large} do {\em{not}} hold as operator equations.

This problem will be resolved in Section~\ref{secgauge} by taking into account the gauge phases
of the potentials~$A_a$. Before explaining how this works, we conclude this section with a few
clarifying considerations. In the next section, we make the just-described no-go result precise.
In Section~\ref{secpos} we complement the picture by computations in position space.

\subsubsection{Shortcoming of the Ansatz} \label{secnogo}
We now explain in more detail why the above ansatz~\eqref{hatBjdefastic} with vectorial
stochastic potentials~\eqref{Bavec}, \eqref{covariancevector} is not quite sufficient for
implementing the CCR. Our argument is more general than the
consideration in the previous section, because we shall not make use of the specific form
of the homogeneous solutions as depicted in Figure~\ref{fighom}.
Beginning from~\eqref{hatA}, we compute the following statistical mean with the help of~\eqref{covariancevector},
\begin{align*}
&\bbra \sum_{a,b=1}^N \hat{A}_a^j(q)\, \hat{L}_a(p_L) \: \hat{A}_b^k(q')\, \hat{L}_b(p_R) \kket = (2 \pi)^4\:
\delta^4(q+q')\: K^{jk}(q,p_L, p_R) 
\end{align*}
with
\beq \label{Kjkform}
K^{jk}(q,p_L, p_R) := \sum_{a,b=1}^N \hat{h}_{a,b}^{jk}(q)\: \hat{L}_a(p_L)\: \hat{L}_b(p_R) \:.
\eeq
Using the notation~\eqref{shortnot}, we thus obtain
\[ \bbra \hat{\B}^j_q|_p \; \hat{\B}^{k}_{q'}|_{p'} \kket
= (2 \pi)^4\: \delta^4(q+q')\:K^{jk} \Big( q, p+ \frac{q}{2}, p' - \frac{q}{2} \Big) \:. \]
Next, similar to~\eqref{CCRtensor} we consider the product
\begin{align*}
&\big[ \hat{\B}^i_q, \hat{\B}^j_{q'} \big]
\big|_p \otimes \hat{\B}^k_r \big|_{k} \otimes \hat{\B}^l_{r'} \big|_{k'} \\
&= \hat{\B}^i_q \big|_{p+q'}\, \hat{\B}^j_{q'} \big|_p
\; \hat{\B}^k_r \big|_{k} \; \hat{\B}^l_{r'} \big|_{k'}
- \hat{\B}^j_{q'} \big|_{p+q}\, \hat{\B}^i_q \big|_p
\; \hat{\B}^k_r \big|_{k} \; \hat{\B}^l_{r'} \big|_{k'}\:.
\end{align*}
Its statistical mean can be computed with the Wick rules to
\begin{align}
&\frac{1}{(2 \pi)^8} \:\bbra \big[ \hat{\B}^i_q, \hat{\B}^j_{q'} \big]
\big|_p \otimes \hat{\B}^k_r \big|_{k} \otimes \hat{\B}^l_{r'} \big|_{k'} \kket \notag \\
&= \delta^4(q+q')\: \delta^4(r+r') \:\bigg( 
K^{ij} \Big( q, p-\frac{q}{2}, p-\frac{q}{2} \Big) \notag \\
&\qquad\qquad\qquad\qquad\quad\;\; - K^{ij} \Big( q, p+\frac{q}{2}, p+\frac{q}{2} \Big) \bigg) K^{kl} \Big( r, k+ \frac{r}{2}, k' - \frac{r}{2} \Big) \label{p1} \\
&\quad + \delta^4(q+r)\: \delta^4(q'+r') \bigg( 
K^{ik} \Big( q, p+q'+\frac{q}{2}, k-\frac{q}{2} \Big)\: K^{jl} \Big( q', p+\frac{q'}{2}, k'-\frac{q'}{2} \Big) \notag \\
&\qquad\qquad\qquad\qquad\qquad\;\; - K^{ik} \Big( q, p+\frac{q}{2}, k-\frac{q}{2} \Big)\: K^{jl} \Big( q', p+q+\frac{q'}{2}, k'-\frac{q'}{2} \Big) \bigg) \label{p2} \\
&\quad + \delta^4(q+r')\: \delta^4(q'+r) \bigg( 
K^{ik} \Big( q, p+q'+\frac{q}{2}, k'-\frac{q'}{2} \Big)\: K^{jl} \Big( q', p+\frac{q'}{2}, k-\frac{q}{2} \Big) \notag \\
&\qquad\qquad\qquad\qquad\qquad\; - K^{ik} \Big( q, p+\frac{q}{2}, k'-\frac{q}{2} \Big)\: K^{jl} \Big( q', p+q+\frac{q'}{2}, k-\frac{q'}{2} \Big) \bigg) . \label{p3}
\end{align}
In~\eqref{p1} the two operators in the commutator are paired. This pairing gives the desired
commutation relations. In~\eqref{p2} and~\eqref{p3}, however, the operators of the commutator
are paired with operators outside. These contributions should vanish.
Since~$q, q'$ and~$k,k'$ can be chosen arbitrarily, it follows that~\eqref{p2} and~\eqref{p3} must
vanish separately. Moreover, using that~$q$ and~$q'$ can be chosen independently, we obtain
the separate conditions
\begin{align*}
K^{ik} \Big( q, p+q'+\frac{q}{2}, k-\frac{q}{2} \Big) &= K^{ik} \Big( q, p+\frac{q}{2}, k-\frac{q}{2} \Big) \\
K^{jl} \Big( q', p+\frac{q'}{2}, k'-\frac{q'}{2} \Big) &= K^{jl} \Big( q', p+q+\frac{q'}{2}, k'-\frac{q'}{2} \Big) \:,
\end{align*}
to be satisfied for all~$q, q'$ and~$k,k'$. The first equation means that~$K^{ik}$ does not depend
on the second variable. Since the function~$K^{jk}$ is symmetric in its second and third variables
(as is obvious from~\eqref{Kjkform}), we conclude that it is also independent of the third variable.
Hence the covariance depends only on the first variable.
But then also~\eqref{p1} vanishes, so that the CCR are violated.

\subsubsection{The Commutator in Position Space} \label{secpos}
Similar as in Section~\ref{secposex}, we now compute the commutator in position space.
We set
\begin{align}
\B^j_x(k_L, k_R) &= \int \frac{d^4q}{(2 \pi)^4}\: \hat{\B}^j_q(k_L, k_R)\: e^{-i q x} \notag \\
&=  \sum_{a=1}^N \hat{A}^j_a(k_L-k_R)\: \hat{L}_a \Big( \frac{k_L+k_R}{2} \Big) \: e^{-i (k_L-k_R) x} \:.
\label{Ajxdef}
\end{align}
Transforming to position space, we obtain (again using the notation~\eqref{secposex})
\begin{align*}
\B^j_x(y_L, y_R) 
&= \int \frac{d^4p}{(2 \pi)^4} \int \frac{d^4q}{(2 \pi)^4}\: \B^j_x\Big(p + \frac{q}{2} , p - \frac{q}{2} \Big)\:
e^{i p \xi + i q \zeta} \\
&= \sum_{a=1}^N \int \frac{d^4p}{(2 \pi)^4} \int \frac{d^4q}{(2 \pi)^4}\: 
\hat{A}^j_a(q)\: \hat{L}_a(p) \: e^{i p \xi + i q \zeta} \,e^{-i q x} \\
&= \sum_{a=1}^N A^j_a\Big( \frac{y_L+y_R}{2} + x \Big) \: L_a(y_R-y_L) \:.
\end{align*}
This formula shows that the argument~$x$ simply describes a translation of the nonlocal potential
in position space.

Next, it is useful to introduce the notation
\[ 
\big( {\mathfrak{D}}^j A \big) (x,y) = (y-x)^j\: A(x,y) \:. \]
This notation is motivated by the fact that the factor~$(y-x)^j$ corresponds to a partial
derivative in momentum space (as one verifies in detail by differentiating~\eqref{Lahat}).
This also explains the product rule
\[ \mathfrak{D}_j (AB) = (\mathfrak{D}_jA)\, B + A\, (\mathfrak{D}_j B) \]
(where~$AB$ is the product of two convolution operators).
\begin{Lemma} \label{lemmacommute} For any~$x,y \in M$, the commutator of the
field operators defined by~\eqref{Ajxdef} is given in position space by
\begin{align*}
\big[ \B^j_x, \B^k_y \big](y_L, y_R)
&= \frac{1}{2} \sum_{a,b=1}^N 
\bigg( \partial_l A^j_a\big( \zeta + x \big) \: A^k_b\big( \zeta + y \big) - \partial_l A^k_a\big( \zeta + y \big)\: A^j_b\big( \zeta + x \big) \bigg) \notag \\
&\qquad\qquad\qquad \times
\big( L_a\:(\mathfrak{D}^l L_b) + (\mathfrak{D}^l L_b)\: L_a \big)(y_R-y_L) + \O \big( \ell^2 \big)
\end{align*}
with
\[ \zeta = \frac{y_L+y_R}{2} \:. \]
\end{Lemma}
\Proof First,
\begin{align*}
& \big[ \B^j_x, \B^k_y \big](y_L, y_R) \\
&= \sum_{a,b=1}^N \int d^4z \:\Big\{
A^j_a\Big( \frac{y_L+z}{2} + x \Big) \: L_a(z-y_L) \:
A^k_b\Big( \frac{z+y_R}{2} + y \Big) \: L_b(y_R-z) \Big\} \\
&\quad\: -\sum_{a,b=1}^N \int d^4z \:\Big\{
A^k_b\Big( \frac{y_L+z}{2} + y \Big) \: L_b(z-y_L) \:
A^j_a\Big( \frac{z+y_R}{2} + x \Big) \: L_a(y_R-z) \Big\} \:.
\end{align*}
In the arguments of the potentials~$A_a$ and~$A_b$ we use the transformations
\[ \frac{y_L+z}{2} = \frac{y_L+y_R}{2} + \frac{z-y_R}{2} \qquad \text{and} \qquad
\frac{z+y_R}{2} = \frac{y_L+y_R}{2} + \frac{z-y_L}{2} \]
and expand in powers of the last summands. We thus obtain
\begin{align*}
& \big[ \B^j_x, \B^k_y \big](y_L, y_R) = \sum_{a,b=1}^N A^j_a\big( \zeta + x \big) \: A^k_b\big( \zeta + y \big) \\
&\qquad \times \int d^4z \:\Big\{
\: L_a(z-y_L) \:L_b(y_R-z) - L_b(z-y_L)\: L_a(y_R-z) \Big\} \\
&\quad\: + \frac{1}{2}
\sum_{a,b=1}^N \partial_l A^j_a\big( \zeta + x \big) \: A^k_b\big( \zeta + y \big) \\
&\qquad \times
\int d^4z \:\Big\{
(z-y_R)^l\: L_a(z-y_L) \:L_b(y_R-z) - L_b(z-y_L)\: (z-y_L)^l\: L_a(y_R-z) \Big\} \\
&\quad\: + \frac{1}{2}
\sum_{a,b=1}^N A^j_a\big( \zeta + x \big) \: \partial_l A^k_b\big( \zeta + y \big) \\
&\qquad \times
\int d^4z \:\Big\{
\: L_a(z-y_L) \:(z-y_L)^l\: L_b(y_R-z) - (z-y_R)^l\: L_b(z-y_L)\: L_a(y_R-z) \Big\} \\
&\quad\: + \O \big( \ell^2 \big) 
\end{align*}
Being complex convolution operators, the operators~$L_a$ and~$L_b$ clearly commute.
Therefore, the zero order terms drop out. The remaining first order terms can be written as
\begin{align*}
\big[ \B^j_x, \B^k_y \big](y_L, y_R)
&= \frac{1}{2}
\sum_{a,b=1}^N \partial_l A^j_a\big( \zeta + x \big) \: A^k_b\big( \zeta + y \big) \\
&\qquad\qquad\qquad \times
\big( L_a\:(\mathfrak{D}^l L_b) + (\mathfrak{D}^l L_b)\: L_a \big)(y_R-y_L)  \\
&\quad\: + \frac{1}{2}
\sum_{a,b=1}^N A^j_a\big( \zeta + x \big) \: \partial_l A^k_b\big( \zeta + y \big) \\
&\qquad\qquad\qquad \times
\big( - (\mathfrak{D}^l L_a)\:L_b-  L_b\: (\mathfrak{D}^l L_a) \big)(y_R-y_L \big) + \O \big( \ell^2 \big)  \:,
\end{align*}
giving the result.
\QED
As is verified by a straightforward computation, the formula of this lemma is indeed the Fourier
transform of~\eqref{comm1}. Also the error terms agree, if one keeps in mind that in the last proof,
every factor~$\ell$ comes with a derivative acting on the potential, which in momentum space
gives rise to a factor~$q$ or~$q'$.

We finally remark that the above argument~\eqref{small} and~\eqref{large} in which we
made use of smoothness properties of~$\hat{L}_a$ and~$\hat{L}_b$ can be recast in position space by saying
that the operator product~$L_a L_b$ is typically of much shorter range than the operator~$L_a^2$.

\section{Holographic Mixing} \label{secgauge}
The analysis of the previous section showed that a nonlocal potential of the form~\eqref{hatBjdefastic}
is not sufficient for realizing the CCR as operator equations.
The general idea in order to overcome this problem is to take into account that, similar to
gauge phases of classical electrodynamics, the potentials~$A_a$ in~\eqref{hatBjdefasticintro}
give rise to phase factors. Having a multitude of~$N$ potentials yields~$N$ different
phase factors. Consequently, relative phase factors should lead to dephasing effects.
In this section, we will work out this mechanism and its consequences in detail.
The mechanism is referred to as {\em{holographic mixing}}.
Holographic mixing is a specific feature of causal fermion systems.
For this reason, we introduce it in detail step by step.
After a few general considerations (Section~\ref{secmixintro}),
we begin under simplifying assumptions with an analysis of momentum-dependent gauge transformations
(Section~\ref{secgaugemom}). We proceed by specifying the form of the dynamical gauge potentials
in the presence of holographic phases (Section~\ref{secBdyn}). Then we will show that these potentials
naturally satisfy the CCR (Sections~\ref{secrealizecomm} and~\ref{secstationaryphase}).
In Section~\ref{secdynholo} the dynamical equations including holographic phases will be
derived.

\subsection{Basic Concepts Behind Holographic Mixing} \label{secmixintro}
The concept of holographic mixing evolved in various steps over several years.
Before entering the detailed constructions, we now give a brief outline of how holographic mixing developed.
It is one of the basic features of causal fermion systems that one can go beyond classical spacetimes and
allow for the description of spacetimes which have a non-smooth, possibly discrete, microstructure
or which on small scales resemble a ``quantum spacetime'' with ``quantum geometric'' structures
(as is made precise in~\cite{lqg}).
Since in the setting of causal fermion systems, the geometric structures of spacetime are encoded in the
family of all physical wave functions, such novel or generalized spacetime structures are implemented by
working with physical wave functions having a non-trivial behavior on small scales.
This concept was first introduced in~\cite[Section~4]{entangle} under the name {\em{microscopic mixing}}.
Refinements of this concept were used in~\cite[Section~3]{qft} for getting
a first connection between causal fermion systems and QFT.
Subsequently, in~\cite[\S1.5.3]{cfs} microscopic mixing came up from a somewhat different perspective
by studying a decomposition of the measure~$\rho$ of the causal fermion system into many components,
with the purpose of decreasing the causal action in the vacuum.
This point of view was followed up in~\cite[Section~5]{positive}, where also the
term {\em{fragmentation}} was introduced. A more systematic study of fragmentation and
the resulting dynamics of the causal fermion system was carried out in~\cite[Section~5]{perturb}.

A simple way of understanding fragmentation is the observation that the measure~$\rho$ of the causal fermion
system does not necessarily need to be the push-forward of the volume measure of a classical spacetime.
One method for constructing more general measures is to take convex combinations of measures,
leading to the ansatz
\beq \label{lincomb}
\rho = \sum_{a=1}^N c_a \,\rho_a \qquad \text{with~$c_a > 0$ and} \quad \sum_{a=1}^N c_a = 1\:.
\eeq
If each measure is constructed from a classical spacetime, then~$M_a := \supp \rho_a$ is diffeomorphic
to Minkowski space (or to a curved spacetime). In this case, the spacetime~$M:= \supp \rho$ of the
causal fermion system is the union of all the sub-spacetimes~$M_a$,
\[ M = \bigcup_{a=1}^N M_a \:. \]
The resulting physical picture is that spacetime consists of~$N$ copies of classical spacetimes,
each with its own physical wave functions and corresponding spacetime structures.
This has similarities with taking a ``superposition'' of classical spacetimes, except that the
coefficients~$c_a$ in the linear combination~\eqref{lincomb} must all be positive
(in order to ensure that~$\rho$ is a positive measure).
The decomposition of spacetime into sub-spacetimes becomes a bit clearer by writing the
corresponding wave evaluation operator~$\Psi : \H \rightarrow C^0(M, SM)$ (see~\eqref{weo}) as
\beq \label{Psia}
\Psi(x) = \sum_{a=1}^N \chi_{M_a}(x)\: \Psi_a(x) \:,
\eeq
where~$\Psi_a : \H \rightarrow C^0(M_a, SM_a)$ are the wave evaluation operators of the sub-spacetimes.

Clearly, the fact that taking convex combinations is mathematically allowed does not necessarily mean that
measures of the form~\eqref{lincomb} should occur in physics. One argument in favor of~\eqref{lincomb}
is that this ansatz makes it possible to decrease the causal action. Indeed, rewriting the causal action as
\begin{align*}
\Sact(\rho) &= \sum_{a,b=1}^N c_a\, c_b\,\iint_{\F \times \F} \L(x,y)\: d\rho_a(x)\: d\rho_b(y) \\
&= \sum_{a=1}^N c_a^2 \, \Sact(\rho_a) + \sum_{a \neq b} 
c_a\, c_b\,\iint_{\F \times \F} \L(x,y)\: d\rho_a(x)\: d\rho_b(y) \:,
\end{align*}
the first sum can be arranged to give a scaling factor~$1/N$
(for example by choosing~$\Sact(\rho_a) =\Sact(\rho_b)$ and setting~$c_a=1/N$
for all~$a,b \in \{1,\ldots, N\}$).
The summands for~$a \neq b$, on the other hand, can be made small using dephasing effects.
This becomes clear by transforming the wave evaluation operators according to
\[ \Psi_a \longrightarrow \Psi_a \, \scrU_a \:, \]
where~$\scrU_a$ are unitary operators on~$\H$. In this case, the kernel of the fermionic projector transforms
according to
\begin{align*}
\chi_{M_a}&(x) \,P(x,y) \, \chi_{M_b}(y) = - \chi_{M_a}(x) \,\Psi_a(x)\, \Psi_b(y)^* \, \chi_{M_b}(y) \\
&\longrightarrow - \chi_{M_a}(x) \,\Psi_a(x)\,\scrU_a\, \scrU_b^*\, \Psi_b(y)^* \, \chi_{M_b}(y) \:.
\end{align*}
In the case~$a =b$, the unitary operators drop out, showing that~$\Sact(\rho_a)$ remains unchanged.
In the case~$a \neq b$, however, the unitary operator~$\scrU_a\, \scrU_b^*$ can be arranged to
be small due to dephasing (for more details see~\cite[\S1.5.3]{cfs}).

Considerations along this line explain why fragmentation occurs naturally when minimizing the causal action principle. This also means that dephasing effects between the physical wave functions in spacetime
should be of physical relevance.
Nevertheless, fragmentation is not a fully convincing concept for the following reasons.
One difficulty is to describe fragmentation dynamically. How do the sub-spacetimes interact with
each other? What is the resulting ``effective'' dynamics of the whole system?
Another, more conceptual problem is that, thinking of a fully interacting situation, it does not seem
to be a sensible concept to speak of individual sub-spacetimes.
Should one not consider~$M$ instead
as a whole, without decomposing it into sub-spacetimes?
Thinking about these questions led to {\em{holographic mixing}} as a preferable method for describing
the above dephasing effects. To this end, we generalize~\eqref{Psia} by an ansatz for the wave
evaluation operator~\eqref{weo} as a linear combination
\beq \label{Psiholo}
\Psi(x) = \sum_{a=1}^N e^{i \Lambda_a(x)}\: \Psi_a(x) \qquad \text{with} \qquad
\Psi_a : \H \rightarrow C^0(M, SM) \:,
\eeq
and~$\Lambda_a$ are real-valued functions.
Note that, in contrast to~\eqref{Psia}, we no longer have sub-spacetimes. Instead, all the
mappings~$\Psi_a(x)$ are defined in the whole spacetime~$x \in M$.
The dephasing effects are described by the phase factors~$e^{i \Lambda_a}$ which may
oscillate on small length scales, thereby implementing the original concept of 
a non-trivial microstructure of spacetime.
The name ``holographic mixing'' is inspired by
the similarity to a hologram in which several pictures are stored, each of which becomes visible only
when looking at the hologram in the corresponding coherent light.

One advantage of holographic mixing~\eqref{Psiholo} is that it fits better to the
structure of the linearized field equations in Minkowski space as unveiled in~\cite{nonlocal},
because the summands $\Psi_a$ can be associated with the physical wave functions
lying in the support of the functions~$\hat{L}_a$ (i.e.\ the shaded regions in Figure~\ref{fighom}).
These summands are no longer localized in separate sub-spacetimes (as in~\eqref{Psia}),
but rather correspond to ``subsystems'' in Minkowski space formed of wave functions
propagating into a common null direction.
Another advantage of this description is that the phase functions~$e^{i \Lambda_a}$
can be associated with local phases corresponding to the potentials~$A_a$ (this is also why we
use the same notation with a subscript~$a$). This also gives an explanation for why these phase
functions come about.
Moreover, it will become possible to derive dynamical equations for the system including holographic
phases. Finally, the concept of holographic mixing ties in nicely with the constructions in~\cite{fockentangle},
where dephasing effects also play a crucial role.

The critical reader may wonder how fragmentation~\eqref{lincomb}
and holographic mixing~\eqref{Psiholo} are related to each other, and how it comes about that
both methods can be used to describe interacting causal fermion systems.
In non-technical terms, this can be understood as follows (a more detailed explanation of this
point from a slightly different perspective can be found in~\cite[Chapter~22]{intro}).
In the limit when~$N$ gets large, the fragmented measure~$\tilde{\rho}$
goes over to a measure with enlarged support (see the gray region on the left of Figure~\ref{figholo2}).
Integrating with respect to this measure does not only involve an integral over
the four-dimensional spacetime (indicated by the black line on the left of Figure~\ref{figholo2}),
but it also involves an integration over the ``internal degrees of freedom''
corresponding to the directions which are transverse to the four-dimensional spacetime.
\begin{figure}
\begin{center}
\psscalebox{1.0 1.0} 
{
\begin{pspicture}(0,28.643261)(9.296764,30.690586)
\definecolor{colour0}{rgb}{0.8,0.8,0.8}
\definecolor{colour1}{rgb}{0.7019608,0.7019608,0.7019608}
\pspolygon[linecolor=colour0, linewidth=0.02, fillstyle=solid,fillcolor=colour0](5.570035,29.212551)(5.658924,29.176996)(5.8233685,29.119217)(6.0189238,29.079218)(6.2322574,29.052551)(6.5033684,29.056995)(6.7744794,29.065884)(6.9967017,29.096996)(7.338924,29.154774)(7.570035,29.208107)(7.778924,29.265884)(8.112257,29.350328)(8.32559,29.399218)(8.583368,29.45255)(8.890035,29.510328)(9.107813,29.54144)(9.272257,29.554773)(9.276702,30.136995)(8.952257,30.203663)(8.605591,30.279217)(8.24559,30.34144)(7.978924,30.372551)(7.5744796,30.372551)(7.281146,30.34144)(6.9478126,30.25255)(6.6189237,30.136995)(6.3344793,30.06144)(5.9567018,29.999218)(5.703368,30.039217)(5.5744796,30.088106)
\pspolygon[linecolor=colour1, linewidth=0.02, fillstyle=solid,fillcolor=colour1](0.020034993,29.162552)(0.10892388,29.126995)(0.27336833,29.069218)(0.4689239,29.029217)(0.68225724,29.00255)(0.9533683,29.006996)(1.2244794,29.015884)(1.4467016,29.046995)(1.7889239,29.104773)(2.020035,29.158106)(2.2289238,29.215885)(2.5622573,29.30033)(2.7755907,29.349218)(3.0333683,29.402552)(3.340035,29.46033)(3.5578127,29.49144)(3.7222571,29.504774)(3.7267017,30.086996)(3.4022572,30.153662)(3.0555906,30.229218)(2.6955905,30.29144)(2.4289238,30.322552)(2.0244794,30.322552)(1.7311461,30.29144)(1.3978127,30.20255)(1.0689238,30.086996)(0.78447944,30.01144)(0.40670165,29.949217)(0.15336832,29.989218)(0.024479438,30.038107)
\psbezier[linecolor=black, linewidth=0.04](0.03670166,29.68255)(0.5844548,29.397636)(1.025005,29.335691)(1.6944795,29.368106553819445)(2.3639538,29.400522)(2.760531,29.695997)(3.7367017,29.802551)
\rput[bl](1.9311461,28.698107){$\supp \rho$}
\psbezier[linecolor=black, linewidth=0.02, arrowsize=0.05291667cm 2.0,arrowlength=1.4,arrowinset=0.0]{->}(1.8762708,28.864773)(1.6849209,28.885693)(1.584782,28.910864)(1.5000342,28.959277288050362)(1.4152865,29.00769)(1.3622024,29.054342)(1.3032857,29.109219)
\rput[bl](3.270035,30.439219){$\F$}
\psline[linecolor=black, linewidth=0.03](5.606702,29.893661)(5.6767015,29.383661)(5.8467016,29.493662)(5.816702,29.833662)(5.9767017,29.203663)(6.0467014,29.913662)(6.0967016,29.533663)(6.2267017,29.273663)(6.3067017,29.753662)(6.1567016,29.843662)(6.5067015,29.953663)(6.5567017,29.593662)(6.3267016,29.563662)(6.3867016,29.283663)(6.7167015,29.203663)(6.6967015,29.993662)(6.4867015,29.163662)(6.936702,29.993662)(6.8467016,29.403662)(6.896702,29.273663)(7.0567017,30.203663)(7.0967016,29.263662)(7.2067018,30.233662)(7.3267016,29.403662)(6.9867015,29.663662)(7.4767017,30.193663)(7.416702,29.383661)(7.666702,30.243662)(7.9067016,30.103662)(7.686702,29.843662)(7.9767017,29.613663)(7.5967016,29.413662)(7.6967015,30.063662)(8.106702,29.963661)(7.7367015,29.593662)(8.126701,29.473661)(8.156702,30.223661)(8.486702,30.213661)(8.346702,30.063662)(8.496701,29.773663)(8.256701,29.523663)(8.246701,30.193663)(8.436702,29.593662)(8.536701,30.053661)(8.756701,30.243662)(8.686702,29.983662)(8.016702,30.083662)(8.786701,29.633661)(8.811702,30.103662)(9.001701,30.123663)(8.531702,29.593662)(9.041701,29.673662)(8.951702,29.883661)(9.171701,30.103662)(9.171701,29.803661)(8.871701,29.733662)(8.621701,29.923662)(7.771702,30.183662)
\rput[bl](7.151146,28.708107){$\supp \rho$}
\psbezier[linecolor=black, linewidth=0.02, arrowsize=0.05291667cm 2.0,arrowlength=1.4,arrowinset=0.0]{->}(7.076271,28.794773)(6.884921,28.815693)(6.784782,28.840864)(6.700034,28.889277288050362)(6.6152864,28.93769)(6.5622025,28.984343)(6.503286,29.039217)
\rput[bl](8.720035,30.429218){$\F$}
\end{pspicture}
}
\end{center}
\caption{A measure obtained by fragmentation (left) and by holographic mixing with fluctuations
(right).}
\label{figholo2}
\end{figure}
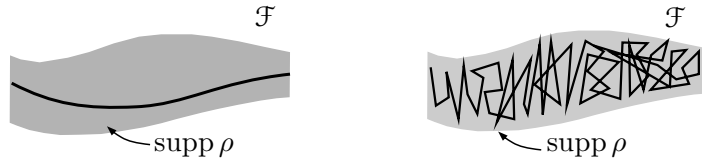
This situation bears similarity to the path integral formulation of QFT
if one identifies the above ``internal degrees of freedom'' with field configurations.
As explained above, in the method of holographic mixing, spacetime~$M:= \supp \rho$ remains
four-dimensional. But, keeping in mind that we allow for high-frequency fluctuations,
the measure could still approximate the measure described by fragmentation
(as shown by the black line on the right of Figure~\ref{figholo2}, noting that high frequencies
correspond to low regularity in position space, as drawn for simplicity by a non-smooth curve).
With this intuitive picture, it becomes clear why, from the mathematical point of view,
fragmentation and holographic mixing can be used equivalently.
However, it seems that the method of holographic mixing is more suitable for the analysis of the
causal action principle.

We finally remark that the concept of holographic mixing is also related to
Everett's multiverse interpretation of quantum theory and the collapse of the wave function.
We refer the interested reader to~\cite{collapse, heatcoll}.

\subsection{Momentum-Dependent Phase Transformations} \label{secgaugemom}
In order to develop our methods, we begin with the behavior of Dirac wave functions under local
gauge transformations. The local gauge transformations of electrodynamics are described by the
joint transformation
\[ 
\psi(x) \rightarrow e^{i \Lambda(x)} \, \psi(x) \]
of all wave functions, with~$\Lambda$ a real-valued function. The Dirac operator of the vacuum
transforms according to
\[ 
i \Pdd \rightarrow e^{i \Lambda} \big( i \Pdd \big) e^{-i \Lambda} = i \Pdd + (\Pdd \Lambda)\:. \]
The additional term~$(\Pdd \Lambda)$ is the standard coupling term to the gauge
potentials~$A_j = \partial_j \Lambda$.
Before going on, we point out that, in order to capture dephasing effects, 
we need to cover the case that the phase functions like~$e^{i \Lambda}$ have rapid oscillations
on small length scales. Such oscillatory phase functions cannot be treated with a simple
perturbation expansion. Therefore, it will not suffice for our purposes to take the
Dirac operator~$i \Pdd + (\Pdd \Lambda)$ as the starting point. Instead, it is preferable to
work with the phase factors~$e^{\pm i \Lambda}$ where the oscillations are manifest.
This observation will serve as a guiding principle for the following constructions.

As summarized in Section~\ref{secdir}, it is the main conclusion of the analysis in~\cite{nonlocal}
that the causal action principle admits a larger class of homogeneous fields. Rather than just the
electromagnetic field, there is a plethora of fields described by nonlocal potentials (see~\eqref{dirnonloc}, 
\eqref{Bnonlocal} and~\eqref{hatBjdefastic}). These fields are related to the corresponding local
gauge invariance~\eqref{dirlocintro} of the causal Lagrangian.
This suggests that nonlocal potentials should give rise to corresponding
phase factors, also leading to dephasing effects. In order to analyze how such effects can be
modeled mathematically, we proceed step by step and begin with the situation that the different
phase factors act on orthogonal subspaces of the wave functions in spacetime. To this end,
we consider a family~$(L_a)_{1=1,\ldots, N}$ of homogeneous scalar operators, i.e.\
\beq \label{Kakernel}
\big( L_a \psi \big)(x) = \int L_a(y-x)\: \psi(y)\: d^4y \qquad \text{and} \qquad L_a(y-x) \in \C \:,
\eeq
which for simplicity we again assume to be smooth. Moreover, we assume that the~$L_a$ form a
complete set of projection operators, i.e.\
\beq \label{Kproj1}
L_a L_b = \delta_{ab} \, L_a \qquad \text{and} \qquad \sum_{a=1}^N L_a = \1 \:.
\eeq
Now for each~$a \in \{1,\ldots, N\}$ we introduce a real-valued phase function~$\Lambda_a$ and introduce the operator~$U$ by
\beq \label{Udef}
(U \psi)(x) := \sum_{a=1}^N e^{i \Lambda_a(x)}\: \big( L_a \psi \big)(x) \:.
\eeq
In order to simplify the situation further, we also assume that the images of the summands are orthogonal
and again complete, i.e.\
\beq \label{Kproj2}
L_a\, e^{-i \Lambda_a}\: e^{i \Lambda_b}\: L_b = \delta_{ab}\: L_a 
\qquad \text{and} \qquad \sum_{a=1}^N e^{i \Lambda_a}\: L_a \:e^{-i \Lambda_a} = \1 \:.
\eeq
Then the above operator~$U$ is unitary on~$L^2(M, \C)$, because
\begin{align}
U U^* &= \sum_{a,b=1}^N e^{i \Lambda_a}\: L_a\: L_b \:e^{-i \Lambda_b} = 
\sum_{a=1}^N e^{i \Lambda_a}\: L_a \:e^{-i \Lambda_a} = \1 \label{unit1} \\
U^* U &= \sum_{a,b=1}^N L_a\: e^{-i \Lambda_a}\: e^{i \Lambda_b}\: L_b  = 
\sum_{a=1}^N L_a = \1 \:. \label{unit2}
\end{align}
Now the vacuum Dirac operator transforms to
\begin{align}
i \Pdd &\rightarrow U (i \Pdd) U^*
= \sum_{a,b=1}^N e^{i \Lambda_a} \:L_a\: (i \Pdd)\: L_b \: e^{-i \Lambda_b}
= \sum_{a=1}^N e^{i \Lambda_a} \:L_a\: (i \Pdd) \: e^{-i \Lambda_a} \label{dirphase0} \\
&= \frac{1}{2}\: i \Pdd \:\bigg( \sum_{a=1}^N e^{i \Lambda_a} \:L_a \: e^{-i \Lambda_a} \bigg) -
\frac{1}{2}\: \sum_{a=1}^N \big[ i \Pdd , e^{i \Lambda_a} \big] \:L_a\: e^{-i \Lambda_a} \notag \\
&\quad\: +\frac{1}{2}\:\bigg( \sum_{a=1}^N e^{i \Lambda_a} \:L_a\: e^{-i \Lambda_a} \bigg) \:  i \Pdd
+ \frac{1}{2}\: \sum_{a=1}^N e^{i \Lambda_a} \:L_a\: \big[ i \Pdd , e^{-i \Lambda_a} \big] \notag \\
&= i \Pdd + \frac{1}{2} \sum_{a=1}^N e^{i \Lambda_a} \,\Big( (\Pdd \Lambda_a) \,L_a + L_a\, (\Pdd \Lambda_a) \Big) \, e^{-i \Lambda_a} \label{dirphase}
\end{align}
(in the last step we again used the completeness relation in~\eqref{Kproj2}).
With this transformation, we introduced an interaction operator into the Dirac operator.
It can be written as in~\eqref{dirnonloc} with a nonlocal potential~$\B=\B_\Lambda$ given by
\beq
\B_\Lambda(x,y) = \frac{1}{2} \sum_{a=1}^N e^{i \Lambda_a(x)} \,\Big( \big( \Pdd \Lambda_a(x) \big) \,L_a(y-x) + L_a(y-x)\, \big( \Pdd \Lambda_a \big)(y) \Big) \, e^{-i \Lambda_a(y)} \:. \label{Bphase}
\eeq

Before going on, we explain the connection to the potential in~\eqref{hatBjdefastic}.
We first point out that the form of the
potential~\eqref{hatBjdefastic} was derived in~\cite{nonlocal} for {\em{linearized}} fields in the Minkowski vacuum.
Therefore, in order to make the connection, we also need to linearize~\eqref{Bphase} to obtain
\[ 
\B_\Lambda(x,y) = \frac{1}{2} \sum_{a=1}^N \Big( \big( \Pdd \Lambda_a(x) \big) \,L_a(y-x) + L_a(y-x)\, \big( \Pdd \Lambda_a \big)(y) \Big) + \O \big( \Lambda_a^2 \big) \:. \]
Next, we need to assume that the phase functions are smooth and
do not oscillate on the length scale~$\ell_{\min}$ of the nonlocality, i.e.\ if
\beq \label{Lamscale}
\big| D^p \Lambda_a \big| \lesssim \frac{1}{\ell_\Lambda^p}\: |\Lambda_a| \qquad \text{with}
\qquad \ell_\Lambda \gg \ell_{\min} \:.
\eeq
Under this condition, we may expand the phase functions in a Taylor series. Indeed, writing
\[ \Lambda_a(x) = \Lambda_a \Big( \frac{x+y}{2} - \frac{\xi}{2} \Big) \:,\qquad
\Lambda_a(y) = \Lambda_a \Big( \frac{x+y}{2} + \frac{\xi}{2} \Big) \]
with~$\xi:= y-x$ and expanding in powers of~$\xi$, the potential~\eqref{Bphase} can be written as
\beq \label{Bgaugelin}
\B_\Lambda(x,y) = \sum_{a=1}^N \sum_\kappa
(\Pdd \Lambda_{a, \kappa}) \Big( \frac{x+y}{2} \Big)\:L_a^\kappa(y-x)
\eeq
with new potentials and kernels
\[ \Lambda_{a, \kappa} = \frac{1}{|\kappa|!}\: \partial^\kappa \Lambda_a \qquad \text{and} \qquad
L_a^\kappa(y-x) := (y-x)^\kappa\: L_a(y-x) \:, \]
where~$\kappa$ runs over all multi-indices.
The first expansion term with~$\kappa=0$ is exactly of the form~\eqref{hatBjdefastic}
with~$A_a = \partial \Lambda_a$.
Combining~$a$ with~$\kappa$ to a new summation index, also the higher order terms
can be written in the form~\eqref{hatBjdefastic}.

We finally point out that summing from~$1$ to~$N$ in~\eqref{Udef} merely is a matter of convenience,
which will later facilitate to get a connection between the dynamical potentials and the gauge phases.
But at this stage, the number of summands in~\eqref{Udef} could be chosen arbitrarily.
Taking the limit where the number of summands tends to infinity, one can also describe situations
where the sum becomes a momentum integral, giving rise to the continuous ansatz
\[ (U \psi)(x) := \int \frac{d^4k}{(2 \pi)^4} \int d^4y \: e^{i \Lambda(x,p)}\: e^{-i k(x-y)}\: \psi(y) \:, \]
involving a momentum-dependent phase function~$\Lambda(x,p)$. This ansatz may be
of advantage for explicit computations.

\subsection{Dynamical Gauge Potentials Including Holographic Phases} \label{secBdyn}
The method for introducing gauge phases introduced in the previous section was too
simple for two reasons. First, choosing the operators~$L_a$ as projection 
operators~\eqref{Kproj1} is not compatible with the form of the operators~$L_a$ in the
nonlocal potential~$\B$ in Section~\ref{secdir} (see~\eqref{hatBjdefastic} and the discussion
of the support of~$\hat{L}_a$ shown in Figure~\ref{fighom}). Second, assuming that the
images of the summands in~\eqref{Udef} are orthogonal~\eqref{Kproj2} seems too restrictive.
In order to improve the situation, we need to drop conditions~\eqref{Kproj1} and~\eqref{Kproj2}.
Thus we consider the operator~$U$ of the form~\eqref{Udef} with general phase functions~$\Lambda_a$
and homogeneous scalar operators~\eqref{Kakernel}.
Then this operator is no longer unitary on~$L^2(M, \C)$, because in~\eqref{unit1} and~\eqref{unit2}
summands with~$a \neq b$ do not drop out.
In more general terms, we must take into account that the different phase components 
interact with each other. In what follows, we will explain step by step how this can be done.
This section is devoted to the first step, which consists in 
specifying the form of the dynamical gauge potential in the presence of holographic phases.

The simplest approach is to proceed similarly to~\eqref{dirphase0} by introducing
the holographic phases into the Dirac operator again by the transformation
\beq \label{dirphase2}
i \Pdd \rightarrow U (i \Pdd) U^* \:.
\eeq
Let us analyze how a nonlocal potential of the form~\eqref{hatBjdefastic} and~\eqref{Bavec}
transforms under the operator~$U$ of the form~\eqref{Udef}. Including the nonlocal potential in
the transformation~\eqref{dirphase2}, we obtain
\[ i \Pdd + \B \rightarrow U (i \Pdd + \B) U^* = i \Pdd + \B_\Lambda + \B_\dyn \:, \]
where~$\B_\Lambda$ is again the holographic gauge potential
(in the simplest case given again by~\eqref{Bphase}) and
\begin{align}
&\B_\dyn(x,y) = \big( U \B U^* \big)(x,y) \notag \\
&= \sum_{a,b,c=1}^N \int_M d^4z_1 \int_M d^4z_2 \notag \\
&\qquad \times e^{i \Lambda_a(x)} \:L_a(z_1-x)\:
\slashed{A}_c \Big( \frac{z_1+z_2}{2} \Big) \:L_c(z_2-z_1)
\: L_b(y-z_2) \: e^{-i \Lambda_b(y)} \:. \label{potprelim}
\end{align}
This consideration shows that the dynamical potentials naturally contain different phase factors
on the left and on the right. Moreover, the momentum change of the
operator sandwiched between these phase factors is determined by the potential~$A_c$.
The potential~\eqref{potprelim} has the disadvantage that it is somewhat complicated.
Therefore, we now generalize it using a compact notation.
First, introducing in position space the notation
\[ \big( \slashed{A} \starD L \big)(x,y) := \slashed{A} \Big( \frac{x+y}{2} \Big)\:
L(y-x) \]
(and likewise in momentum space by taking the Fourier transform of this expression),
the integral operator corresponding to the nonlocal potential~\eqref{potprelim} can be written as
\beq \label{potprelim2}
\B_\dyn = \sum_{a,b,c=1}^N e^{i \Lambda_a} \:L_a\: \big( \slashed{A}_c \starD L_c \big)
\: L_b \: e^{-i \Lambda_b} \:.
\eeq
In order to further simplify the notation, we take the ansatz
\beq \label{Bansatz}
\B_\dyn = \sum_{a,b,c=1}^N e^{i \Lambda_a} \:\big( \slashed{A}_{c} \starD L^c_{a,b} \big)\: e^{-i \Lambda_b} \:,
\eeq
which includes potentials~$A_c$ as well as homogeneous operators~$L^c_{ab}$ with~$a,b,c \in \{1, \ldots, N\}$.
Clearly, the potential~\eqref{Bansatz} should again be symmetric, meaning that
\[ 
\overline{A^j_c(z)} = A^j_c(z) \qquad \text{and} \qquad \overline{L^c_{a,b}(\xi)}=L^c_{b,a}(-\xi) \:. \]

We conclude this section with a discussion of the new ansatz~\eqref{Bansatz}.
We begin by pointing out that it is very general and
covers a large class of nonlocal potentials.
The following constructions will not depend on how precisely the potential~$\B_\dyn$ is chosen.

In order to related the new ansatz~\eqref{Bansatz} to~\eqref{potprelim2}, one can choose~$L^c_{b,a}$ as the
product~$L_a L_b L_c$. However, this does not quite give back~\eqref{potprelim2}, because
in general
\beq \label{triangles}
\slashed{A}_{c} \starD \big( L_a\, L_c\, L_b \big) \neq 
L_a \,\big( \slashed{A}_{c} \starD L_c \big)\, L_b \:.
\eeq
Nevertheless, the mathematical structure of these expressions is quite similar,
so much so that their difference will be of no relevance for what follows.
Indeed, all our considerations and computations apply also to~\eqref{potprelim2}, in which case
one can take the right side of~\eqref{triangles} as the definition of the operation~$\slashed{A}_{c} \starD L^c_{a,b}$.

Another class of potentials which seems sensible is to choose
\[ L^c_{a,b} = d(c,a,b)\: \,L_a\, L_b \]
with real coefficients~$d(c,a,b)$. This choice has the desirable property that for large momenta
(i.e.\ if the frequency~$p$ in Figure~\ref{fighom}
is chosen as~$p^0 \ll -\omega_{\min}$) the product~$L_a L_b$ vanishes unless~$a=b$,
so that the potential simplifies considerably.

We finally note that, in the case without holographic phases, our general ansatz~\eqref{Bansatz} reduces to
\[ \B_\dyn = \sum_{c=1}^N \slashed{A}_c \starD L_c \qquad \text{with} \qquad
L_c := \sum_{a,b=1}^N L^c_{a,b} \:, \]
giving back~\eqref{hatBjdefastic}. With this in mind, we can regard~\eqref{Bansatz} as a
natural generalization of~\eqref{hatBjdefastic}, obtained by suitably inserting the holographic phases.

\subsection{Realizing the Canonical Commutation Relations} \label{secrealizecomm}
We now define the field operators in analogy to~\eqref{hatAdef} by
\beq \label{hatBjdef}
\hat{\B}^j_q = \sum_{a,b,c=1}^N \hat{\B}^{j,c}_{a,b,q} \qquad \text{with} \qquad
\hat{\B}^{j,c}_{a,b,q} := e^{i \Lambda_a} \:\big( A^j_{c, q} \starD L^c_{a,b} \big)\: e^{-i \Lambda_b} \:,
\eeq
where~$A^j_{c, q}$ is the plane-wave component, i.e.\
\beq \label{Ajcqdef}
A^j_{c, q}(x) := \hat{A}^j_c(q)\: e^{-i q x} \:.
\eeq
When taking products of these operators,
\beq \label{toapprox}
\hat{\B}^{j,c}_{a,b,q}\, \hat{\B}^{k,f}_{d,e,q'} 
= e^{i \Lambda_a} \:\big( A^j_{c,q} \starD L^c_{a,b} \big)\: e^{-i \Lambda_b + i \Lambda_d}
\:\big( A^k_{f,q'} \starD L^f_{d,e} \big)\: e^{-i \Lambda_e} \:,
\eeq
the intermediate phase factor~$e^{-i \Lambda_b + i \Lambda_d}$ oscillates rapidly unless~$b=d$.
Therefore, we expect that the leading contributions are obtained in the case~$b=d$, i.e.,\
\beq \label{approxdef}
\hat{\B}^{j,c}_{a,b,q}\, \hat{\B}^{k,f}_{d,e,q'} \approx \delta_{b,d}\: \hat{\B}^{j,c}_{a,b,q}\, \hat{\B}^{k,f}_{b,e,q'}  \:.
\eeq
This approximation will be justified in Section~\ref{secstationaryphase} using a stationary phase analysis,
which will also give a precise scaling of the error terms. In order not to distract from the main construction,
we here restrict attention to the leading contributions, disregarding the error terms with the notation~``$\approx$''.
We thus obtain
\[ \hat{\B}^{j,c}_{a,b,q}\, \hat{\B}^{k,f}_{d,e,q'} 
\approx \delta_{bd} \: e^{i \Lambda_a} \:\big( A^j_{c,q} \starD L^c_{a,b} \big)
\:\big( A^k_{f,q'} \starD L^f_{b,e} \big)\: e^{-i \Lambda_e} \]
and
\begin{align*}
&\big[ \hat{\B}^j_{q},\, \hat{\B}^k_{q'} \big] \\
&\approx \sum_{a, b,c,e,f=1}^N \!\!\!\!
e^{i \Lambda_a} \:\Big( \big( A^j_{c,q} \starD L^c_{a,b} \big) \:\big( A^k_{f,q'} \starD L^f_{b,e} \big)
- \big( A^k_{c,q'} \starD L^c_{a,b} \big) \:\big( A^j_{f,q} \starD L^f_{b,e} \big)
\Big)\: e^{-i \Lambda_e} \\
&= \sum_{a, b,c,e,f=1}^N \!\!\!\! \hat{A}^j_c(q)\:\hat{A}^k_f(q') \:
e^{i \Lambda_a} \:\Big[ \: (E_q \starD L^c_{a,b}) \,(E_{q'} \starD L^f_{b,e})
- \:(E_{q'} \starD L^f_{a,b}) \,(E_q \starD L^c_{b,e}) \Big]
\:e^{-i \Lambda_e} \:,
\end{align*}
where we used~\eqref{Ajcqdef} and defined~$E_q$ as the operator of multiplication by a plane wave
of momentum~$q$, i.e.\
\beq \label{Eqdef}
(E_q \psi)(x) := e^{-iqx}\: \psi(x)\:.
\eeq

In order to write these formulas in a more compact form, it is useful to introduce a matrix notation by setting
\beq \label{notmatrix}
\mathbf{L}^c := (L^c_{a,b})_{a,b=1,\ldots, N} \qquad \text{and} \qquad
\mathbf{D} := \big( e^{i \Lambda_a}\:\delta_{ab}\ \big)_{a,b=1,\ldots, N} \:,
\eeq
making it possible to write the above commutator more compactly as
\beq \label{commgen}
\big[ \hat{\B}^j_{q},\, \hat{\B}^k_{q'} \big] \approx \sum_{a,c,e,f=1}^N \hat{A}^j_c(q)\:\hat{A}^k_f(q') \;
\Big( \mathbf{D} \:\big[ \: (E_q \starD \mathbf{L}^c), \,(E_{q'} \starD \mathbf{L}^f) \big] \:\mathbf{D}^{-1} \Big)_{a,e} \:.
\eeq
We remark that, at this point, one could expand the operators~$E_q \starD \mathbf{L}^c$
and~$E_{q'} \starD \mathbf{L}^f$ in a Taylor series in~$q$ using that~$q \ell_{\min} \ll 1$.
This would lead to a formalism similar to that used in Section~\ref{secstoch}
(see for example Lemma~\ref{lemmacommute}). We prefer not to work with this expansion,
also in order to clarify that such an expansion is not really essential for the CCR.

We are now ready to state and prove the main result of this section.
We again assume that the potentials~$\hat{A}^j_a(q)$ are Gaussian with mean zero and
covariance given by~\eqref{covariancevector}.
\begin{Thm} {\bf{(Realizing the canonical commutation relations)}} \label{thmccr}
Considering a nonlocal potential involving holographic phases~\eqref{Bansatz},
the bosonic field operators defined by~\eqref{hatBjdef} can be arranged 
by a suitable choice of the operators~$L^c_{a,b}$ in~\eqref{potprelim2}
and of the covariance in~\eqref{covariancevector} to satisfy the
CCR in the operator sense, i.e.\ (in analogy to~\eqref{CCRtensor})
\beq \label{CCRB}
\bbra \big[ \hat{\B}^j_q, \hat{\B}^k_{q'}  \big] \otimes \hat{\B}^l_{r} \otimes \hat{\B}^{l'}_{r'} \kket
\approx (2 \pi)^4\: \delta^4(q+q')\: \hat{K}^{jk}(q) \; 
\1 \otimes \bbra \hat{\B}^l_{r} \otimes \hat{\B}^{l'}_{r'} \kket \:,
\eeq
where~$\hat{K}^{ij}(q)$ is the causal fundamental solution of the Maxwell field.
Here~``$\approx$'' refers to the approximation~\eqref{approxdef}
(the errors of this approximation will be specified in Theorems~\ref{thmerr1} and~\ref{thmerr2} below).
\end{Thm} \noindent
In the example of the Feynman gauge, the causal fundamental solution takes the form
\[ 
\hat{K}^{jk}(q) = g^{jk}\: \delta(q^2)\: \epsilon(q^0) \:. \]
Our methods apply similarly also to other choices of gauge.
We note that, in view of the nonlocality of the potential~$\B$, also the causal fundamental solution
in the CCR will be regularized on the length scale~$\ell_{\min}$.
We do not need to make this explicit here, because the regularization
can be absorbed into the error terms.
\Proof[Proof of Theorem~\ref{thmccr}]
Taking the statistical mean of~\eqref{commgen}, we obtain
\begin{align}
& \bbra \big[ \hat{\B}^j_{q},\, \hat{\B}^k_{q'} \big] \kket \notag \\
& \approx
(2 \pi)^4\: \delta^4(q+q') \!\!\sum_{a,c,e,f=1}^N \!\!\hat{h}^{jk}_{c,f}(q)
\Big( \mathbf{D} \:\big[ \: (E_q \starD \mathbf{L}^c), \,(E_{-q} \starD \mathbf{L}^f) \big]
\:\mathbf{D}^{-1} \Big)_{a,e} \:. \label{covariance3}
\end{align}

This formula can be further simplified. We recall that the covariance
has the symmetry and positivity properties~\eqref{covsymm} and~\eqref{covpos}.
The symmetry property~\eqref{covsymm} implies that the matrix~$\hat{h}(q)$ is Hermitian.
Therefore, it can be diagonalized, having positive eigenvalues~$\lambda_d$ and
corresponding eigenvectors~$\phi_d \in \C^{4N}$ with complex components, i.e.\
\[ \hat{h}^{jk}_{a,b}(q) = \sum_{d=1}^{4N} \lambda_d(q)\: (\phi_d(q))^j_a\, \overline{(\phi_d(q))^k_b}\:. \]
Using this formula in~\eqref{covariance3}, we obtain
\[ 
\bbra \big[ \hat{\B}^j_{q},\, \hat{\B}^k_{q'} \big] \kket \\
\approx (2 \pi)^4\: \delta^4(q+q') \sum_{a,e=1}^N \sum_{d=1}^{4N}
\Big( \mathbf{D} \:\big[ \mathbf{M}^j_d(q), \,\mathbf{M}^k_d(-q) \big]
\:\mathbf{D}^{-1} \Big)_{a,e} \]
with
\begin{align}
\mathbf{M}^j_d(q) &:= \sqrt{\lambda_d(q)} \:\sum_{c=1}^N (\phi_d(q))^j_c\: (E_q \starD \mathbf{L}^c)  \label{g1} \\
\mathbf{M}^k_d(-q) &:= \sqrt{\lambda_d(q)} \:\sum_{f=1}^N 
\overline{(\phi_d(q))^k_f}\: (E_{-q} \starD \mathbf{L}^f) \:.
\end{align}
We note that, using again the symmetry properties~\eqref{covsymm}, 
by suitably ordering the eigenvectors of~$\hat{h}(q)$ and~$\hat{h}(-q)$ we can arrange that
\beq \label{Madjoint}
\big( \mathbf{M}^k_d(-q) \big)^* = \mathbf{M}^k_d(q)
\eeq
(where the star is the Hermitian adjoint of an $N \times N$-matrix).

In this formulation, the stochastic field and its coupling to the fermions is described
by the family of matrices~$\mathbf{M}^j_d(q)$ with~$j$ a tensor index, $d \in \{1, \ldots, 4N\}$
and~$q \in \hat{M}$. Apart from the symmetry condition~\eqref{Madjoint}, these matrices
can be chosen arbitrarily. This brings us into the position to satisfy the CCR,
which (in analogy to~\eqref{ccr} and~\eqref{hatccr}) we write as~\eqref{CCRB}.
In order to satisfy the CCR in the statistical mean, we must satisfy the relation
\beq
\sum_{a,e=1}^N \sum_{d=1}^{4N}
\Big( \mathbf{D} \:\big[ \mathbf{M}^j_d(q), \,\mathbf{M}^k_d(-q) \big]
\:\mathbf{D}^{-1} \Big)_{a,e} = 
\hat{K}^{jk}(q)\:\1_\K \:. \label{condn1}
\eeq
Moreover, in order to ensure that the outer parings are negligible
(as defined and explained after~\eqref{comm1}), it suffices to satisfy for all~$q \neq -q'$
the conditions
\beq
\big[ \mathbf{M}^j_d(q), \mathbf{M}^k_e(q') \big]
\ll \big[ \mathbf{M}^j_d(q), \mathbf{M}^k_d(-q) \big]  . \label{condn2}
\eeq

In order to satisfy conditions~\eqref{condn1} and~\eqref{condn2}, we proceed as follows.
Noting that~\eqref{condn2} involves the commutator of an operator with its own adjoint
(see~\eqref{Madjoint}), the commutator can be arranged to be large, even if the operators
have a small rank. Using this fact, for every~$d$ and~$q$ one chooses the operators~$\mathbf{M}^j_d(q)$
in such a way that~\eqref{condn1} holds. This can be done in agreement with the concept
described in Section~\ref{secdir} that the operators~$L_a$ are projection operators to
subspaces of~$\K$ (see also Figure~\ref{fighom}).
Next, the condition~\eqref{condn2} can be
satisfied by choosing the covariance in such a way that
the images of the operators~$\mathbf{M}^j_d(q)$ and~$\mathbf{M}^k_e(q')$ for~$d \neq e$ 
or~$q \neq q'$ are orthogonal, up to errors which can be made arbitrarily small by choosing~$N$
large.

We finally point out that the positivity condition~\eqref{covpos} can be satisfied
by adding to~$\hat{h}^{jk}_{ab}$ a sufficiently large multiple of the identity matrix~$\delta^{jk} \delta_{ab}$.
This has no effect on the CCR, because the identity matrix drops out of the commutator.
\QED

\begin{Remark} \label{remnatural} {\em{
The above arguments even show that {\em{the CCR arise naturally}}, as we now explain.
It is easiest to argue directly with the operators~$\mathbf{M}^j_d(q)$ introduced in~\eqref{g1},
which we regard as a family of stochastic operators on the Krein space~$(\K, \bra.|.\ket)$
(as introduced after~\eqref{stip}).
It is important to observe that these operators act as matrices on the holographic components
labeled by~$a \in \{1,\ldots, N\}$
(as one sees from the matrix notation introduced in~\eqref{notmatrix}).
The typical situation is that a holographic component~$a$ is mapped to itself
or to other holographic components in a commuting way (like the contributions
obtained in~\eqref{potprelim2} to zero order in an expansion in~$q \ell_{\min}$).
The non-commuting character of these operators can typically be regarded as a small correction
(obtained either by higher orders in an expansion in~$q \ell_{\min}$ or by
a coupling of the holographic components described in the next paragraph below).
Clearly, in the commutators in~\eqref{condn1} and~\eqref{condn2} only these small
corrections contribute. The images of these corrections are typically low-dimensional subspaces
of~$\K$ (as can be understood from the factors~${\mathbf{L}}^c$ in~\eqref{g1}; see
again Figure~\ref{fighom}).
Consequently, in a stochastic description, the images of two such operators
will be orthogonal up to errors which tend to zero if~$N$ tends to infinity.
The commutator on the right of~\eqref{condn2}, on the other hand,
is typically {\em{not}} small because one operator is the adjoint of the other
(see~\eqref{Madjoint}), meaning that these two operators are not statistically independent.
This explains~\eqref{condn2}. Moreover, it becomes clear why the left of~\eqref{condn1}
is typically {\em{not}} small. 
It remains to explain why the right side in~\eqref{condn1} has this specific form.
Here we can argue with causality and symmetries:
Combining~\eqref{covariancevector}
and~\eqref{condn1} with the
fact that the propagator~$\hat{K}^{jk}$ is causal, these conditions mean that these stochastic operators
should satisfy causality in the sense that they are uncorrelated for points with spatial separation.
Moreover, the precise form of the propagator in~\eqref{condn1} follows from the assumption of local
Lorentz covariance. Clearly, here the notions ``causality'' and ``Lorentz covariance'' are to be understood
modulo the effects of the regularization, which break Lorentz symmetry on the length scale~$\ell_{\min}$.

We finally explain how this qualitative picture can be understood from the constructions in~\cite{nonlocal}.
In this description, the holographic components correspond to null directions along the mass
cone (as depicted in Figure~\ref{fighom}).
In~\cite[Section~9.3]{nonlocal}, the linearized field equations are analyzed using the so-called
mass cone expansion. The zero order in this expansion is local in rays along the mass cone,
meaning that the linearized field equations decouple into independent equations for each holographic
component. This corresponds to the above-mentioned commuting contributions to the
operators~$\mathbf{M}^j_d(q)$.
The higher orders in the mass cone expansion, however, describe a coupling of the holographic components,
giving rise to non-commuting contributions to these operators. }} \QEDrem
\end{Remark}

\subsection{Stationary Phase Analysis of the Error Terms} \label{secstationaryphase}
In~\eqref{approxdef} we used the approximation where in the operator products~\eqref{toapprox}
we disregarded the terms involving oscillatory factors~$e^{i \Lambda_b - i \Lambda_a}$ with~$a \neq b$.
We now justify this approximation by working out the corresponding correction terms.
To this end, we choose two compactly supported wave functions~$\psi, \phi$ in spacetime and
consider the Krein inner product
\begin{align}
&\bra \psi \:|\: \hat{\B}^{j,c}_{a,b,q}\, \hat{\B}^{k,f}_{d,e,q'} \: \phi \ket \notag \\
&= \int_M d^4x \int_M d^4z \int_M d^4y\; \Sl \psi(x) \:|\: \hat{\B}^{j,c}_{a,b,q}(x,z)\:
\hat{\B}^{k,f}_{d,e,q'}(z,y) \: \phi(y) \Sr \:. \label{opprod}
\end{align}
We are interested in the situation where the phase factors oscillate on a length scale which is much
smaller than~$\ell_{\min}$. In this limiting case, we can compute the operator products
in the {\em{stationary phase approximation}}, giving the following result:
\begin{Thm} {\bf{(Error estimates, method I)}} \label{thmerr1} The approximation ``$\approx$''
introduced in~\eqref{approxdef} holds up to relative errors with the scaling behavior~\eqref{err1}.
\end{Thm}
The remainder of this section is devoted to the proof of this theorem.
Before entering the detailed computations, we determine the
scaling behavior and the errors of this approximation. We begin with the case~$b=d$ where the
operator product involves {\em{no}} intermediate phase factors, giving rise to the integral
\[ \int_M e^{-i q \frac{x+z}{2} }\: L^c_{a,b}(z-x) \:e^{-i q' \frac{z+y}{2} }\: L^f_{b,e}(y-z)\: d^4z \:. \]
In this case, the scaling of the integral is given simply
by the value of the integrand times the volume of the integration domain, i.e.\
\beq \label{nophase}
\sim L^c_{a,b}(y-x)\,  L^f_{b,e}(y-x) \, \ell_{\min}^4 \:.
\eeq
In the case~$b \neq d$, however, intermediate phase factors arise, giving integrals of the form
\beq \label{saddlephase}
\int_M e^{-i q \frac{x+z}{2} }\: L^c_{a,b}(z-x)\: e^{-i \lambda \big( \Lambda_b(z) - \Lambda_d(z) \big)}\: e^{-i q' \frac{z+y}{2} }\: L^f_{d,e}(y-z) \: d^4z \:.
\eeq
Now the main contribution to the integral comes from the critical points of the phase function~$ \Lambda_b - \Lambda_d$. In order to determine the scalings, let us assume that this function has an extremal point
at~$z_0$, i.e.\
\[ \frac{\partial}{\partial x^j} \big( \Lambda_b - \Lambda_d \big)(z_0) = 0 \:. \]
Then the phase is stationary provided that
\[ \big| z^j -z_0^j \big| \lesssim \ell_\Lambda \:, \]
where~$\ell_\Lambda$ is again the length scale on which the phase functions oscillate~\eqref{Lamscale}.
Therefore, the integral scales like
\beq \label{saddle}
\sim L^c_{a,b}(z_0-x)\: L^f_{d,e}(y-z_0) \, \ell_\Lambda^4 \:.
\eeq
Comparing~\eqref{nophase} with~\eqref{saddle}, one sees that the dephasing
decreases the size of the operator products for each stationary point by a scaling factor
\beq \label{saddlescale}
\Big( \frac{\ell_\Lambda}{\ell_{\min}} \Big)^4 \:.
\eeq
The errors of the stationary phase approximation can be quantified by expanding the other functions in~\eqref{saddlephase} in a Taylor series about the stationary points,
giving scaling factors~$\ell_\Lambda/\ell_{\min}$.
Therefore, the error of the stationary phase approximation is of the desired order~\eqref{err1}.
We conclude that, under the assumption~$\ell_\Lambda \ll \ell_{\min}$, the stationary phase approximation
is justified. The dephasing effect can be described by scaling factors~\eqref{saddlescale}, which are typically
very small.

The analysis of the error terms becomes more involved because we also must take into account
their combinatorics. In particular, the operator products for~$b \neq d$ may still be relevant if
the number of stationary points is very large.
The remainder of this section is devoted to a detailed analysis of the resulting counting.
Keeping in mind that the results of this analysis will not be used later in this paper,
it may be skipped by the reader more interested in the improved scaling in Section~\ref{secimproved}.
Before quantifying the counting, we collect the relevant formulas of the stationary phase approximation.
\begin{Lemma} \label{lemmasaddle}
Evaluating the inner product~\eqref{opprod} in the stationary phase approximation
gives the following results, up to errors of the order~\eqref{err1}. In the case~$b=d$, 
\begin{align}
\bra \psi \:|\: \hat{\B}^{j,c}_{a,b,q}\, \hat{\B}^{k,f}_{b,e,q'} \: \phi \ket
&= c  \int d^4z \sum_{k=1}^{K_2} \frac{1}{\sqrt{|\det D^2 \Lambda_a(x_k)|}}\:\frac{1}{\sqrt{|\det D^2 \Lambda_e(y_k)|}} 
\notag \\
&\quad\: \times \Sl \psi(x_k) \:|\: 
e^{i \Lambda_a(x_k)}\: \hat{A}^j_c(q) \: e^{-i q \frac{x_k+z}{2} }\: L^c_{a,b}(z-x_k) \notag \\
&\qquad\qquad \times  \: \hat{A}^k_f(q')\: e^{-i q' \frac{z+y_k}{2} }\: L^f_{b,e}(y_k-z) \: e^{-i \Lambda_e(y_k)} \: \phi(y_k) \Sr
\label{saddle2}
\end{align}
with a numerical constant~$c$. Here~$(x_k, y_k)$ with~$k \in \{1, \ldots, K_2\}$ denote the critical
points of the phase functions determined by the equations
\begin{align}
\frac{\partial}{\partial x^j} \bigg( \Lambda_a(x) - \frac{q}{2}\, x + \im \log \Big( \overline{\psi(x)}\, L^c_{a,b}(z-x) \Big)
\bigg) &= 0 \label{ps1} \\
\frac{\partial}{\partial y^j} \bigg( -\Lambda_e(y) - \frac{q'}{2}\, y + \im \log \Big( L^f_{d,e}(y-z)\, \phi(y) \Big) \bigg) &= 0 \:. \label{ps2}
\end{align}

Likewise, in the case~$b \neq d$, 
\begin{align}
&\bra \psi \:|\: \hat{\B}^{j,c}_{a,b,q}\, \hat{\B}^{k,f}_{b,e,q'} \: \phi \ket \notag \\
&= c \sum_{k=1}^{K_3} \frac{1}{\sqrt{|\det D^2 \Lambda_a(x_k)|}}\:
\frac{1}{\sqrt{|\det D^2 (\Lambda_b-\Lambda_d)(z_k)|}}\:\frac{1}{\sqrt{|\det D^2 \Lambda_e(y_k)|}} \notag \\
&\quad\: \times \Sl \psi(x_k) \:|\: 
e^{i \Lambda_a(x_k)}\: \hat{A}^j_c(q) \: e^{-i q \frac{x_k+z_k}{2} }\:
L^c_{a,b}(z_k-x_k) \:e^{-i \big( \Lambda_b(z_k) - \Lambda_d(z_k) \big)} \notag \\
&\qquad\qquad \times  \: \hat{A}^k_f(q')\: e^{-i q' \frac{z_k+y_k}{2} }\: L^f_{d,e}(y_k-z_k) \: e^{-i \Lambda_e(y_k)} \: 
\phi(y_k) \Sr \:, \label{saddle3}
\end{align}
where the critical points~$(x_k, z_k, y_k)$ with~$k \in \{1, \ldots, K_3\}$ are determined by the equations
\begin{align}
\frac{\partial}{\partial x^j} \bigg( \Lambda_a(x) - \frac{q}{2}\, x + \im \log \Big( \overline{\psi(x)}\, L^c_{a,b}(z-x) \Big)
\bigg) &= 0 \label{ps3} \\
\frac{\partial}{\partial z^j} \bigg( -\Lambda_b(z) + \Lambda_d(z) - \frac{q+q'}{2}\, z + \im \log \Big( L^c_{a,b}(z-x)\, L^f_{d,e}(y-z) \Big) \bigg) &= 0 \\
\frac{\partial}{\partial y^j} \bigg( -\Lambda_e(y) - \frac{q'}{2}\, y + \im \log \Big( L^f_{d,e}(y-z)\, \phi(y) \Big) \bigg) &= 0 \:.
\label{p5}
\end{align}
\end{Lemma}
\Proof
Using~\eqref{hatBjdef} in~\eqref{opprod}, we obtain
\begin{align*}
&\bra \psi \:|\: \hat{\B}^{j,c}_{a,b,q}\, \hat{\B}^{k,f}_{d,e,q'} \: \phi \ket \\
&= \int_M d^4x \int_M d^4z \int_M d^4y\; \Sl \psi(x) \:|\: 
e^{i \Lambda_a(x)}\: \hat{A}^j_c(q) \: e^{-i q \frac{x+z}{2} }\:
L^c_{a,b}(z-x) \\
&\qquad \times e^{-i \big( \Lambda_b(z) - \Lambda_d(z) \big)}\: \hat{A}^k_f(q') \: e^{-i q' \frac{z+y}{2} }\:
L^f_{d,e}(y-z) \: e^{-i \Lambda_e(y)} \: \phi(y) \Sr \:.
\end{align*}
In the case~$b=d$, the intermediate phase factor vanishes,
\begin{align*}
&\bra \psi \:|\: \hat{\B}^{j,c}_{a,b,q}\, \hat{\B}^{k,f}_{b,e,q'} \: \phi \ket \\
&= \int_M d^4x \int_M d^4z \int_M d^4y\; \Sl \psi(x) \:|\: 
e^{i \Lambda_a(x)}\: \hat{A}^j_c(q) \: e^{-i q \frac{x+z}{2} }\:
L^c_{a,b}(z-x) \\
&\qquad \times \hat{A}^k_f(q') \: e^{-i q' \frac{z+y}{2} }\:
L^f_{b,e}(y-z) \: e^{-i \Lambda_e(y)} \: \phi(y) \Sr \:.
\end{align*}
In the case~$b \neq d$, we can also apply the stationary phase approximation to the $z$-integral.
\QED

We now describe the method for evaluating the sums over the stationary points.
Here we need to take into account that the formulas for the expectation values in Lemma~\ref{lemmasaddle}
involve phase factors, which means that summing over the stationary points may lead to cancellations
due to dephasing. This effect can be treated systematically as follows.

In preparation, we specify the number of stationary points further.
Clearly, the two equations~\eqref{ps1} and~\eqref{ps2}
(and similarly the three equations~\eqref{ps3}--\eqref{p5}) are coupled because both equations
involve both~$x$ and~$y$. However, they decouple when only the functions~$\Lambda_a$ and~$\Lambda_e$
are taken into account. 
Therefore, the solutions of these coupled equations can be obtained by first computing the
stationary points of~$\Lambda_a$ and~$\Lambda_e$ and then treating the other functions
in~\eqref{ps1} and~\eqref{ps2} perturbatively (giving again an expansion in powers of~$\ell_\Lambda/\ell_{\min}$).
We do not need to enter the details of this construction.
This procedure shows that the total number of stationary points is given by               
\[ K_2 \simeq K^2 \qquad \text{and} \qquad K_3 \simeq K^3 \:, \]
where the parameter~$K$ denotes the average number of stationary points of the functions~$\Lambda_a$.

We begin with the formula~\eqref{saddle2} in the case~$b=d$. First, we recall that evaluating
each stationary point gives a scaling factor~\eqref{saddlescale}. It is convenient to introduce the abbreviation
\beq \label{alphadef}
\alpha := \Big( \frac{\ell_\Lambda}{\ell_{\min}} \Big)^4 \ll 1 \:.
\eeq
Therefore, each summand in~\eqref{saddle2} has the scaling
\[ \sim \alpha^2\:\big| L^\circ_{\circ,\circ} \big|^2\: \ell_{\min}^{12} \]
(where for ease of notation, we leave out the factors~$\psi$, $\phi$, $\hat{A}^j_c(q)$ and~$\hat{A}^k_f(q')$
and do not specify the arguments and indices of the operator~$L^\circ_{\circ,\circ}$).
Next, one must keep in mind that the product of the kernels~$L^\circ_{\circ,\circ}(z-x_k)$ and~$L^\circ_{\circ,\circ}(y_k-z)$ vanishes
unless the points~$x_k$ and~$y_k$ are close on the scale~$\ell_{\min}$ in the sense that
\beq \label{mincond}
\big| x_k^j -y_k^j \big| \lesssim \ell_{\min} \:.
\eeq
This reduces the number of stationary points to be taken into account by a factor which we denote by
\beq \label{betarel}
\beta \ll 1 \:.
\eeq
If the stationary points~$x_k$ and~$y_k$ are close even on the scale~$\ell_\Lambda$, i.e.\ if
\beq \label{Lamcond}
\big| x_k^j -y_k^j \big| \lesssim \ell_\Lambda \:,
\eeq
then the phase function~$\Lambda_a(x_k) - \Lambda_a(y_k)$ is close to zero. As a consequence,
there are no relative phases, and no dephasing occurs. 
Therefore, the total contribution
of these stationary points to~\eqref{saddle2} is obtained simply by counting the number of such stationary
points. For this counting, we must take into account that the condition~\eqref{Lamcond}
reduces the number of stationary points by an additional factor~$\alpha$
(because, given~$x_k$, the four-volume of the ball in which~$y_k$ may lie is reduced by a factor~$\alpha$).
Moreover, we must keep in mind that, to leading order in~$\ell_\Lambda/\ell_{\min}$, the equations in~\eqref{ps1}
and~\eqref{ps2} have the same solutions, implying that there are at least~$K$ stationary points which satisfy~\eqref{Lamcond}. Therefore, the total number of stationary points satisfying~\eqref{mincond} is given
by~$(1 + \alpha \beta K)K$. We thus obtain the scaling
\beq \label{nophasei}
\bra \psi \:|\: \hat{\B}^{j,c}_{a,b,q}\, \hat{\B}^{k,f}_{b,e,q'} \: \phi \ket \simeq
(1+\alpha \beta K)\,K\: \alpha^2\: \big| L^\circ_{\circ,\circ} \big|^2\: \ell_{\min}^{12} \:.
\eeq
Before going on, we point out that the number~$\alpha \beta K$ tells us about the average
number of stationary points inside a ball of radius~$\ell_\Lambda$. In order for the stationary phase
approximation to be sensible, the distance of the stationary phases must be much larger than~$\ell_\Lambda$.
This leads to the condition
\beq \label{alphabetarel}
\alpha \beta K \ll 1 \:.
\eeq
Therefore, we may leave out the summand~$\alpha \beta K$ in~\eqref{nophasei} to obtain
\beq \label{nophase2}
\bra \psi \:|\: \hat{\B}^{j,c}_{a,b,q}\, \hat{\B}^{k,f}_{b,e,q'}\: \phi \ket \simeq
K \alpha^2\: \big| L^\circ_{\circ,\circ} \big|^2\: \ell_{\min}^{12} \qquad \text{without dephasing} \:.
\eeq

If the stationary points do {\em{not}} satisfy~\eqref{Lamcond}, we can treat the factors~$e^{i (\Lambda_a(x_k)-\Lambda_a(y_k)}$ stochastically as random phases. Therefore, the amplitude of the sum is
obtained by multiplying the amplitude of one summand by the square root of the number of summands.
This gives the scaling
\beq \label{dephase2}
\bra \psi \:|\: \hat{\B}^{j,c}_{a,b,q}\, \hat{\B}^{k,f}_{b,e,q'} \: \phi \ket \simeq
\sqrt{\beta K^2} \: \alpha^2\:\big| L^\circ_{\circ,\circ} \big|^2\: \ell_{\min}^{12} \qquad \text{with dephasing} \:.
\eeq
Note that, in view of~\eqref{betarel}, this contribution is much {\em{smaller}} than~\eqref{nophase2}.

In the case~$b \neq d$, we can proceed similarly. Yet, we must keep in mind that also the $z$-integral
is evaluated in the stationary phase approximation. Therefore, each summand in~\eqref{saddle3} has the
scaling
\[ \sim \alpha^3\:\big| L^\circ_{\circ,\circ} \big|^2\: \ell_{\min}^{12} \]
(for ease of notation, we again leave out the factors~$\psi$, $\phi$, $\hat{A}^j_c(q)$ and~$\hat{A}^k_f(q')$).
If all three points~$x_k$, $y_k$ and~$z_k$ are close to the scale~$\ell_\Lambda$, we have no dephasing.
When counting the number of corresponding stationary points, we must keep in mind that, similar as explained
above, there are at least~$K$ stationary points satisfying~\eqref{Lamcond}, and in this case
there is also a solution~$z_k$ which is close to~$x_k$ and~$y_k$ on the scale~$\ell_\Lambda$.
We thus obtain the scaling
\[ \bra \psi \:|\: \hat{\B}^{j,c}_{a,b,q}\, \hat{\B}^{k,f}_{b,e,q'} \: \phi \ket \simeq
(1 + \alpha^2 \beta^2 K^2)\, K \: \alpha^3\:\big| L^\circ_{\circ,\circ} \big|^2\: \ell_{\min}^{12} \:. \]
In view of~\eqref{alphabetarel}, we can leave out the summand~$\alpha^2 \beta^2 K^2$, i.e.\
\beq \label{nophase3}
\bra \psi \:|\: \hat{\B}^{j,c}_{a,b,q}\, \hat{\B}^{k,f}_{b,e,q'} \: \phi \ket \simeq
K \alpha^3\:\big| L^\circ_{\circ,\circ} \big|^2\: \ell_{\min}^{12} \qquad \text{without dephasing} \:.
\eeq
If none of the points~$x_k$, $y_k$ and~$z_k$ are close to the scale~$\ell_\Lambda$, we obtain
\beq \label{dephase3}
\bra \psi \:|\: \hat{\B}^{j,c}_{a,b,q}\, \hat{\B}^{k,f}_{b,e,q'} \: \phi \ket \simeq
\sqrt{\beta^2 K^3} \: \alpha^3\:\big| L^\circ_{\circ,\circ} \big|\: \ell_{\min}^{12} \qquad \text{with dephasing} \:.
\eeq
Finally, there is an additional contribution if two of the points~$x_k$, $y_k$ and~$z_k$ are close to the
scale~$\ell_\Lambda$, but not the third point. They have the scaling
\[ \bra \psi \:|\: \hat{\B}^{j,c}_{a,b,q}\, \hat{\B}^{k,f}_{b,e,q'} \: \phi \ket \simeq
\sqrt{\alpha \beta^2 K^3} \: \alpha^3\:\big| L^\circ_{\circ,\circ} \big|^2\: \ell_{\min}^{12} \:. \]
Since this is much smaller than~\eqref{dephase3}, we may disregard these contributions in what follows.

The precise scaling behavior of the above stationary phase contributions~\eqref{nophase2}, \eqref{dephase2}
and~\eqref{nophase3}, \eqref{dephase3} will be important when working out corrections to the
Fock space dynamics (see Section~\ref{secoutlook}). For the purpose of the present paper, it suffices to note that,
apart from the constraints in~\eqref{alphadef}, \eqref{betarel} and~\eqref{alphabetarel}, the
parameters~$\alpha$, $\beta$ and~$K$ can be chosen arbitrarily. Therefore, all the stationary phase
contributions can be made arbitrarily small by choosing~$\alpha$ sufficiently small.
In this way, we can arrange that the error terms are indeed of the desired form~\eqref{err1}.

\section{The Dirac Dynamics with Holographic Mixing} \label{secdynholo}
\subsection{Preparatory Considerations} \label{secdynholoprep}
We now briefly outline the path taken by the authors to arrive at the formulation
of the dynamical equations with holographic mixing.
Although not quite straightforward, these considerations
may nevertheless explain and motivate the subsequent construction of the holographic Green's operator
(in Section~\ref{secngenholo}). At the beginning of Section~\ref{secBdyn}
we argued that the conditions~\eqref{Kproj1}
and~\eqref{Kproj2} are too restrictive and must be relaxed. If this is done, the operator~$U$
defined by~\eqref{Udef} is no longer unitary.
Introducing the holographic phases again by~\eqref{dirphase2} has the disadvantage
that, rewriting the Dirac operator similar to~\eqref{dirphase} as the sum~$i \Pdd + \B_\Lambda$
of the vacuum Dirac operator and a perturbation, the operator~$\B_\Lambda$
involves a differential operator whose coefficients are convolution operators
including the holographic phases. As a consequence, the operator~$\B_\Lambda$ can no longer be regarded
as a small perturbation, and it cannot be treated perturbatively in a straightforward way.
In order to bypass this difficulty, one can try to construct a {\em{unitary}} operator~$V$ on the
Krein space~$(\K, \bra .|. \ket)$ (introduced after~\eqref{stip}) which again involves the holographic phases.
Then the procedure
\begin{align*}
i \Pdd \rightarrow V (i \Pdd) V^*
&= \frac{1}{2}\: (i \Pdd) \:V V^* - \frac{1}{2}\: \big[ i \Pdd, V \big]\: V^*
+  \frac{1}{2}\: V V^*\:(i \Pdd) + \frac{1}{2}\: V\: \big[ i \Pdd, V^* \big] \\
&= i \Pdd + \B_\Lambda
\end{align*}
with
\beq \label{BLamdef}
\B_\Lambda := -\frac{1}{2}\: \big[ i \Pdd, V \big]\: V^*  + \frac{1}{2}\: V\: \big[ i \Pdd, V^* \big] \:,
\eeq
gives rise to a potential~$\B_\Lambda$ which, due to its commutator structure, is no longer a
differential operator. We refer to~$V$ as the {\em{unitary holographic mixing operator}}.
The obvious question is how to choose or construct this operator. One strategy is to begin with
a potential of the form~\eqref{Bgaugelin}; in the simplest case
\beq \label{Bsimp}
\B_\Lambda(x,y) = \sum_{a=1}^N (\Pdd \Lambda_a) \Big( \frac{x+y}{2} \Big)\:L_a(y-x) \:,
\eeq
and to construct a unitary operator~$V$ with the property that
\[ V (i \Pdd) V^* = i\Pdd + \B_\Lambda \:. \]
A perturbative treatment in powers of~$\B_\Lambda$ is not suitable because the
oscillatory phase factors must be treated non-perturbatively. But, as is worked out in detail in
Appendix~\ref{apppert}, a resummation technique makes it possible to compute~$V$ in an
\beq \label{expansion}
\text{expansion in powers of~$\ell_{\min}/\ell_\Lambda$}\:.
\eeq
Here all phase factors~$e^{\pm i \Lambda_a}$ are taken into account non-perturbatively.
However, there remains the limitation that the length scale~$\ell_\Lambda$ on which the
phase factors oscillate (see~\eqref{Lamscale}) must be much larger than~$\ell_{\min}$.
At present, it is not clear whether this condition is satisfied in physically realistic situations.
For this reason and for the sake of mathematical generality,
it seems preferable not to rely on an expansion of the form~\eqref{expansion}.

In order to avoid the expansion~\eqref{expansion}, one can construct the operator~$V$ non-perturbatively,
for example by taking the polar decomposition of the operator~$U$. To this end, one merely needs to assume that
\[ 
U \::\: L^2(M,\C) \rightarrow L^2(M,\C) \quad \text{is bounded and has a bounded inverse} \]
(note that, for technical simplicity, we restrict attention to mappings which act trivially on the spinor indices,
making it possible to work in the Hilbert space~$L^2(M,\C)$ rather than the Krein space~$(\K, \bra .|. \ket)$
introduced after~\eqref{stip}).
Then the operator~$V$ defined by
\beq \label{Vdef}
V := U\, \big(U^* U\big)^{-\frac{1}{2}} \::\: L^2(M,\C) \rightarrow L^2(M,\C)
\eeq
is unitary, as desired. This abstract method has the disadvantage that the operator~$V$
can no longer be given explicitly. Yet, as illustrated in Appendix~\ref{appholo}, it can still be
computed using microlocal techniques, but again under the assumption~$\ell_\Lambda \gg \ell_{\min}$.

Above we outlined various methods for defining and analyzing a unitary holographic mixing operator~$V$.
It turns out that these methods are not yet sufficient, because there is the general problem
that the fermionic Green's operator is incompatible with the bosonic commutation relations. We now explain the basic difficulty, which will be resolved by the constructions in Section~\ref{secngenholo}.
We write the Dirac equation as
\beq \label{discunit}
0 = \big( i \Pdd + \B - m \big) \tilde{\psi}  = V (i \Pdd - m) V^* \tilde{\psi} + \B_\dyn \tilde{\psi} \:,
\eeq
where~$V$ is a unitary operator on the Krein space~$(\K, \bra .|. \ket)$
(for example the operator~\eqref{Vdef}), and~$\B_\dyn$ is the dynamical gauge potential
(for example of the form~\eqref{Bansatz}).
Considering the operator~$V (i \Pdd - m) V^* \tilde{\psi}$ as the unperturbed operator,
the corresponding Green's operator is given explicitly by~$V s_m V^*$,
where~$s_m$ is a Dirac Green's operator (like for example the retarded Green's operator as
given in momentum space by~\eqref{smwedge}). Now we can treat~$\B_\dyn$ perturbatively,
giving rise to the perturbation series
\begin{align}
\tilde{\psi} &= \sum_{n=0}^\infty \Big(- V s_m V^* \B_\dyn \Big)^n \, V\,\psi \notag \\
&= V \psi - \big(V s_m \,V^* \big) \B_\dyn \, V \psi 
+ \big( V s_m V^* \big) \B_\dyn \big( V s_m \,V^* \big) \B_\dyn \, V \psi - \cdots \:. \label{BpertV}
\end{align}
This perturbation series describes the dynamics of the system completely. Therefore, it should
correspond to the Dyson series in perturbative QFT.

It is useful to consider~$V s_m \,V^*$ as the effective Green's operator; we also refer to it as
the {\em{holographic Green's operator}} and denote it by
\[ s^\hol := V s_m \,V^* \:. \]
In Sections~\ref{secrealizecomm} and~\ref{secstationaryphase}
we showed that the potential~$\B_\dyn$ can be regarded as being composed of the bosonic field operators,
which act on the Krein space~$(\K, \bra .|. \ket)$ and satisfy the CCR.
In order to make use of these commutation relations in~\eqref{BpertV}, the bosonic field operators must
also be commuted with the holographic Green's operators. In the setting of QFT, this
is unproblematic, because the bosonic and fermionic field operators act on different spaces
(the bosonic and fermionic Fock space, respectively), and therefore they commute by construction.
In our setting, however, all field operators act on the same space: the Krein space~$(\K, \bra .|. \ket)$
formed of one-particle wave functions in Minkowski space. Therefore, the bosonic field operators
will in general {\em{not}} commute with the holographic Green's operator. Arranging such commutation relations
poses non-trivial conditions for the choice of the unitary holographic mixing operator.

In order to satisfy these conditions, we need more freedom for the choice of the holographic
Green's operator. A first step in this direction is to drop the condition that~$V$ be unitary.
Thus we let~$V$ be an operator on the Krein space~$(\K, \bra .|. \ket)$
which is merely assumed to be invertible, i.e.\
\[ V, V^{-1} \::\: \K \rightarrow \K \]
(more precisely, we assume that~$V$ and~$V^{-1}$ are continuous with respect to the Krein topology).
Writing the Dirac equation again in the form~\eqref{discunit}, the only difference compared to the
above setting is that we need to carefully distinguish between~$V^*$ and~$V^{-1}$ as well as
between~$V$ and~$(V^{-1})^*$. In particular, the holographic Green's operator now becomes
\beq \label{sholV}
s^\hol := \big( V^{-1} \big)^*\, s_m \,V^{-1} \:.
\eeq
Likewise, in the perturbation expansion~\eqref{BpertV} we need to replace the factors~$V s_m V^*$
by this holographic Green's operator.

The next and final step is to observe that it is not crucial for our constructions that the
Dirac operator and the Green's operator are of the form~\eqref{discunit} and~\eqref{sholV}.
Instead, we can work more generally with a pair of operators~$\Dir^\hol$ and~$s^\hol$ with the following
properties:
\begin{align}
\big( \Dir^\hol \big)^* &= \Dir^\hol && \text{(symmetry of the Dirac operator)} \label{dirsymm} \\
\big( \Dir^\hol - m \big)\, s^\hol &= \1 && \text{(defining property of the Green's operator)}\:. \label{greendef}
\end{align}
Here the symmetry is needed in order to obtain a conserved inner product
which generalizes~\eqref{c11}--\eqref{c3}. More precisely, this inner product can be
constructed as follows. In case that the operator~$\Dir^\hol-m$ can be written as an integral operator
with a locally integrable kernel, denoting this kernel by~$Q(x,y)$ one can work directly with~\eqref{cip}.
In case that the integral kernel is a singular distribution on the diagonal~$x=y$, one needs
to first regularize, then compute the surface layer integral~\eqref{cip} and finally remove the regularization.
For example, for the singular kernel~$Q(x,y) = (i \Pdd_x-m)\, \delta^4(y-x)$, this procedure
gives back, up to a prefactor, the usual Dirac scalar product~\eqref{c11}\footnote{More precisely, choosing
a test function~$\eta \in C^\infty_0(M, \R^+_0)$ which is symmetric (i.e.~$\eta(-\xi)=\eta(\xi)$
for all~$\xi \in M$), setting
\[ Q^\delta(x,y) := \frac{1}{\delta^4}\: (i \Pdd_x-m) \:\eta \Big( \frac{x-y}{\delta} \Big) \]
and choosing~$\Omega$ as the past of the Cauchy surface at time~$t$,
a direct computation using integration by parts yields for any smooth~$\psi$ and~$\phi$
\[ -2i \lim_{\delta \searrow 0} \bigg( \int_{\Omega} \!d\rho(x) \int_{M \setminus \Omega} \!\!\!\!\!\!\!\!d\rho(y)
- \int_{M \setminus \Omega} \!\!\!\!\!\!\!\!d\rho(x) \int_{\Omega} \!d\rho(y) \bigg)
\;\Sl \psi(x) \:|\: Q^\delta(x,y)\, \phi(y) \Sr_x 
= c \int \Sl \psi \,|\, \gamma^0\, \phi \Sr_{(t,\vec{x})} \: d^3x \]
with~$c$ a nonzero constant.}.

Of course, working in the general setting~\eqref{dirsymm} and~\eqref{greendef},
one faces the computational difficulty that
the solutions of the holographic Dirac equation~$(\Dir^\hol - m) \psi=0$ can no longer
be obtained from standard Dirac solutions by a perturbative treatment.
From the conceptual point of view, however, the holographic Dirac equation provides a
clean and simple way of describing the dynamics including holographic phases.
This will be worked out in more detail in the next section.

\subsection{The Holographic Green's and Dirac Operators} \label{secngenholo}
We now return to our ansatz for the dynamical potential involving holographic phases~\eqref{Bansatz}.
We have the situation in mind that the considered wave function~$\psi$ can be decomposed into a
sum of holographic components
\beq \label{psiholo}
\psi = \sum_{a=1}^N e^{i \Lambda_a}\, \psi_a
\eeq
with wave functions~$\psi_1, \ldots, \psi_N$ whose momenta are much smaller than~$1/\ell_\Lambda$.
Then, using again the approximation~\eqref{approxdef} of disregarding dephased contributions,
the potential~\eqref{Bansatz} acts componentwise, i.e.\
\[ \B_\dyn \psi \approx \sum_{a,b,c=1}^N e^{i \Lambda_a} \:\big( \slashed{A}_{c} \starD L^c_{a,b} \big)\: \psi_b \:. \]
This corresponds to the concept that, similar to classical gauge phases, the holographic phase factors
cannot be observed but merely describe a mixing of internal degrees of freedom.
Following this concept, the Green's operator should also act on each holographic component
irrespective of the holographic phases, i.e.\
\beq \label{sholapprox}
s^\hol \psi \approx \sum_{a=1}^N e^{i \Lambda_a}\, (s_m \psi_a) \:,
\eeq
where~$s_m$ is again the vacuum Dirac Green's operator. 
Following up on the explanation in the last paragraph of Remark~\ref{remnatural},
we note that the relation~\eqref{sholapprox} is also motivated by the constructions in~\cite{nonlocal}
which show that the Green's operator~$s^\hol$ should act on
each holographic component like the vacuum Dirac Green's operator.

A simple and direct way of implementing the relation~\eqref{sholapprox}
is to define the holographic Green's operator by
\beq \label{sholdef}
s^\hol := \sum_{a=1}^N e^{i \Lambda_a}\: s_m \:e^{-i \Lambda_a} \:.
\eeq
This Green's operator can be arranged to be {\em{advanced}} or {\em{retarded}} simply
by choosing~$s_m = s_m^\vee$ or~$s_m = s_m^\wedge$, respectively.
It acts on wave functions by
\[ s^\hol\, \psi = \sum_{a=1}^N e^{i \Lambda_a}\, (s_m \psi_a)
+ \sum_{a,b = 1,\ldots, N \text{ with } a \neq b} e^{i \Lambda_a}\, s_m\,
\big( e^{-i \Lambda_a + i \Lambda_b}\: \psi_b \big) \:. \]
The summands for~$a \neq b$ are small for the following reason.
The phase factor typically gives rise to high-frequency wave functions.
Due to the decay of the Green's operator for large momenta, these high-frequency
contributions become very small when acted upon by the Green's operator. More precisely, they give
rise to errors of the order
\[ \times \Big( 1 + \O \big( \ell_\macro / \ell_\lambda \big) \Big) \:. \]
This can be made precise with a stationary phase analysis as worked out in Section~\ref{secstationaryphase}.
We conclude that the holographic Green's operator~\eqref{sholdef} indeed satisfies the condition~\eqref{sholapprox}.

Having introduced the holographic Green's operator, we can arrange~\eqref{greendef} by
defining the {\em{holographic Dirac operator}} by
\[ \Dir^\hol := \big( s^\hol \big)^{-1} + m \]
(here we need to assume that the inverse exists).
Using~\eqref{sholdef}, this operator is again symmetric~\eqref{dirsymm}.
This symmetry property is needed in order for current conservation to hold, and it is indeed the
only assumption needed for this conservation law.
In order to see this, we write the holographic Dirac equation as
\[ \int_M Q^\hol(x,y)\: \psi(y)\: d^4y = 0 \:, \]
where~$Q^\hol$ is a distributional kernel which is symmetric in the sense that
\[ Q^\hol(x,y)^* = Q^\hol(y,x) \]
(where the star is the adjoint with respect to the spin inner product).
Then the commutator inner product~\eqref{c11}--\eqref{c3} can be generalized and
written again in the form~\eqref{cip} with the kernel~$Q^\text{dyn}$ replaced by~$Q^\hol(x,y)$
(as was explained in more detail after~\eqref{greendef}).
Although this conservation law will not enter the subsequent constructions,
it is important conceptually because it yields a scalar product on
the physical wave functions even in the interacting regime.

\subsection{The Holographic Perturbation Expansion}
We now insert the dynamical gauge potentials~\eqref{Bansatz} into the holographic Dirac equation,
\[ 
\big( \Dir^\hol + \B_\dyn - m \big)\, \psi = 0 \:. \]
The corresponding perturbation expansion becomes
\beq \label{pertholo}
\tilde{\psi} = \sum_{n=0}^{\infty} \big( -s^\hol\, \B_\dyn \big)^n \,\psi \:.
\eeq
Using~\eqref{psiholo}, \eqref{Bansatz} and~\eqref{sholdef}, we can proceed as in
Section~\ref{secrealizecomm} and work with the approximation~\eqref{approxdef},
leaving out all terms involving rapidly oscillating phase factors. We again use a matrix notation in
the components~\eqref{notmatrix}. Moreover, we use a vector notation for the fermionic waves
\[ 
|\psi\ket := (\psi)_{a=1,\ldots, N} \]
and similarly with the tilde. We thus obtain the {\em{holographic perturbation expansion}}
\beq \label{tildeapprox1}
| \tilde{\psi} \ket \approx \sum_{n=0}^{\infty} \Big[ -s_m\, \sum_{b=1}^N \slashed{A}_b \starD \mathbf{L}^b
\Big]^n \,|\psi \ket \:.
\eeq
The error of this approximation will be worked out in detail in the next section.

\subsection{Improved Scaling of the Error Terms} \label{secimproved}
We now adapt the stationary phase analysis of the error terms in Section~\ref{secstationaryphase}
to the perturbation expansion~\eqref{pertholo}. Our main finding will be that, as a consequence of the
intermediate Green's operators, the scaling behavior of the error terms improves considerably:
\begin{Thm} {\bf{(Error estimates, method II)}} \label{thmerr2} The approximation ``$\approx$''
introduced in~\eqref{approxdef} holds up to relative errors with the scaling behavior~\eqref{err2}.
\end{Thm}
The major improvement compared to~\eqref{err1} is that it becomes
unnecessary to assume that~$\ell_\Lambda \ll \ell_{\min}$.
Instead, we can work with the opposite limiting case~$\ell_{\min} \ll \ell_{\Lambda}$,
which also makes it possible to apply the perturbative and
microlocal techniques outlined in Appendices~\ref{apppert} and~\ref{appholo}.

The remainder of this section is devoted to the proof of this theorem.
Our starting point is the expectation value of an operator product of the general form
\beq \label{exgen}
\bra \psi \:|\: e^{i \Phi_0} \,K_1 \,e^{i \Phi_1}\, K_2 \,e^{i \Phi_2}  \cdots e^{i \Phi_{p-1}} \,K_p\,
e^{i \Phi_p} \: \phi \ket \:,
\eeq
where the factors~$K_\ell$ are homogeneous operators (either~$s_m$ or~$L^\circ_{\circ,\circ}$),
and~$e^{i \Phi_\ell}$ are products of two adjacent phase factors,
which we write as
\beq \label{Phidef}
\Phi_\ell := -\Lambda_{a_\ell} + \Lambda_{b_\ell}
\eeq
with parameters~$a_\ell, b_\ell \in \{0, \ldots, p\}$.
We write the operator products as multiple integrals in position space,
\begin{align*}
&\bra \psi \:|\: e^{i \Phi_0} \,K_1\, e^{i \Phi_1} \,K_2\, \cdots e^{i \Phi_{p-1}} \,K_p\, e^{i \Phi_p}\:
\phi \ket \\
&= \int_M d^4x_0 \cdots \int_M d^4x_p \;\Sl \psi(x_0) \:|\: e^{i \Phi_0(x_0)}
\:K_1(x_1-x_0)\: e^{i \Phi_1(x_1)}\: K_2(x_2-x_1) \\
&\qquad\qquad\qquad\qquad\qquad\qquad \: \cdots\: e^{i \Phi_{p-1}(x_{p-1})}
\:K_p(x_p-x_{p-1})\: e^{i \Phi_{p}(x_{p})}\: \phi(x_p) \Sr \:.
\end{align*}
We now transform to the new integration variables
\[ x_0 \qquad \text{and} \qquad \xi_\ell := x_\ell - x_{\ell-1} \text{ with } \ell \in \{1,\ldots, p\} \]
to obtain
\begin{align*}
&\bra \psi \:|\: e^{i \Phi_0} \,K_1\, e^{i \Phi_1} \,K_2\, \cdots e^{i \Phi_{p-1}} \,K_p\, e^{i \Phi_p}\:
\phi \ket \\
&= \int_M d^4x_0 \int_M d^4 \xi_1 \cdots \int_M d^4 \xi_p \;\Sl \psi(x_0) \:|\: e^{i \Phi_0(x_0)}
\:K_1(\xi_1)\: e^{i \Phi_1(x_1)}\: K_2(\xi_2) \\
&\qquad\qquad\qquad\qquad\qquad\qquad \: \cdots\: e^{i \Phi_{p-1}(x_{p-1})}
\:K_p(\xi_p)\: e^{i \Phi_{p}(x_{p})}\: \phi(x_p) \Sr \:,
\end{align*}
where the variables~$x_1, \ldots, x_p$ are expressed in terms of~$x_0$ and~$\xi_1, \ldots, \xi_p$ by
\[ x_\ell = x_0 + \xi_1 + \cdots +\xi_\ell \:. \]
We now carry out the integrals successively. The $x_0$-integral involves all the phase factors
and the wave functions~$\psi$ and~$\phi$ but not the kernels~$K_\ell$    
(note that these kernels do not depend on~$x_0$).
Therefore, we get a contribution only if all the phases cancel. Using the form of the phase factors~\eqref{Phidef},
we conclude that there must be a permutation~$\sigma \in S_p$ with
\beq \label{abcond}
a_\ell = b_{\sigma(\ell)} \qquad \text{for all~$\ell=0,\ldots, p$} \:.
\eeq
The error of this approximation can again be specified with the stationary phase analysis.
Similar to~\eqref{saddlescale} the error terms involve scaling factors
\[ 
\Big( \frac{\ell_\Lambda}{\ell_\macro} \Big)^4 \:. \]

We next carry out the integrals over~$\xi_p$, $\xi_{p-1}$, \ldots, $\xi_1$ in this order.
The integral over~$\xi_p$ takes the form
\[ \int_M K_p(\xi_p)\: e^{i \Phi_p(\xi_p + \zeta_p)}\: \phi(\xi_p + \zeta_p)\: d^4 \xi_p \]
with~$\zeta_p:= x_0 + \xi_1 + \cdots + \xi_{p-1}$.
We need to consider the cases~$K_p = s_m$ and~$K_p=L^\circ_{\circ,\circ}$ separately. In the first case, we
integrate the Dirac matrices by parts,
\[ \int_M s^\wedge_m(\xi)\: e^{i \Phi_p(\xi + \zeta)}\: \phi(\xi + \zeta)\: d^4 \xi 
= \int_M S^\wedge_{m^2}(\xi)\: (i \Pdd_{\xi} +m) \big( e^{i \Phi_p(\xi + \zeta)}\: \phi(\xi + \zeta) \big)\: d^4\xi \:, \]
where the factors~$S^\wedge_{m^2}$ are the Klein-Gordon Green's operators,
which in position space take the form (see for example~\cite[eq.~(2.2.6)]{cfs}),
\begin{align*}
S^\wedge_{m^2}(x,y) &= -\frac{1}{2 \pi} \:\delta \big( (y-x)^2 \big) \:
\Theta \big( x^0 - y^0 \big) \notag \\
&\quad\:+ \frac{m^2}{4 \pi} \:\frac{J_1 \Big( \sqrt{m^2 
\:(y-x)^2} \Big)}{\sqrt{m^2 \:(y-x)^2}} \:\Theta\big( (y-x)^2 \big) \:\Theta \big(x^0 - y^0 \big) \:.
\end{align*}
Being smooth, the Bessel function can be treated as in Section~\ref{secstationaryphase}.
In the contribution involving the $\delta$-distribution, however, we can carry out the time integral to obtain
\begin{align*}
&\int_M S^\wedge_{0}(\xi)\: (i \Pdd_{\xi} +m)\: e^{i \Phi_p(\xi + \zeta)}\: \phi(\xi + \zeta)\: d^4 \xi \\
&= -\frac{1}{2\pi} \int_{\R^3} \frac{1}{2\, |\vec{\xi}|}\: (i \Pdd_{\xi} +m) \big( e^{i \Phi_p(\xi + \zeta)}\:
\phi(\xi + \zeta) \big) \Big|_{\xi^0 = -|\vec{\xi}|}\: d^3 \xi \:.
\end{align*}
Now we can again use the stationary phase method, but in three dimensions. Every stationary point gives
rise to a scaling factor
\[ \Big( \frac{\ell_\Lambda}{\ell_{\macro}} \Big)^3 \:. \]
Moreover there may be a stationary point near the origin. Here the factor~$1/|\vec{\xi}|$ effectively
reduces the dimension by one (as is obvious in polar coordinates). We thus get the scaling factor
\[ \Big( \frac{\ell_\Lambda}{\ell_{\macro}} \Big)^2 \:. \]
But the stationary points near the origin are combinatorially suppressed because there is at most one
of them. One should also keep in mind that the derivative~$i \Pdd_\xi$ gives 
scaling factor~$\ell_\Lambda^{-1}$. Nevertheless, at least one factor~$\ell_\Lambda$ remains,
giving rise to error terms of the form
\beq \label{saddleerror3}
\times \Big( 1 + \O \Big( \frac{\ell_\Lambda}{\ell_\macro} \Big) \Big) \:.
\eeq

We next consider the case~$K_p=L^\circ_{\circ,\circ}$. In this case, an explicit analysis is possible
in two complementary limiting cases.
If~$\ell_\Lambda \ll \ell_{\min}$, we can proceed with the stationary phase analysis of
Section~\ref{secstationaryphase}. In the opposite limiting case~$\ell_{\min} \ll \ell_\Lambda$,
on the other hand, we may expand the phase factor in a Taylor series. As a consequence, we do not get dephasing
effects. In general (in particular, if we are in none of the above limiting cases), no computational tools
are available for analyzing the integral explicitly.

The other integrals over~$\xi_{p-1}, \ldots, \xi_1$ can be carried out similarly.
The result on the phase factors~$e^{i \Phi_\ell}$ with phases~$\Phi_\ell$ given by~\eqref{Phidef}
can be described systematically as follows.
We first note that, writing the perturbation expansion~\eqref{pertholo} in the form~\eqref{exgen},
the factors~$K_\ell$ are alternating Green's operators and factors~$L^\circ_{\circ,\circ}$. More precisely,
\[ K_\ell = \left\{ \begin{array}{cl} s_m & \text{if~$\ell$ is odd} \\
L^\circ_{\circ,\circ} & \text{if~$\ell$ is even}\:. \end{array} \right. \]
Moreover, the total number~$p$ of such factors is odd.
As explained above, the stationary phase method applies to all factors~$s_m$, i.e.\
to all the integrals
\[ \xi_1, \xi_3, \ldots, \xi_p \]
(with odd subscripts).
Carrying out the last integral over~$\xi_p$ gives  (as shown in detail above) the condition that the
phase factor~$\Phi_p$ is trivial,
\beq \label{Phip}
e^{i \Phi_p} = 1 \:.
\eeq
According to~\eqref{Phidef}, this means that
\beq \label{condp}
a_p=b_p \:.
\eeq
Next, integrating over~$\xi_{p-2}$ gives the condition that
\[ e^{i \Phi_{p-2} + i \Phi_{p-1} + i \Phi_p} = 1 \:. \]
Using~\eqref{Phip} and again~\eqref{Phidef}, we conclude that
\[ \Lambda_{a_{p-2}} - \Lambda_{b_{p-2}} + \Lambda_{a_{p-1}} - \Lambda_{b_{p-1}} = 0 \:. \]
This condition can be satisfied in two ways, namely
\[ \left\{ \begin{array}{cl} \text{either$\qquad$} &  a_{p-2} = b_{p-2}, \quad a_{p-1} = b_{p-1} \\
\text{or$\qquad$} & a_{p-2} = b_{p-1}, \quad a_{p-1} = b_{p-2}\:. \end{array} \right. \]
Proceeding inductively, one obtains for each~$\ell$ the alternative conditions
\beq \label{condl}
\left\{ \begin{array}{cl} \text{either$\qquad$} &  a_{\ell-2} = b_{\ell-2}, \quad a_{\ell-1} = b_{\ell-1} \\
\text{or$\qquad$} & a_{\ell-2} = b_{\ell-1}, \quad a_{\ell-1} = b_{\ell-2} \end{array} \right.
\qquad \text{for all~$\ell=3,5,\ldots, p$}\:.
\eeq
Finally, using all these relations, the condition~\eqref{abcond} obtained by carrying out
the $x_0$-integral simplifies to the condition
\beq \label{cond0}
a_0 = b_0 \:.
\eeq
We point out that, at this stage, we applied the stationary phase method only to the Green's operators~$s_m$.
Therefore, the error terms scale like~\eqref{saddleerror3}. We also note that, so far,
the parameter~$\ell_{\min}$ has not yet come into play.

We next write out parts of the operator products in the perturbation expansion~\eqref{pertholo} in detail
\[ \sum_{a,\ldots, f} \cdots e^{i \Lambda_a} \,s_m\, e^{-i \Lambda_a}\: e^{i \Lambda_e} \: \big( \slashed{A}_c \starD L^c_{e,f} \big)\: e^{-i \Lambda_f}\: e^{i \Lambda_b} \,s_m\, e^{-i \Lambda_b} \cdots \:. \]
In the first case in~\eqref{condl}, this simplifies to
\beq \label{chol1}
\sum_{a,b,c} \cdots e^{i \Lambda_a} \,s_m\: \big( \slashed{A}_c \starD L^c_{a,b} \big)\: s_m\, e^{-i \Lambda_b}
\cdots \:.
\eeq
This gives back precisely the approximation used in~\eqref{tildeapprox1}.
The interesting point is that now we also have the second case in~\eqref{condl}, which gives
\[ \sum_{a,b,c} \cdots e^{i \Lambda_a} \,s_m\, e^{-i (\Lambda_a-\Lambda_b)}
\:\big( \slashed{A}_c \starD L^c_{b,b} \big)\:
e^{-i (\Lambda_b-\Lambda_a)} \,s_m\, e^{-i \Lambda_a} \cdots \:. \]
Due to the additional phase factors~$e^{\pm i (\Lambda_a-\Lambda_b)}$, this operator product can in general
not be simplified further. But it simplifies in the limiting case~$\ell_\Lambda \gg \ell_{\min}$.
Namely, in this limiting case, the operator~$L^c_{a,b}$ acts on the phase factors like a multiplication operator,
so that we obtain
\beq \label{chol2}
\sum_{a,b,c} \cdots e^{i \Lambda_a} \,s_m
\:\big( \slashed{A}_c \starD L^c_{b,b} \big)\: s_m\, e^{-i \Lambda_a} \cdots
\bigg( 1 + \O\Big( \frac{\ell_\Lambda}{\ell_{\min}} \Big) \bigg) \:.
\eeq
Writing both contributions~\eqref{chol1} and~\eqref{chol2} in the matrix notation~\eqref{notmatrix},
they can be combined to
\beq \label{upshot}
\sum_c \cdots s_m \:\Big( \slashed{A}_c \starD \mathbf{L}^c + \slashed{A}_c \starD \Tr \big(\mathbf{L}^c \big)\: \1 \Big)\: s_m \cdots \bigg( 1 + \O\Big( \frac{\ell_\Lambda}{\ell_{\min}} \Big) \bigg) \:.
\eeq

After these preparations, we are ready to analyze the commutation relations, following the procedure
in Section~\ref{secrealizecomm}. The difference of~\eqref{upshot} compared to the formalism in Section~\ref{secrealizecomm} is the additional term involving the trace in~\eqref{upshot}.
This difference can be implemented by modifying the matrices~$\mathbf{L}^c$ by a multiple
of the identity matrix, as is made precise by the replacement rule
\[ \mathbf{L}^c \rightarrow \mathbf{L}^c + \Tr \big(\mathbf{L}^c \big)\: \1 \:. \]
Since the identity matrix drops out of all commutators, all the methods and results in Section~\ref{secrealizecomm}
remain valid. In particular, we can satisfy the conditions~\eqref{condn1} and~\eqref{condn2}.
In this way, we have realized the CCR~\eqref{CCRB}, up to
errors of the form~\eqref{err2}.
We finally remark that the combinatorics of the stationary points could be analyzed similarly to what
described in Section~\ref{secstationaryphase}.

\section{Description with Bosonic and Fermionic Fock Spaces} \label{secfock}

\subsection{Separating the Fermionic and Bosonic Degrees of Freedom} \label{secfermibose}
In Section~\ref{secgauge} it was shown that decomposing the nonlocal perturbation operator~$\B_\dyn$
of the form~\eqref{Bansatz} according to their momentum transfer~\eqref{hatBjdef},
the resulting operators~$\hat{\B}^j_q$ can be identified with the bosonic field operators
which satisfy the commutation relations~\eqref{CCRB}.
Clearly, starting from the bosonic field operators, we obtain the dynamical gauge potential by
integrating over~$q$,
\[ \B_\dyn = \int \frac{d^4q}{(2 \pi)^4}\: \gamma_j \hat{\B}^j_q \:. \]
The goal of this section is to study the bosonic field operators in the perturbation expansion~\eqref{pertholo}.
Before entering the construction, we point out that the operator~$\B_\dyn$ in~\eqref{pertholo}
has two tasks. On the one hand, in view of the 
CCR~\eqref{CCRB}, it plays the role of the bosonic field operators.
On the other hand, in the operator products in~\eqref{pertholo} it changes the momenta of the
fermionic wave functions. In order to get into the standard formalism, we need to disentangle these
two tasks such as to obtain separate operators acting on the bosonic and fermionic degrees of freedom.

Using again the approximation of dropping all rapidly oscillating contributions~\eqref{tildeapprox1}, we obtain
\begin{align*}
|\tilde{\psi}\ket &\approx  \sum_{n=0}^{\infty} (-1)^n
\int_{\R^4} \frac{d^4q_1}{(2 \pi)^4} \cdots \int_{\R^4} \frac{d^4q_n}{(2 \pi)^4} \\
&\quad\: \times \sum_{b_1,\ldots, b_n=1}^N   s_m\, \big( \,\, \hat{\!\!\slashed{A}}_{b_1}(q_1)\: E_{q_1} \starD \mathbf{L}^{b_1} \big)
\cdots s_m\, \big( \,\,\hat{\!\!\slashed{A}}_{b_n}(q_n) \:E_{q_n} \starD \mathbf{L}^{b_n} \big) |\psi\ket \:,
\end{align*}
where~$E_q$ again denotes the operator of multiplication by a plane wave~\eqref{Eqdef}.
Here the symbol ``$\approx$'' again means that we allow for error terms of the form~\eqref{err1}
or~\eqref{err2}. Working with this approximation gives us the conservation of energy-momentum
within the perturbation expansion. This fact is crucial for the following construction steps.
For clarity, we write out the momenta at each operator,
\begin{align*}
|\tilde{\psi}(x)\ket &\approx \sum_{n=0}^{\infty} (-1)^n
\int_{\R^4} \frac{d^4q_1}{(2 \pi)^4} \cdots \int_{\R^4} \frac{d^4q_n}{(2 \pi)^4} \int_{\R^4} \frac{d^4p}{(2 \pi)^4} \\
&\qquad \times
\sum_{b_1,\ldots, b_n=1}^N e^{-i (q_1+ \cdots + q_n+p)\, x}\: \hat{\psi}_{b_1, \ldots, b_n}
(q_1, \ldots, q_n, p)
\end{align*}
with
\begin{align*}
\hat{\psi}_{b_1, \ldots, b_n}(q_1, \ldots, q_n, p)
&:= s_m\big|_{p+q_1+\cdots + q_n}\, \,\, \hat{\!\!\slashed{A}}_{b_1}(q_1)\: (E_{q_1} \starD \mathbf{L}^{b_1})\big|_{p+q_2+\cdots+q_n} \\
&\qquad \times 
\cdots s_m\big|_{p+q_n}\, \,\,\hat{\!\!\slashed{A}}_{b_n}(q_n) \:(E_{q_n} \starD \mathbf{L}^{b_n})\big|_p\, |\hat{\psi}(p)\ket \:.
\end{align*}
Using that the kernels~$E_{q_\ell} \starD \mathbf{L}^{b_\ell}(p+q_\ell+ \cdots + q_n, p+q_{\ell+1}+ \cdots + q_n)$
are complex-valued and thus act trivially on the spinors, we may factor them out to obtain
\begin{align}
&\hat{\psi}_{b_1, \ldots, b_n}(q_1, \ldots, q_n, p) 
= s_m\big|_{p+q_1+\cdots + q_n}\, \gamma^{i_1}\,
s_m \big|_{p+q_2+\cdots + q_n} \, \gamma^{i_2}
 \cdots s_m\big|_{p+q_n}\, \gamma^{i_n} \label{pertf} \\
&\qquad \times \big(\hat{A}^{i_1}_{b_1}(q_1)\, E_{q_1} \starD \mathbf{L}^{b_1} \big)\big|_{p+q_2+\cdots+q_n}
\cdots \:\big( \hat{A}^{i_n}_{b_n}(q_n)\,E_{q_n} \starD \mathbf{L}^{b_n} \big) \big|_p\, |\hat{\psi}(p) \ket \:. \label{pertb}
\end{align}
Now~\eqref{pertf} corresponds to the usual fermionic tree diagram
involving the Dirac matrices and the momentum transfer~$q_\ell$ at each leg.
The factors in~\eqref{pertb}, on the other hand, can be written as a bosonic operator product
acting on~$|\psi \ket$,
\beq \label{bosprod}
\big(\hat{A}^{i_1}_{b_1}(q_1)\, E_{q_1} \starD \mathbf{L}^{b_1} \big)\:
\big( \hat{A}^{i_2}_{b_2}(q_2)\, E_{q_2} \starD \mathbf{L}^{b_2}
\big)\: \cdots \:\big(\hat{A}^{i_n}_{b_n}(q_n)\, E_{q_n} \starD \mathbf{L}^{b_n} \big)\, |\psi \ket \:.
\eeq
Now we can commute the bosonic operators using the commutation relations as worked out in Section~\ref{secrealizecomm}, keeping the fermionic diagram~\eqref{pertf} and the momenta therein fixed.
In this way, we have separated the bosonic and fermionic degrees of freedom in the desired way.

The factorization~\eqref{pertf} and~\eqref{pertb} also give a direct understanding of how the
fermionic and bosonic degrees of freedom are encoded in the causal fermion system:
The fermionic degrees of freedom describe the low-frequency behavior of the wave function
(modulo gauge phases); this is why in~\eqref{pertf} the momenta are shifted by~$q_1, \ldots, q_n$.
The high-frequency behavior of the wave function, however, encodes the bosonic degrees of freedom,
as is obvious from the fact that the bosonic operators in~\eqref{pertb} act on the
components~$a \in \{1, \ldots, N\}$ of the holographic components.

In order to formulate these considerations in a mathematically clean way, it is helpful to let the operators 
in the above perturbation series act on a pair of functions, one being a standard spinorial wave function, and
the other consisting of~$N$ complex-valued functions in spacetime,
\beq \label{doublepsi}
| \psi \otimes \phi \ket \in \K \otimes C^\infty(M, \C)^N \:.
\eeq
We then replace the operator~$\hat{\!\!\slashed{A}}_b(q)\, E_q \starD \mathbf{L}^b$ by
\beq \label{double}
\gamma_j E_q \otimes \big( \hat{A}^j_b(q)\, E_q \starD \mathbf{L}^b \big) \:.
\eeq
Now the Dirac matrices appear in the first factor, acting on the spinorial wave function.
The second factor, on the other hand, contains a matrix acting on~$\C^N$.
Note that the operator~$E_q$ describing the momentum shift acts both on the first and the second factor.
This has the effect that, if~$\psi$ and~$\phi$ have the same momentum, then this property is preserved,
in agreement with the momenta in~\eqref{pertf} and~\eqref{pertb}.
But now the operator~\eqref{double} may act more generally on a tensor product~$\psi \otimes \phi$
where~$\psi$ and~$\phi$ have different momenta. This is needed once the bosonic operators
in~\eqref{bosprod} are commuted.

Finally, we simplify the notation by setting
\begin{align*}
\scrA^j(q) &:= \sum_{b=1}^N \hat{A}^j_b(q)\, E_q \starD \mathbf{L}^b \\
\gamma_j E \otimes \scrA^j &:=
\int \frac{d^4q}{(2 \pi)^4}\: \gamma_j E_q \otimes \scrA^j(q) \:.
\end{align*}
Using the above results, the operators~$\scrA^j$ can be identified with the usual bosonic field operators.

\subsection{Incorporating the Fermionic Fock Space}
In order to get into the setting of QFT, the remaining task is to also rewrite the
spinorial wave functions and Green's operators in terms of field operators acting on a fermionic Fock space.
Moreover, we want to formulate the whole dynamics in the standard formalism of QED.
To this end, we again drop all rapidly oscillating contributions and begin
with the perturbation series~\eqref{tildeapprox1}, again allowing for error terms of the form~\eqref{err1}
or~\eqref{err2}. Acting with the Dirac operator
and working again with pairs of functions~\eqref{doublepsi}, we obtain the Dirac equation
\[ (i \Pdd \otimes \1 + \gamma_j E \otimes \scrA^j - m)\,|\psi \otimes \phi \ket = 0 \:. \]
This Dirac equation can be written in the Hamiltonian form as
\beq \label{dynHamilton}
\big( i \partial_t - H \big) |\psi \otimes \phi \ket = 0 \: ,
\eeq
where the Hamiltonian is of the form
\[ 
H = H_0 + V \]
with the standard Dirac Hamiltonian~$H_0 := -i \gamma^0 \vec{\gamma} \vec{\nabla} + m \gamma^0$
and the nonlocal operator~$V$ given by
\[ V = -\gamma^0\, \big( \gamma_j E \otimes \scrA^j \big)
=  -\big( \alpha^j E \otimes \scrA_j \big) \:, \]
where we used the standard notation~$\alpha^j := \gamma^0 \gamma^j$ (see for example~\cite{thaller}).
The operator~$H$ acts on the fermionic component as a purely spatial operator (indeed, $H_0$
is the usual Dirac Hamiltonian, whereas~$\gamma_j E$ is a multiplication operator in position space).
Therefore, we can consider the Hamiltonian equation~\eqref{dynHamilton} as an evolution equation
on the function space
\beq \label{spatial}
| \psi \otimes \phi \ket \in L^2(\R^3, \C^4) \otimes C^\infty(M, \C)^N \:.
\eeq

Next, we want to rewrite the dynamics in terms of fermionic Fock spaces. 
For clarity, we proceed in two steps. The first step is to rewrite~\eqref{dynHamilton}
equivalently as the evolution equation for a one-fermion state on the Fock space.
In the second step we will move on to many-fermion states.
In the first step we can proceed in the standard way.
Working for convenience in position space, we introduce the fermionic field
operators~$\Psi$ and~$\Psi^\dagger$ by their equal-time anti-commu\-ta\-tion relations
\beq \label{CAR}
\big\{ \Psi^\alpha(\vec{x}), \Psi^\beta(\vec{y})^\dagger \big\} =
\delta^{\alpha \beta}\: \delta^3 \big(\vec{x}-\vec{y} \big) \qquad \text{and} \qquad
\big\{ \Psi^\alpha(\vec{x}), \Psi^\beta(\vec{y}) \big\} = 0  \:.
\eeq
Now we replace the function~$|\psi \otimes \phi \ket$ in~\eqref{spatial} by the Fock vector
\[ |\Psi\ket := \Psi^\dagger(\psi)\, |0 \ket \otimes \phi \:, \]
where~$|0\ket$ denotes the fermionic vacuum vector and
\[ \Psi^\dagger(\psi) := \int_{\R^3} \Psi^\dagger(\vec{x}) \:\psi(\vec{x})\: d^3x \:. \]
Then the dynamics of~$|\Psi \ket$ is described by the Schr\"odinger equation
\beq \label{Hfock}
i \partial_t |\Psi \ket = H |\Psi \ket
\eeq
with the ``second-quantized'' Hamiltonian computed by
\begin{align*}
H &= \int_{\R^3} d^3x\: \bigg( \Psi(t, \vec{x})^\dagger
H_0 \Psi(t, \vec{x}) + \int \frac{d^4q}{(2 \pi)^4} \Psi(t, \vec{x})^\dagger
\,\alpha_j\,e^{-i q x} \Psi(t, \vec{x}) \otimes \scrA^j(q) \bigg)\\
&= \int_{\R^3}
\Psi(t, \vec{x})^\dagger \, (H_0 - \alpha^j \scrA_j(t, \vec{x}) )\Psi(t, \vec{x})\: d^3x \:.
\end{align*}
This is the standard Hamiltonian of QED.

Before moving on to many-fermion states, it is convenient to describe the fermions
instead of equal-time anti-commutation relations by covariant CAR for general spacetime
points~$x=(t,\vec{x})$ and~$y=(t', \vec{y})$ as
\beq \label{CARcov}
\begin{split}
\{ \Psi^\alpha(x), \Psi^\beta(y)^\dagger \} &= 2 \pi\,
\big( k_m(x,y) \,\gamma^0 \big)^\alpha_\beta \\
\{ \Psi^\alpha(x), \Psi^\beta(y) \} =&\; 0 =
\{ \Psi^\alpha(x)^\dagger, \Psi^\beta(y)^\dagger \} \:,
\end{split}
\eeq
where~$k_m$ is the causal fundamental solution of the vacuum Dirac equation
(defined as the difference of the advanced and retarded Dirac Green's operators divided by~$2 \pi i$).
This has the advantage that the bosonic CCR and the fermionic CAR are described in a similar formalism (cf.~\eqref{CCRB} and~\eqref{CARcov}). In more physical terms, the free fields are described in the
Heisenberg picture (where the field operators are time dependent and satisfy the homogeneous
field equations). Accordingly, in the Hamiltonian we need to omit the free Hamiltonians
(of both the bosonic and fermionic fields), giving rise to the Hamiltonian of QED in the interaction picture
\beq H = - \int_{\R^3}
\Psi(t, \vec{x})^\dagger \, \alpha^j \scrA_j(t,x)\,\Psi(t, \vec{x})\: d^3x \:. \label{Hint}
\eeq

We now come to the question of how a many-fermion state is to be described in our setting.
This question is rather subtle, as we now explain in detail.
We saw that the nonlocal Dirac equation~\eqref{dirnonloc}
can be written in the Hamiltonian form~\eqref{dynHamilton}. In this description, the
microscopic structure of the physical wave function is described by the function~$\phi \in C^\infty(M, \C)^N$,
whereas its macroscopic form is described by a usual Dirac wave function~$\psi$.
Clearly, the causal fermion system is composed of many physical wave functions.
The naive idea is to describe each of them as in~\eqref{dynHamilton} and to take their
anti-symmetrized product, i.e.\
\beq \label{HFneu}
|\Psi \ket := |\psi_1 \otimes \phi_1 \ket  \wedge \cdots \wedge |\psi_L \otimes \phi_L \ket
\eeq
(here~$L$ denotes the total number of physical wave functions, including those describing the Dirac sea;
here for ease in presentation we choose~$L$ to be finite).
Considering the anti-symmetrized product~\eqref{HFneu} as the quantum state of the system is problematic,
as we now explain. Suppose we want to let a fermionic field operator~$\Psi^\dagger(\psi)$ act on~$|\Psi \ket$.
Then~$\psi$ is a fermionic wave function. The naive ansatz
\[ \Psi^\dagger(\psi) \:|\Psi \ket
= |\psi \ket \;\wedge\; |\psi_1 \otimes \phi_1 \ket  \wedge \cdots \wedge |\psi_L \otimes \phi_L \ket \]
does not work, because the first factor does not have a bosonic component, so that anti-symmetrization
is ill-defined. Building in a bosonic component~$\phi$ makes mathematical sense,
\beq \label{phintro}
\Psi^\dagger(\psi) \:|\Psi \ket
= |\psi \otimes \phi \ket \wedge |\psi_1 \otimes \phi_1 \ket \wedge \cdots \wedge |\psi_L \otimes \phi_L \ket \:,
\eeq
but this raises the question how~$\phi$ is to be chosen. As the vacuum? Or as the physical bosonic state?
This is unclear, showing that also the ansatz~\eqref{phintro} is not a sensible concept.
More generally, it is essential to anti-symmetrize only the fermionic degrees of freedom.
The resulting totally anti-symmetric fermionic wave function should then be tensored with the bosonic
degrees of freedom.

In order to implement this concept in our construction, we need a simplifying assumption,
which we refer to as the {\em{bosonic Fock space approximation}}. It states that the bosonic
field operator can be approximated by a matrix acting on the index~$a=1,\ldots, N$ 
(similar as in the matrix notation~\eqref{notmatrix}) which does not
change the spacetime dependence of the bosonic component~$\phi$. Thus, using again the notation~\eqref{spatial},
we assume that
\beq \label{fockspaceapprox}
\scrA_j \, | \psi \otimes \phi \ket = | \psi \otimes \scrA_j \phi \ket 
\qquad \text{and} \qquad
(\scrA_j \phi)^a(x) = \sum_{b=1}^N \big( \scrA_j \big)^a_b\: \phi^b(x) \:.
\eeq
\Felix{Should this be discussed a bit more?}%
Under this assumption, the spacetime dependence of the bosonic component is irrelevant
and can be dropped. Then the bosonic component is described only by the index~$a \in \{1,\ldots, N\}$,
making it possible to write
\[ |\psi \otimes \phi>  =  \sum_{a=1}^N | \psi^a \otimes \phi_a \ket \:, \]
where~$(\phi_a)_{a=1,\ldots, N}$ can be identified with the canonical basis of~$\C^N$.
Now we define the Fock state by
\beq \label{Fockdefnew}
|\Psi\ket := \sum_{a=1}^N \psi_1^a \wedge \cdots \wedge \psi_L^a\: \otimes\: \phi_a \:.
\eeq
We point out that the fermionic wave functions all carry the same bosonic index~$a$.
Therefore, as desired, the anti-symmetrization takes into account only the fermionic degrees of freedom.
The fermionic field operator act on this Fock state in an obvious way.
The action of the bosonic field operators, on the other hand, is defined by
\[  \scrA_j |\Psi \ket := 
\sum_{a, b=1}^N \psi_1^a \wedge \cdots \wedge \psi_L^a\: \otimes\: (\scrA_j)_a^b \:\phi_b \:. \]
This definition has the advantage that the commutation relations are respected, because
for example
\[  \scrA_k \scrA_j |\Psi \ket := 
\sum_{a, b, c=1}^N \psi_1^a \wedge \cdots \wedge \psi_L^a\: \otimes\: (\scrA_k)^c_b (\scrA_j)^b_a \:\phi_c \:. \]

Having specified how the fermionic and bosonic field operators act on the state~$|\Psi \ket$
defined by~\eqref{Fockdefnew}, it is also clear how the Hamiltonian~$H$ of QED given by~\eqref{Hint}
acts on~$|\Psi \ket$.

Likewise, the Hamiltonian acts on the Fock space according to
\[ H |\Psi \ket := 
\sum_{a, b=1}^N \Big( \big(\alpha^j E \psi_1^a \big) \wedge \cdots \wedge \psi_L^a
+ \cdots + \psi_1^a \wedge \cdots \wedge \big(\alpha^j E \psi_L^a \big)
 \Big)\: \otimes\: (\scrA_j)_a^b \:\phi_b \:. \]
We note that, as a consequence of the bosonic Fock space approximation, the product structure
in~\eqref{Fockdefnew}
is not preserved by the time evolution. Instead, the state is described by a general vector
in~$\Fock^\fermi \otimes \Fock^\bose$, where~$\Fock^\bose$ effectively describes all the
``bosonic'' factors~$\phi_\ell \in \C^N$ in~\eqref{Fockdefnew}.
In this way, we also obtain fermionic entanglement.

With the Schr\"odinger equation~\eqref{Hfock} with the Hamiltonian~\eqref{Hint} we have
rewritten the dynamics in the notions of standard QED.
For clarity and completeness, we now outline the construction steps 
needed for the perturbative description and point
out how the standard constructions need to be modified in each step.
\bitem
\item[(i)] First, one needs to choose the quantum state~$\omega$ at an initial time.
One way of doing so is to specify~$\omega$ as a positive linear functional on the
algebra generated by the bosonic and fermionic field operators.
More concretely, one can make use of the fact that we introduced the field
operators as acting on tensor products of wave functions, giving a natural representation
on a Fock space~$\Fock$.
Then~$\omega$ can be represented by a density operator on~$\Fock$. In the simplest case
of a pure state, $\omega$ can be written as~$\omega = |\Psi \ket \bra \Psi|$ with a unit vector~$\Psi \in \Fock$.

When describing a scattering process, the situation is particularly simple because we may assume
that the system is non-interacting initially. Then the causal fermion system is described by
solutions of the vacuum Dirac equation. In non-technical terms, the initial fermionic state is chosen
as the Hartree-Fock state formed of all physical wave functions, i.e.\
\[ \Psi^\fermi := \psi_1 \wedge \cdots \wedge \psi_f \]
where~$(\psi_\ell)_{\ell=1,\ldots, f}$ is an orthonormal basis of~$(\H, \la .|. \ra_\H)$. 
This equation becomes mathematically meaningful in the infinite-dimensional case~$f=\infty$
by choosing~$\omega$ as a regularized quasi-free Hadamard state.
\item[(ii)] The Schr\"odinger equation~\eqref{Hfock} can be solved perturbatively with the
Dyson series. The only modification in our setting is that the bosonic field operators~$\scrA^j$
contained in the potential~$V$ are nonlocal in space and time on the scale~$\ell_{\min}$.
The reader interested in more details on this Dyson series is referred to~\cite{collapse}.
\item[(iii)] Next, one needs to specify the initial state~$|\Psi(t_0)\ket$. The easiest choice is
the vacuum~$|0\ket$ determined as usual by the condition that it vanishes when acted upon by the
annihilation operators. In preparation, for the bosonic operators, one implements the
usual frequency splitting by writing~$\scrA^j = a^j + (a^j)^\dagger$, where the operators
\[ 
a_j := \int \frac{d^4q}{(2 \pi)^4}\: \Theta(q^0)\: \hat{\scrA}^j_q \qquad \text{and} \qquad
(a^j)^\dagger := \int \frac{d^4q}{(2 \pi)^4}\: \Theta(-q^0)\: \hat{\scrA}^j_q \]
are composed of all positive and negative frequencies, respectively
(note that the dagger can be regarded as the adjoint on the Krein space~$(\K, \bra .|. \ket)$).
Then the vacuum is characterized
by the conditions
\beq \label{vac}
\Psi^\alpha(x) \,|0\ket = 0 = a^j(x)\, |0\ket \qquad \text{for all~$\alpha, j$ and~$x \in M$}\:.
\eeq
Clearly, instead of the vacuum one can also choose an initial state~$|\Psi(t_0) \ket$ involving particles and anti-particles and/or
photons. This state is obtained by acting on the vacuum~$|0\ket$ with a finite number of creation operators.
\item[(iv)] Expectation values of the state~$|\Psi(t)\ket$ at a later time~$t$ can be computed
to every order in perturbation theory by expanding the time-ordered exponential in the Dyson series and
using the canonical commutation and anti-commutation relations
together with the relations~\eqref{vac} characterizing the vacuum.
This gives the standard {\em{Feynman diagrams}}, involving both fermionic and bosonic loops.
The time ordering in the Dyson series has the effect that the diagrams are formed
of the {\em{Feynman propagators}}, characterized by the property that positive frequencies propagate
to the future, whereas negative frequencies propagate to the past.
\item[(v)] In view of the ultraviolet cutoff on the scale~$\varepsilon$, all obtained Feynman diagrams
are well-defined and finite. Using the standard renormalization techniques, one could analyze the limiting
case~$\varepsilon \searrow 0$ when the ultraviolet regularization is removed.
\eitem
We conclude with a remark concerning {\em{Haag's theorem}}~\cite{haag-nogo, hall-wightman, powers},
which states that that an interacting quantum field theory in the interaction picture
cannot be unitarily equivalent to a free field theory. The critical reader may wonder whether
our description is meaningful in view of this no-go theorem. The point is that, in our setting, the interaction
potential is non-local, thus breaking Poincar{\'e} invariance.
Therefore, we are not abiding by the hypotheses of this theorem and, more
generally, of the local quantum physics approach.

\section{The Dynamics of the Quantum State} \label{secstate}
In~\cite{fockbosonic, fockfermionic, fockentangle} an interacting causal fermion system
was described at any time~$t$ by a quantum state~$\omega^t$.
We now explain how the above description of the dynamics on Fock spaces can be
related to the dynamics of this quantum state.
Before beginning, we point out that the construction of the quantum state
applies in a more general setting than the constructions in the present paper.
In particular, the separation of the fermionic and bosonic degrees of freedom
in Section~\ref{secfermibose} is possible only under additional assumptions and involves error terms.
In case that these assumptions are not satisfied, the constructions in~\cite{fockfermionic, fockentangle}
still give a dynamics of a quantum state. However, it will differ from the standard unitary time evolution
on Fock spaces.

We recall that, in the setting of algebraic QFT, a {\em{quantum state}}
is defined as a positive linear mapping from the algebra of observables~$\A$
(defined from linear fields in the vacuum) to the complex numbers.
For convenience we choose a state which can be written as the expectation value of a density operator~$\sigma^t$ 
acting on the Fock space~$\Fock$,
\[ 
\omega^t(A) = \tr_{\Fock}(\sigma^t A) \qquad \text{for all~$A \in \A$} \:. \]
The density operator, in turn, can be written in {\em{bra/ket}} notation as
\[ 
\sigma^t = \sum_{a \in {\mathfrak{S}}} c_a \;\Big| \sum_{\alpha \in {\mathfrak{T}}_a} \Psi^\Fock_{a \alpha} \Big\ra
\Big\la \sum_{\beta \in {\mathfrak{T}}_a} \Psi^\Fock_{a \beta}  \Big| \]
with~${\mathfrak{T}}_a$ an index set depending on~$a$.
In the case of a {\em{pure state}}, the set~${\mathfrak{S}}$ has only one
element, so that the density operator can be written as
\[ 
\sigma^t = \Big| \sum_{\alpha =1}^L \Psi^\Fock_\alpha \Big\ra
\Big\la \sum_{\beta=1}^L \Psi^\Fock_\beta  \Big| \:, \]
where~$L$ denotes the number of Fock components.

The methods employed for the construction of the quantum state in~\cite{fockfermionic, fockentangle}
are quite different from the techniques used here. The general idea behind the construction
is to ``compare'' the interacting measure~$\tilde{\rho}$ at a given time~$t$ with the vacuum measure.
The freedom in identifying the Hilbert spaces of the interacting and the vacuum
causal fermion systems is described by a unitary operator~$\scrU \in \G$, where~$\G$ is a
compact group~$\G \simeq \U(N)$ of unitary operators on the Hilbert space~$\H$.
In order to obtain information independent of this unitary freedom, we integrate over the unitary group.
More precisely, the {\em{refined partition function}} is defined by
\[ Z^t_V\big( \alpha, \beta, \tilde{\rho} \big) = 
\fint_\G d\mu_\G \big(\scrU_< \big)  \fint_\G d\mu_\G \big( \scrU_> \big) \:
e^{\alpha N \T^t_V \big(\tilde{\rho}, T_{\scrU_<, \scrU_>}  \rho \big)} \:, \]
where~$V$ is the spacetime region under consideration, and~$\T_V$ is a certain nonlinear
surface layer integral. The {\em{refined state}} is introduced by
\Felix{Sage hier etwas genauer, wie das zusammenhängt?}%
\beq \label{statedef}
\omega^t_V \big( \cdots \big) = \frac{1}{Z^t_V\big( \alpha, \beta, \tilde{\rho} \big)} \fint_\G d\mu_\G \big(\scrU_< \big)  \fint_\G d\mu_\G \big( \scrU_> \big) \:
e^{\alpha N \T^t_V \big(\tilde{\rho}, T_{\scrU_<, \scrU_>}  \rho \big)} \,\big( \cdots \big) \:,
\eeq
where the dots on the left stand for an operator in the observable algebra~$\A$ formed of the
linearized fields in the vacuum spacetime, whereas the
dots on the right stand for suitable surface layer integrals which again involve the
linearized fields in the vacuum.

This construction yields a quantum space at any time~$t$. However, it is a shortcoming of
this approach that it does not tell us about the time evolution of the quantum state.
This shortcoming is overcome with the constructions of the present paper, which make it possible
to rewrite the dynamics in terms of a unitary time evolution on Fock spaces.
Nevertheless, the present results do not make the constructions in~\cite{fockfermionic, fockentangle}
obsolete. On the contrary, it seems that these constructions complement those in the present paper,
giving a different perspective and thereby giving a more complete picture of the quantum dynamics.

On a technical level, at present it is unclear how to relate the refined state~\eqref{statedef}
to the constructions in the present paper. But on a conceptual level, the constructions fit together
nicely, as we now briefly explain. The unitary operators~$\scrU_<$ and~$\scrU_>$ in~\eqref{statedef}
may compensate for the dephasing of the wave functions in the interacting spacetime relative
to the vacuum. More concretely, with specific choices of these unitary operators one may
``detect'' the different summands of the dephasing operator~$U$ in~\eqref{Udef}
and similarly of the unitary holographic mixing operator~$V$ (as defined by~\eqref{Vdef}).
Likewise, the entanglement as found in~\cite[Section~6]{fockentangle} by evaluating
the low-energy saddle points should be related to and arise dynamically 
as a consequence of the bosonic Fock space approximation introduced in~\eqref{fockspaceapprox}.
Making these relations and correspondences mathematically precise seems an interesting topic for future research.

\section{Remarks and Outlook} \label{secoutlook}
In this paper, QED was derived in a well-defined limiting case from a more fundamental physical theory.
{\em{Corrections}} to this limiting case are physical predictions by the theory of causal fermion systems
to be tested in future experiments. Therefore, it is of utmost importance to work out and
analyze different correction terms, which at present are subsumed in the error terms~\eqref{err1}
or~\eqref{err2}. We now discuss a few of these corrections, leaving the detailed analysis
as future research projects.

Before beginning, we point out that the description of the dynamics by a nonlocal
operator~\eqref{hatBjdefastic} in the Dirac equation involves many unknowns. It leaves us with
the freedom in choosing the operators~$L_a$ as well as the covariances of the corresponding
Gaussian stochastic fields~$B_a$.
This freedom was partially removed by the requirement of satisfying the linearized field equations
and of getting covariant commutation relations on scales much larger than~$\ell_{\min}$.
But even then, we are left with a many freedoms in choosing the unknowns.
This situation also made our analysis rather involved, because we had to carefully analyze
to what extent these unknowns affect our end results.
In order to improve the situation, one needs to derive stronger structural results
which would impose further constraints on the form of the nonlocal potential.
As long as such stronger results are not available, one should consider~\eqref{hatBjdefastic}
as a suitable ansatz for describing the microscopic structure of spacetime.
Our lack of knowledge on the microstructure of spacetime is reflected in the
freedom in choosing the operators~$L_a$ and the stochastic fields~$B_a$.
Taking the statistical mean amounts to taking averages of the unknown microstructure.
Our analysis shows that, doing so in a suitable way, does indeed give rise to an effective
description by bosonic quantum fields. The fact that this limiting case comes with error terms
gives the hope that, by quantifying the errors, one can get corrections to standard quantum field theory.
But clearly, these corrections will again depend on the above-mentioned unknowns.
The goal is to find corrections which, in our stochastic description, take a simple and quantifiable form.
One strategy is to consider the high-precision measurements of QFT (like the Lamb shift or the
spectrum of the hydrogen atom) and try to work out corresponding corrections.
Alternatively, one could consider cosmological phenomena.
Going into details seems premature and should better be the objective of future research.

Clearly, the first question in this context concerns the values
of the parameters~$\ell_\Lambda$ and~$\ell_{\min}$. Thus, how large are the corrections to be expected?
At present, not much is known about these parameters, except that~$\ell_{\min}$ lies between the
Planck scale and the length scale of macroscopic physics~\eqref{ellmindef}.
This lack of knowledge is also the reason why we kept our analysis as general as possible by allowing
for error terms of the form~\eqref{err1} or~\eqref{err2}.
Thinking of the phase factors~$e^{i \Lambda_a}$ as being generated by the stochastic potentials~$A_a$,
the assumption~$\ell_\Lambda \gg \ell_{\min}$ seems favorable. But at present, even this is not clear.
The number~$N$ of the stochastic potentials is determined by~$\ell_{\min}$ (see~\eqref{Nscaleintro}).
Another unknown is the strength of the stochastic fields as described by the
covariances~\eqref{covariancevector}. Clearly, these covariances are partly determined by the
CCR.

A promising approach to reduce the number of unknowns is to make use of the fact that the
parameter~$\ell_{\min}$ as well as the covariances also arise in
the collapse phenomena as worked out in the non-relativistic limit in~\cite{collapse}.
In this case, the nonlocality of the potential in time gives rise to non-symmetric
potentials in the non-relativistic limit, which are crucial for the reduction of the wave function
in the measurement process. These connections give
the hope that the effective parameters in the resulting collapse model will
give information on~$\ell_{\min}$ and the strength of the stochastic fields, which will in turn
pose constraints or even give predictions for the corrections to QED.

In addition to the corrections presently contained in our error terms, there are also corrections to QED
of a different nature due to the {\em{nonlocality in time}} and the {\em{nonlinearity}} of the
EL equations. These effects are closely related to the collapse phenomena as analyzed and discussed in~\cite{collapse}. Finally, corrections arise when going beyond the {\em{bosonic Fock space
approximation}} introduced in~\eqref{fockspaceapprox}. The exact dynamics
cannot be formulated on a tensor product of a bosonic and a fermionic Fock space.
Working out what this means quantitatively is a challenging
problem for the future.

Apart from these corrections to QED, it is also an important open problem to
extend our methods to {\em{non-abelian gauge fields}}. Generalizing the methods to the gravitational field
would lead to a mathematically precise formulation of {\em{quantum gravity}}.

\appendix
\section{A Perturbation Expansion in Powers of~$\ell_{\min}/\ell_\Lambda$} \label{apppert}
This section provides a detailed construction of the unitary holographic mixing operator.
Our starting point is the Dirac operator including holographic gauge potentials of the form~\eqref{Bsimp}, i.e.\
\beq \label{DirB}
\Dir := i \Pdd + \B \qquad \text{with} \qquad
\B(x,y) = \sum_{a=1}^L (\Pdd \Lambda_a)\Big( \frac{x+y}{2} \Big)\: L_a(y-x) \:.
\eeq
Our goal is to construct a unitary operator on the Krein space,
\beq \label{Vunitary}
V \::\: \K \rightarrow \K \qquad \text{unitary on~$(\K, \bra .|. \ket)$} \:,
\eeq
which has the property that the Green's operator~$\tilde{s}_m$ of
as defined by
\[ (\Dir-m) \,\tilde{s}_m = \1 \]
has the representation
\beq \label{Vrep}
\tilde{s}_m = V s_m V^*
\eeq
(where the star denotes the adjoint with respect to the Krein inner product~$\bra .|. \ket$ in spacetime).

Our method is to apply perturbative techniques as developed in~\cite{sea, grotz, norm}
(see also~\cite[Section~2.1]{cfs} or~\cite[Chapter~18]{intro}).
Before stating our result, we must introduce the spectral decomposition of the Dirac operator in the vacuum.
Since we want to compute the Green's operators, we must take into account all the eigenspaces of the
Dirac operator, including the imaginary eigenvalues (as first done in~\cite[Section~2.1]{endlich}).
\begin{Def}
For~$a \in \R$ and~$m \in \R\cup i \R$, $m \neq 0$, 
we define the following tempered distributions in momentum space,
\begin{align}
P_a (k) &:= \delta(k^2-a) \label{Padef} \\
p_m (k) &:= \frac{|m|}{m} \: (\slashed{k} + m) \: \delta(k^2-m^2) \:. \label{pmdef}
\end{align}
Likewise, for~$m=0$ we set
\[ p_0(k) := \slashed{k} \: \delta(k^2) \:. \]
We also regard these distributions as multiplication operators in momentum space.
\end{Def}
By direct computation one verifies that these distributions are solutions of the Klein-Gordon
and Dirac equations, respectively. More precisely,
\beq \label{eqns}
(k^2-a)\, P_a (k) = 0 \qquad \text{and} \qquad (\slashed{k}-m)\: p_m(k) = 0 \:.
\eeq

Products of the above operators are well-defined if we work with a $\delta$-normalization in the mass.
Indeed, by direct computation one finds that
\begin{align}
P_a \: P_b &= \delta(k^2-a) \; \delta(k^2-b) \;=\; \delta(a-b) \; P_a \\
p_m \: p_{m'} &= \frac{|mm'|}{mm'} \; (\slashed{k}+m)(\slashed{k}+m') \;
	\delta(k^2-m^2) \: \delta\big(k^2-(m')^2 \big) \notag \\
&= \delta \big( m^2-(m')^2 \big) \; \frac{|mm'|}{mm'}
	\; \big( k^2 + (m+m') \: \slashed{k} + m m' \big) \; \delta \big( k^2-(m')^2 \big) \notag \\
&= \delta \big( m^2-(m')^2 \big) \; \frac{|mm'|}{mm'}
	\; (m+m') \:\big( \slashed{k} + m' \big) \; \delta \big( k^2-(m')^2 \big) \notag \\
&= \frac{1}{2 |m|} \:\delta(m-m') \; \frac{|mm'|}{mm'}
	\; (m+m') \:\big( \slashed{k} + m' \big) \; \delta \big( k^2-(m')^2 \big) \notag \\
&= \delta(m-m') \; \frac{|m'|}{m'}\:\big( \slashed{k} + m' \big) \; \delta \big( k^2-(m')^2 \big) 
= \delta \big(m-m' \big) \: p_m \:. \label{pmpmp}
\end{align}
Moreover, a straightforward computation using the symmetry properties under
reflections~$k \rightarrow -k$ yields the following completeness relations,
\begin{align*}
\int^\infty_{-\infty} P_a \: da &= \int^\infty_{-\infty} \delta(k^2-a) \: da = \1  \\
\int_{\R \cup i \R} p_m \: dm &= \int_{\R^+ \cup i \R^+}
\frac{|m|}{m} \;2m \: \delta(k^2-m^2) \: dm \notag \\
&= \int^\infty_{-\infty} \delta(k^2-m^2) \: d(m^2) = \1 \:,
\end{align*}
where~$dm$ denotes the Lebesgue measure on~$\R \cup i \R$.
In view of~\eqref{eqns} and the completeness relations, the distributions~$P_a$ and~$p_m$
can be viewed as the spectral projection operators of the Klein-Gordon and Dirac equations,
respectively. Note that, in contrast to the situation for symmetric operators on Hilbert spaces,
the spectrum of the Dirac operator is complex and the corresponding spectral projections are not symmetric but
\[ p_m^* = p_{\overline{m}} \]
(as is obvious from~\eqref{pmdef}).
The corresponding functional calculus is obtained by integrating over the spectral parameter. For example,
\begin{align*}
\int^\infty_{-\infty} a \; P_a \: da &= \int^\infty_{-\infty} a \; \delta(k^2-a) \: da = k^2  \\
\int_{\R \cup i \R} m \; p_m \: dm &= \int_{\R^+ \cup i \R^+} 2|m|
\: \slashed{k} \; \delta(k^2-m^2) \: dm = \slashed{k}= i \Pdd_x \:,
\end{align*}
and using that multiplication in momentum space corresponds to differentiation in position
space, one recovers the Klein-Gordon and Dirac operators.
Moreover, the {\em{symmetric Dirac Green's operator}} can be defined by
\beq \label{smdef}
s_m = \int_{\R \cup i \R} \frac{\text{PP}}{\mu-m}\: p_\mu\: d\mu \:.
\eeq
Here ``symmetry'' refers to the fact that
\[ 
s_m^* = s_{\overline{m}} \:. \]
Before going on, we make two remarks. We first point out that the above spectral theorem does not have
an abstract underpinning, because there is no general spectral theorem for symmetric operators on Krein
spaces. Moreover, we note that, in the case that~$m$ is real, the symmetric Green's operator
as defined by~\eqref{smdef} coincides with the mean of the advanced and retarded Green's operators.
In the case that~$m$ is imaginary, however, the connection to the causal Green's operators is not clear.
The advanced and retarded Green's operator can still be constructed in position space using
the theory of linear symmetric hyperbolic systems (as explained for example in~\cite[Chapter~13]{intro}).
However, these solutions increase exponentially in time. Consequently, they are ill-defined as
tempered distributions, making it impossible to take their Fourier transforms.
In what follows, we bypass these issues by restricting attention to the symmetric Green's operators.

After these definitions, we are in the position to state the main result of this appendix.
\begin{Thm} \label{thmholo}
Let~$\Gamma$ be the integral operator on the Krein space~$(\K, \bra .|. \ket)$ with integral kernel
\beq \label{Gammadef}
\Gamma(x,y) := \sum_{a=1}^L \Lambda_a\Big( \frac{x+y}{2} \Big)\: L_a(y-x) \:.
\eeq
We introduce the following operator involving the double commutator~$[ [\Pdd, \Gamma], \Gamma]$,
\beq \label{Cdef}
{\mathscr{C}} := i \int_0^1 s\: e^{-i s\Gamma} \: \big[ [\Pdd, \Gamma], \Gamma \big]\: e^{i s \Gamma} \:ds \:.
\eeq
Then the unitary holographic mixing operator~$V$ satisfying~\eqref{Vunitary} and~\eqref{Vrep}
can be written as
\[ V = \int_{\R \cup i \R} V_m\: dm \]
where the operators~$V_m$ are defined in terms of the following perturbation series in~${\mathscr{C}}$,
\begin{align*}
V_m &:= e^{i \Gamma} \:\sum_{n=0}^\infty \big(-s_m \,{\mathscr{C}} \big)^n \,p_m  \bigg( \1 + 
\sum_{p=1}^\infty (-1)^p\: \frac{(2p-1)!!}{p!\,2^p}\: \big(B_m \big)^p \bigg) \\ 
B_m &:= \pi^2 \:\epsilon(m^2) \sum_{n,n'=0}^\infty \big( - {\mathscr{C}}\, s_m \big)^{n'}\:{\mathscr{C}}\:p_m\:{\mathscr{C}}\:
\big( - s_m \,{\mathscr{C}} \big)^n \,p_m 
\end{align*}
(where~$\epsilon(m^2)$ denotes the sign function).
\end{Thm} \noindent
We note for clarity that the operator~$e^{i \Gamma}$ is defined via a spectral calculus on~$(\K, \bra .|. \ket)$
or alternatively by the power series
\[ e^{i \Gamma} := \sum_{k=0}^\infty \frac{(i \Gamma)^k}{k!} \]
(thus one takes powers of the nonlocal operator, not the exponential of the integral kernel).
Using that~$\Gamma$ is symmetric on~$(\K, \bra .|. \ket)$, it follows that~$(e^{i \Gamma})^* = e^{-i \Gamma}$,
showing that the operator~$e^{i \Gamma}$ is unitary on the Krein space~$(\K, \bra .|. \ket)$.

The remainder of this appendix is devoted to the proof of this theorem.
Having Green's operators to our disposal, the potential~$\B$ in~\eqref{DirB} can be treated
perturbatively. However, using the specific structure of this potential, we can even treat
generalized phase factors non-perturbative, as we now explain. By direct computation, one verifies
that~$\B$ can be written as a commutator,
\[ \B = [\Pdd, \Gamma] \:, \]
where~$\Gamma$ is the operator defined in~\eqref{Gammadef}. Moreover, we set
\beq \label{checksdef}
\check{s}_m = e^{i \Gamma}\, s_m \, e^{-i \Gamma} \:.
\eeq
This operator is an approximate Green's operator of the Dirac operator with nonlocal gauge potentials,
as is made precise in the next lemma.
\begin{Lemma} For any~$m \in \R \cup i \R$,
\beq \label{Err}
(i \Pdd + \B)\: \check{s}_m = E\, \check{s}_m \:,
\eeq
where the operator~$E$ is given by
\[ 
E = i \int_0^1 (1-s)\: e^{i s \Gamma} \: \big[ [\Pdd, \Gamma], \Gamma \big]\: e^{-i s \Gamma} \:ds \:. \]
\end{Lemma}
\Proof Using~\eqref{smdef}, we can rewrite~\eqref{checksdef} as
\[ \check{s}_m = \int_{\R \cup i \R} \frac{\text{PP}}{\mu - m} \: e^{i \Gamma}\, p_\mu \, e^{-i \Gamma} \:. \]
Therefore, it suffices to prove that
\[ (i \Pdd + \B - \mu)\: \big( e^{i \Gamma}\, p_\mu \big) = E\,  \big( e^{i \Gamma}\, p_\mu \big) \:. \]
In order to derive this relation, we make use of the well-known formula for the commutator with an
exponential\footnote{For the proof, one uses that, for any~$N \in \N$,
\[ \big[ \Pdd, e^{i \Gamma} \big] = 
\big[ \Pdd, \big( e^{i \Gamma/N}\big)^N \big] = \sum_{k=0}^{N-1}  
e^{ i \Gamma\,\frac{k}{N}} \: \big[ \Pdd, e^{i \Gamma/N} \big]\: e^{ i \Gamma\, \frac{N-k-1}{N}} \:.\]
Expanding for large~$N$, the commutator on the right becomes~$[\Pdd, i \Gamma]/N + \O(N^{-2})$.
Viewing the sum as a Riemann sum and taking the limit~$N \rightarrow \infty$ gives~\eqref{commexp}.}
\beq \label{commexp}
\big[ \Pdd, e^{i \Gamma} \big] = \int_0^1 e^{i \tau \Gamma}\: [\Pdd, i \Gamma]\: e^{i (1-\tau) \Gamma} \: d\tau\:.
\eeq
We thus obtain
\begin{align*}
(i \Pdd-\mu) \big( e^{i \Gamma} \,p_\mu \big) &= \big[i \Pdd, e^{i \Gamma}\big]\: p_\mu 
= -\int_0^1 e^{i \tau \Gamma}\: [\Pdd, \Gamma]\: e^{i (1-\tau) \Gamma} \: p_\mu\: d\tau \\
&= -[\Pdd, \Gamma]\: e^{i \Gamma}\, p_\mu +
\int_0^1 \big[  [\Pdd, \Gamma], e^{i \tau \Gamma} \big] \: e^{i (1-\tau) \Gamma} \: p_\mu\: d\tau \:,
\end{align*}
showing that
\[ (i \Pdd+ \B -\mu) \big( e^{i \Gamma} \,p_\mu \big)
=  \int_0^1 \big[  [\Pdd, \Gamma], e^{i \tau \Gamma} \big] \: e^{-i \tau \Gamma} \:
\big( e^{i \Gamma} \,p_\mu \big)\: d\tau \:. \]
Again using~\eqref{commexp}, the last integral can be simplified as follows,
\begin{align*}
&\int_0^1 \big[  [\Pdd, \Gamma], e^{i \tau \Gamma} \big] \: e^{-i \tau \Gamma} \: d\tau
= \int_0^1 d\tau \int_0^1 d\tau'\: e^{i \tau' \tau \Gamma}
\big[  [\Pdd, \Gamma], i \tau \Gamma \big] \: e^{-i \tau' \tau \Gamma} \\
&= \bigg\{ \begin{array}{c} \tau \tau' =: s \\
\tau\, d\tau' = ds  \end{array} \bigg\} 
= i \int_0^1 d\tau \int_0^\tau ds \: e^{i s \Gamma}
\big[  [\Pdd, \Gamma], \Gamma \big] \: e^{-i s \Gamma} \\
&= i \int_0^1 ds \int_s^1 d\tau \: e^{i s \Gamma}
\big[  [\Pdd, \Gamma], \Gamma \big] \: e^{-i s \Gamma} 
= i \int_0^1 (1-s) \: e^{i s \Gamma}
\big[  [\Pdd, \Gamma], \Gamma \big] \: e^{-i s \Gamma} \:ds \:.
\end{align*}
This concludes the proof.
\QED

Now we can treat the operator~$E$ in~\eqref{Err} perturbatively, giving the following result.
\begin{Lemma} \label{lemmastilde}
For any~$m \in \R \cup i \R$, 
the Green's operator~$\tilde{s}_m$ corresponding to the operator~$(i \Pdd + \B)$ has the form
\[ \tilde{s}_m = \sum_{n=0}^\infty \big( - \check{s}_m \,E \big)^n \,\check{s}_m \:. \]
\end{Lemma}
\Proof A direct computation using~\eqref{Err} gives
\begin{align*}
(i \Pdd + \B) \big( - \check{s}_m \,E \big)^n \,\check{s}_m = E\, 
\big( - \check{s}_m \,E \big)^n \,\check{s}_m 
+ \bigg\{ \begin{array}{cl} \1 & \text{if~$n=0$} \\
-E\, \big( - \check{s}_m \,E \big)^{n-1} \,\check{s}_m & \text{if~$n>0$} \:. \end{array}
\end{align*}
Summing over~$n$ gives the result.
\QED

The following lemma generalizes formulas from~\cite[Lemma~2.1]{grotz} to the case of imaginary mass
parameters.
\begin{Lemma} For any~$m, m' \in \R \cup i \R$,
\begin{align}
p_m \,s_{m'}&=s_{m'}\,p_m=\frac{\text{\rm{PP}}}{m-m'}\:p_m\label{eq:ps-p} \\
s_m\,s_{m'}&=\frac{\text{\rm{PP}}}{m-m'}\:(s_m-s_{m'})+\pi^2\:\epsilon(m^2)\:\delta(m-m')\:p_m\:. \label{eq:ss-sp}
\end{align}
\end{Lemma}
\Proof The relation~\eqref{eq:ps-p} follows immediately from~\eqref{smdef} and~\eqref{pmpmp}.
The proof of~\eqref{eq:ss-sp} is more subtle. In the case~$m, m' \in \R$ one can argue with
the support properties of the causal Green's operators (for details see~\cite[proof of Lemma~2.1]{grotz}).
Here instead we carefully analyze regularized distributions in the complex plane. If~$m$ is real, we
we write the principal part in~\eqref{smdef} as
\[ s_m = \frac{1}{2} \lim_{\varepsilon \searrow 0} \sum_{s = \pm}
\int_{\R \cup i \R} \frac{\text{1}}{\mu-m + i s \varepsilon}\: p_\mu\: d\mu \:. \]
Now the two two summands can be treated individually by using the distributional relation
(for details see for example~\cite[eqs~(1.2.33), (1.2.33) and Exercises~1.10--1.12]{cfs})
\beq \label{disrel}
\lim_{\varepsilon \searrow 0} \frac{1}{x \pm i \varepsilon} = \frac{\text{PP}}{x} \mp i \pi \: \delta(x) \:.
\eeq
In order to extend this method to the case where~$\mu$ and~$m$ can be both either real or purely imaginary,
we must regularize such as to avoid poles on both the real and the imaginary axes. This can be
achieved for example by setting
\[ s_m = \frac{1}{2} \lim_{\varepsilon \searrow 0} \sum_{s = \pm}
\int_{\R_\varepsilon \cup i \R_\varepsilon} \frac{\text{1}}{\mu-m + (1-i) s \varepsilon}\: p_\mu\: d\mu \:, \]
where we used the abbreviation
\[ R_\varepsilon := \R \setminus (-2 \varepsilon, 2 \varepsilon) \]
(cutting out a neighborhood of the origin has the purpose of avoiding poles if~$m$ and~$\mu$ are
close to zero).
If~$m$ is real, we can again use~\eqref{disrel}. If~$m$ is imaginary, however, one must take into account
that the Lebesgue measure~$dm$ differs from the contour integral~$dz$ along the imaginary axis by a
factor~$i$. We thus obtain the distributional relation
\beq \label{disrelcomplex}
\lim_{\varepsilon \searrow 0} \frac{\text{1}}{\mu-m \pm (1-i) \varepsilon} = \frac{\text{PP}}{\mu-m} \mp i \pi \:
\sigma(m)\: \delta(\mu-m)
\eeq
with
\[ \sigma(m) := \Big\{ \begin{array}{cl} 1 & \text{if~$m \in \R$} \\ i & \text{if~$m \in i \R\:.$} \end{array} \]

Using~\eqref{pmpmp}, we write the left side of~\eqref{eq:ss-sp} as
\beq \label{smsmp}
s_m\, s_{m'} = \frac{1}{4} \lim_{\varepsilon, \varepsilon' \searrow 0} \sum_{s,s' = \pm}
\int_{\R_\varepsilon \cup i \R_\varepsilon} \frac{\text{1}}{\mu-m + (1-i) s \varepsilon}\; \frac{\text{1}}{\mu-m' + (1-i) s' \varepsilon'}\: p_\mu\: d\mu
\eeq
Next, we use the partial sum decomposition
\begin{align}
&\frac{\text{1}}{\mu-m + (1-i) s \varepsilon}\; \frac{\text{1}}{\mu-m' + (1-i) s' \varepsilon'} \notag \\
&= \bigg( \frac{\text{1}}{\mu-m + (1-i) s \varepsilon} - \frac{\text{1}}{\mu-m' + (1-i) s' \varepsilon'} \bigg)\:
\frac{1}{m-m' - (1-i) (s \varepsilon - s' \varepsilon')} \:. \label{reg0}
\end{align}
Now we first take the limit~$\varepsilon \searrow 0$ (of course, taking the limits in another order gives
the same end result). Then in the last factor in~\eqref{reg0} we can leave out~$\varepsilon$.
Applying~\eqref{disrelcomplex} gives
\begin{align*}
&\frac{1}{2} \lim_{\varepsilon \searrow 0} \sum_{s=\pm} \frac{\text{1}}{\mu-m + (1-i) s \varepsilon}\; \frac{\text{1}}{\mu-m' + (1-i) s' \varepsilon'} \notag \\
&= \bigg( \frac{\text{PP}}{\mu-m} - \frac{\text{1}}{\mu-m' + (1-i) s' \varepsilon'} \bigg)\:
\frac{1}{m-m' + (1-i) s' \varepsilon'} \:.
\end{align*}
Next, we take the limit~$\varepsilon' \searrow 0$ and apply again~\eqref{disrelcomplex}. We thus obtain
\begin{align*}
&\frac{1}{4} \lim_{\varepsilon' \searrow 0} \lim_{\varepsilon \searrow 0} \sum_{s,s'=\pm} \frac{\text{1}}{\mu-m + (1-i) s \varepsilon}\; \frac{\text{1}}{\mu-m' + (1-i) s' \varepsilon'} \notag \\
&= \frac{\text{PP}}{\mu-m}\; \frac{\text{PP}}{m-m'} \\
&\quad\: - \frac{1}{2} \sum_{s'=\pm} \bigg( \frac{\text{PP}}{\mu-m'} - i \pi s'\:\sigma(m')\, \delta \big(\mu-m' \big) \bigg)
\bigg( \frac{\text{PP}}{m-m'} - i \pi s'\: \sigma(m)\,\delta \big(m-m' \big) \bigg) \\
&= \frac{\text{PP}}{\mu-m}\; \frac{\text{PP}}{m-m'} - \frac{\text{PP}}{\mu-m'}\; \frac{\text{PP}}{m-m'}
+ \pi^2 \: \sigma(m)^2\, \delta \big(\mu-m' \big)\: \delta \big(m-m' \big) \:.
\end{align*}
Using this identity in~\eqref{smsmp} gives~\eqref{eq:ss-sp}, concluding the proof.
\QED

We next want to compute an operator which maps the unperturbed to the perturbed solutions.
Our first ansatz is
\[ U_m := \sum_{n=0}^\infty \big( - \check{s}_m \,E \big)^n \,e^{i \Gamma}\: p_m \:. \]
Exactly as in the proof of Lemma~\ref{lemmastilde}, one verifies that this operator maps to
solutions, i.e.\
\[ (i \Pdd + \B - m)\: U_m = 0 \:. \]
However, the normalization is not correct, because
(for more details see similar computations in~\cite{grotz})
\begin{align*}
&(U_{\overline{m'}})^* \:U_m = \sum_{n,n'=0}^\infty p_{m'} \,e^{-i \Gamma}\: \big( - E\, \check{s}_{m'} \big)^{n'}
\big( - \check{s}_m \,E \big)^n \,e^{i \Gamma}\: p_m \\
&= \delta(m-m')\: \bigg( p_m + \pi^2\: \epsilon(m^2)
\sum_{n,n'=0}^\infty p_{m} \,e^{-i \Gamma}\: \big( - E\, \check{s}_m \big)^{n'}\:E\:\check{p}_m\:E\:
\big( - \check{s}_m \,E \big)^n \,e^{i \Gamma}\: p_m \bigg)
\end{align*}
with
\beq \label{checkpdef}
\check{p}_m := e^{i \Gamma} p_m e^{-i \Gamma} \:.
\eeq
We write this result in the short form
\[ (U_{\overline{m}})^* \:U_m = \delta(m-m')\:  p_m \big(1 + A_m \big) \]
with
\[ A_m := \pi^2\: \epsilon(m^2)
\sum_{n,n'=0}^\infty e^{-i \Gamma}\: \big( - E\, \check{s}_m \big)^{n'}\:E\:\check{p}_m\:E\:
\big( - \check{s}_m \,E \big)^n \,e^{i \Gamma}\: p_m \:. \]
Following the rescaling procedure in~\cite{grotz}, we set
\[ V_m := U_m\:  \big(\1 + A_m\big)^{-\frac{1}{2}} \:, \]
where the last factor is defined by the perturbation series given by the Taylor series, i.e.
\[ \big(\1 + A_m\big)^{-\frac{1}{2}} := \1 + \sum_{n=1}^\infty (-1)^n\: \frac{(2n-1)!!}{n!\,2^n}\: 
\big(A_m \big)^n \:. \]
Then
\[ (V_{\overline{m'}})^*\, V_m = \delta(m-m')\: \big(1 + (A_{\overline{m}})^*\big)^{-\frac{1}{2}}\:
p_m \big(1 + A_m \big)\: \big(1 + A_m\big)^{-\frac{1}{2}} = \delta(m-m')\: p_m \:, \]
as desired. As a consequence, the operator
\[ V := \int_{\R \cup i \R} V_m \: dm \]
is unitary, because
\[ V^* V = \int_{\R \cup i \R} dm' \int_{\R \cup i \R} dm\: (V_{\overline{m}})^* \: V_m \\
= \int_{\R \cup i \R} p_m\: dm = \1\:. \]
Therefore, the perturbed spectral projectors can be written as
\[ \tilde{p}_m = V p_m V^* \:, \]
and the corresponding symmetric Green's operators are given in analogy to~\eqref{smdef} by
\[ \tilde{s}_m = \int_{\R \cup i \R} \frac{\text{PP}}{\mu-m}\: \tilde{p}_m\: d\mu
= \int_{\R \cup i \R} \frac{\text{PP}}{\mu-m}\: V p_m V^*\: d\mu = V s_m V^*\:. \]
The statement of Theorem~\ref{thmholo} is obtained by rewriting~$\check{s}_m$ and~$\check{p}_m$
in terms of~$s_m$ and~$p_m$ using~\eqref{checksdef} and~\ref{checkpdef}, and noting that
\begin{align*}
e^{-i \Gamma} \,E\, e^{i \Gamma} &= 
i \int_0^1 (1-s)\: e^{-i (1-s) \Gamma} \: \big[ [\Pdd, \Gamma], \Gamma \big]\: e^{i (1-s) \Gamma} \:ds = 
{\mathscr{C}} \:,
\end{align*}
where in the last step we transformed the integration variable according to~$s \rightarrow 1-s$
and used~\eqref{Cdef}.

\section{Microlocal Expansion of the Unitary Holographic Mixing Operator} \label{appholo}
In this appendix we explain how the unitary holographic mixing operator~$V$ defined by~\eqref{Vdef}
can be expanded in powers of~$\ell_{\min}/\ell_\Lambda$. Clearly, this expansion is sensible only
if~$\ell_\Lambda \gg \ell_{\min}$, as is made precise in~\eqref{Lamscale}.
Our method is inspired by the pseudo-differential calculus and microlocal analysis;
for the general context see for example~\cite{grigis}.
In preparation, we expand the phase factors in~\eqref{Udef} in a Taylor expansion
about the arithmetic mean of the left and right arguments of the kernel. Thus, setting
\[ \xi := y-x \qquad \text{and} \qquad \zeta := \frac{y+x}{2} \:, \]
we expand
\begin{align*}
U(x,y) &= \sum_{a=1}^N e^{i \Lambda_a(x)}\: L_a(x,y)
= \sum_{a=1}^N e^{i \Lambda_a(\zeta - \xi/2)}\: L_a(x,y) \\
&= \sum_{\kappa} \sum_{a=1}^N 
\frac{1}{|\kappa|!}\: \big( \partial_\kappa e^{i \Lambda_a(\zeta)} \big)\: \Big( -\frac{\xi}{2} \Big)^\kappa \: L_a(x,y)
\end{align*}
(where~$\kappa$ is a multi-index). We also write this expansion as
\[ U = \sum_{p=0}^\infty U^{(p)} \]
with
\beq \label{Upker}
U^{(p)}(x,y) := \sum_{\text{$\kappa$ with~$|\kappa|=p$}} \;\sum_{a=1}^N 
\frac{1}{p!}\: \big( \partial_\kappa e^{i \Lambda_a(\zeta)} \big)\: \Big( -\frac{\xi}{2} \Big)^\kappa \: L_a(x,y) \:.
\eeq
This has the advantage that the index~$p$ tells us about the scaling power in~$\ell_\Lambda/\ell_{\min}$.
Indeed, counting orders starting from~$L_a$ with the order zero, we have
\[ U^{(p)} = \O \Big( \big( \ell_{\min} / \ell_\Lambda \big)^p \Big) \:. \]

For what follows, it is crucial that these operator commute to leading order in~$\ell_\Lambda/\ell_{\min}$, i.e.\
\beq \label{Upqcomm}
\big[ U^{(p)}, U^{(q)} \big] = \O \Big( \big( \ell_{\min} / \ell_\Lambda \big)^{p+q+1} \Big)
\eeq
and similarly for~$U^*$ and mixed commutators.
In order to see how this comes about, one should keep in mind that, if the kernels in~\eqref{Upker}
do not depend on~$\zeta$, then the operators~$U^{(p)}$ and~$U^{(q)}$ are multiplication operators in
momentum space, which clearly commute. Therefore, the commutator in~\eqref{Upqcomm} is nonzero
as a consequence of a linear expansion of the $\zeta$-dependent factors in~\eqref{Upker}, giving an
additional scaling factor~$\ell_{\min}/\ell_\Lambda$.

Clearly, the above formalism also applies to~$U^*$ or to composite operators. In general terms,
the contributions to the resulting expansions can be written as
\[ 
A(x,y) = f(\zeta)\: K(\xi) \:, \]
where the kernel~$K(\xi)$ decays on the scale~$\ell_{\min}$, and the derivatives of~$f(\zeta)$
scale in powers of~$1/\ell_\Lambda$. Under these assumptions, one can introduce a
an approximate spectral calculus and one can compute approximate inverses,
where ``approximate'' means that we allow for errors of the order~$\ell_{\min}/\ell_\Lambda$.
This is made precise in the next lemma.
\begin{Lemma} {\bf{(microlocal spectral calculus)}} Given a function~$g \in C^1(\C, \C)$
 with~$g \circ f \in C^1(M, \C)$ and~$g \circ \hat{K} \in L^1(\hat{M}, \C)$, the operator~$g_\ml(A)$
defined by
\[ \big( g_\ml(A) \big)(x,y) := (g \circ f)(\zeta)\: 
\int \frac{d^4k}{(2 \pi)^4}\: g \big(\hat{K}(k) \big) \: e^{-i k \xi} \]
satisfies the approximate spectral calculus
\[ g_\ml(A)\, \tilde{g}_\ml(A) = \big( g \tilde{g} \big)_\ml(A) + \O \big( \ell_{\min} / \ell_\Lambda \big) \:. \]
The microlocal inverse
\[ A^{-1}_\ml(x,y) :=  \frac{1}{f(\zeta)}\: 
\int \frac{d^4k}{(2 \pi)^4}\: \frac{1}{\hat{K}(k)}\: e^{-i k \xi} \]
is an approximate inverse even up to second order
corrections, i.e.\
\[ A^{-1}_\ml\: A = \1 + \O \Big( \big( \ell_{\min} / \ell_\Lambda \big)^{2} \Big) \:. \]
\end{Lemma}
\Proof We write the operator product with integral kernels,
\[ \big(g_\ml(A)\, \tilde{g}_\ml(A) \big)(x,y) = \int_M \big(g \circ f \big)\Big( \frac{x+z}{2} \Big)\: 
\big(\tilde{g} \circ f \big)\Big( \frac{z+y}{2} \Big)\: \big(g \circ \hat{K} \big)(k)\: \big(\tilde{g} \circ \hat{K} \big)(k)\:d^4z \:. \]
Replacing the arguments of~$g\circ f$ and~$\tilde{g} \circ f$ by~$\zeta:=(x+y)/2$ gives an error of order~$\ell_{\min} / \ell_\Lambda$. Then we can carry out the~$z$-integral, giving the result.

For the inverse, we obtain the formula

\beq \label{noz}
\big(A^{-1}_\ml\: A\big)(x,y) = \int_M \frac{f((z+y)/2)}{f(x+z)/2} \: K^{-1}(x,z)\: K(z,y) \:d^4z \:,
\eeq
where~$K^{-1}(x,y)$ is the Fourier transform of~$1/\hat{K}(k)$.
We now expand the quotient in the integrand about~$x$,
\begin{align*}
\frac{f((z+y)/2)}{f(x+z)/2} &= 1 + \frac{\partial_j f(x)}{f(x)} \:\Big( \frac{z+y-2x}{2} - \frac{x+z-2x}{2} \Big)
+ \O \Big( \big( \ell_{\min} / \ell_\Lambda \big)^{2} \Big) \\
&= 1 + \frac{\partial_j f(x)}{2 f(x)} \:\xi^j + \O \Big( \big( \ell_{\min} / \ell_\Lambda \big)^{2} \Big) \:.
\end{align*}
Since this expansion does not involve~$z$, we can carry out the $z$-integral in~\eqref{noz}. Using the relation
\[ \int_M  K^{-1}(x,z)\: K(z,y) \: d^4z = \delta^4(x-y) \:, \]
the factor~$\xi$ in the linear expansion term drops out, giving the result.
\QED

Using the above formalism, we can compute~$V$ and the potential~$\B_\Lambda$ in~\eqref{BLamdef}
order by order. We now illustrate this expansion by a computation to zeroth order.
We begin with the operator~$U$ defined in~\eqref{Udef},
\begin{align*}
U(x,y) &= \sum_{a=1}^N e^{i \Lambda_a(\zeta)}\: L_a(x,y) + \O \big( \ell_{\min} / \ell_\Lambda \big) \\
(U^* U)(x,y) &= \sum_{a,b=1}^N e^{-i \Lambda_a(\zeta)+ i \Lambda_b(\zeta)}\: \big(L_a \, L_b\big)(x,y) + \O \big( \ell_{\min} / \ell_\Lambda \big) \\
\big( (U^* U)^{-\frac{1}{2}} \big)(x,y) &= \int \frac{d^4k}{(2 \pi)^4}\:
\bigg( \sum_{a,b=1}^N e^{-i \Lambda_a(\zeta)+ i \Lambda_b(\zeta)}\: \hat{L}_a(k) \: \hat{L}_b(k) \bigg)^{-\frac{1}{2}}
\: e^{-ik(x-y)} \\
&\quad\:+ \O \big( \ell_{\min} / \ell_\Lambda \big) \:.
\end{align*}
Consequently, the operator~$V$ given by~\eqref{Vdef} takes the form
\begin{align*}
V(x,y) &= 
\int \frac{d^4k}{(2 \pi)^4}\: \bigg( \sum_{d=1}^N e^{i \Lambda_d(\zeta)}\: \hat{L}_d(k) \bigg) \\
&\qquad \times
\bigg( \sum_{a,b=1}^N e^{-i \Lambda_a(\zeta)+ i \Lambda_b(\zeta)}\: \hat{L}_a(k) \: \hat{L}_b(k) \bigg)^{-\frac{1}{2}}
\: e^{-ik(x-y)} + \O \big( \ell_{\min} / \ell_\Lambda \big) \\
V^*(x,y) &= 
\int \frac{d^4k}{(2 \pi)^4}\: \bigg( \sum_{d=1}^N e^{-i \Lambda_d(\zeta)}\: \hat{L}_d(k) \bigg) \\
&\qquad \times
\bigg( \sum_{a,b=1}^N e^{i \Lambda_a(\zeta)- i \Lambda_b(\zeta)}\: \hat{L}_a(k) \: \hat{L}_b(k) \bigg)^{-\frac{1}{2}}
\: e^{-ik(x-y)} + \O \big( \ell_{\min} / \ell_\Lambda \big) \:.
\end{align*}
Using these formulas in~\eqref{BLamdef}, we obtain
\begin{align*}
&\big[i \Pdd, V^* \big](x,y) \\
&= -\frac{1}{2}
\int \frac{d^4k}{(2 \pi)^4}\: \bigg( \sum_{d=1}^N e^{-i \Lambda_d(\zeta)}\: \hat{L}_d(k) \bigg) 
\bigg( \sum_{a,b=1}^N e^{i \Lambda_a(\zeta)- i \Lambda_b(\zeta)}\: \hat{L}_a(k) \: \hat{L}_b(k) \bigg)^{-\frac{3}{2}}\\
&\qquad \times
\bigg( \sum_{a,b=1}^N \Big( -(\Pdd \Lambda_a)(\zeta) + (\Pdd \Lambda_b)(\zeta) \Big)\:
e^{i \Lambda_a(\zeta)- i \Lambda_b(\zeta)}\: \hat{L}_a(k) \: \hat{L}_b(k) \bigg)
\: e^{-ik(x-y)} \\
&\quad\: + \int \frac{d^4k}{(2 \pi)^4}\: \bigg( \sum_{d=1}^N (\Pdd \Lambda_d)(\zeta)\: e^{-i \Lambda_d(\zeta)}\: \hat{L}_d(k) \bigg) \\
&\qquad \times
\bigg( \sum_{a,b=1}^N e^{i \Lambda_a(\zeta)- i \Lambda_b(\zeta)}\: \hat{L}_a(k) \: \hat{L}_b(k) \bigg)^{-\frac{1}{2}}
\: e^{-ik(x-y)} + \O \big( \ell_{\min} / \ell_\Lambda \big) \\
&\B_\dyn(x,y) = V\, \big[i \Pdd, V^* \big](x,y) \\
&= -\frac{1}{2}
\int \frac{d^4k}{(2 \pi)^4}\: 
\bigg( \sum_{a,b=1}^N e^{i \Lambda_a(\zeta)- i \Lambda_b(\zeta)}\: \hat{L}_a(k) \: \hat{L}_b(k) \bigg)^{-1}\\
&\qquad \times
\bigg( \sum_{a,b=1}^N \Big( -(\Pdd \Lambda_a)(\zeta) + (\Pdd \Lambda_b)(\zeta) \Big)\:
e^{i \Lambda_a(\zeta)- i \Lambda_b(\zeta)}\: \hat{L}_a(k) \: \hat{L}_b(k) \bigg)
\: e^{-ik(x-y)} \\
&\quad\: + \int \frac{d^4k}{(2 \pi)^4}\: \bigg( \sum_{d,e=1}^N (\Pdd \Lambda_d)(\zeta)\: e^{-i \Lambda_d(\zeta)
+ i \Lambda_d(\zeta)}\: \hat{L}_d(k)\: \hat{L}_e(k) \bigg) \\
&\qquad \times
\bigg( \sum_{a,b=1}^N e^{i \Lambda_a(\zeta)- i \Lambda_b(\zeta)}\: \hat{L}_a(k) \: \hat{L}_b(k) \bigg)^{-1}
\: e^{-ik(x-y)} + \O \big( \ell_{\min} / \ell_\Lambda \big) \\
&= \frac{1}{2}
\int \frac{d^4k}{(2 \pi)^4}\: \bigg( \sum_{a,b=1}^N \Big( (\Pdd \Lambda_a)(\zeta) + (\Pdd \Lambda_b)(\zeta) \Big)\:
e^{i \Lambda_a(\zeta)- i \Lambda_b(\zeta)}\: \hat{L}_a(k) \: \hat{L}_b(k) \bigg) \\
&\qquad \times
\bigg( \sum_{a,b=1}^N e^{i \Lambda_a(\zeta)- i \Lambda_b(\zeta)}\: \hat{L}_a(k) \: \hat{L}_b(k) \bigg)^{-1} 
\: e^{-ik(x-y)} + \O \big( \ell_{\min} / \ell_\Lambda \big) \\
&= \frac{1}{2}
\int \frac{d^4k}{(2 \pi)^4}\: 
\frac{\displaystyle \sum\nolimits_{a,b=1}^N \Big( (\Pdd \Lambda_a)(\zeta) + (\Pdd \Lambda_b)(\zeta) \Big)\:
e^{i \Lambda_a(\zeta)- i \Lambda_b(\zeta)}\: \hat{L}_a(k) \: \hat{L}_b(k)}
{\displaystyle \sum\nolimits_{a,b=1}^N e^{i \Lambda_a(\zeta)- i \Lambda_b(\zeta)}\: \hat{L}_a(k) \: \hat{L}_b(k)} \; e^{-ik(x-y)} \\
&\quad\:+ \O \big( \ell_{\min} / \ell_\Lambda \big) \:.
\end{align*}
If we expand this potential to first order in the phase functions~$\Lambda_a$, assuming that
\[ \label{Kone}
\sum_{a=1}^n L_a = \1 \:, \]
we recover the potential~\eqref{Bsimp}, giving a connection to the perturbative treatment in Appendix~\ref{apppert}.
However, we point out that the above formula for~$\B_\dyn(x,y)$ is much more general due to the
additional phase factors evaluated at~$\zeta$.
  
\Thanks{{{\em{Acknowledgments:}}
We would like to thank Marvin Becker and the referees for helpful comments on the manuscript.
We are grateful to the ``Universit\"atsstiftung Hans Vielberth'' for support.
N.K.'s research was also supported by the NSERC grant RGPIN~105490-2018.
M.R.\ was supported by CityU Start-up Grant 7200748, by CityU Strategic Research Grant 7005839, and by General Research Fund ECS 21306524.
C.D.\ is grateful for the support of Indam, in particular that of the Gruppo Nazionale di Fisica Matematica
(GNFM).
%

\bibliographystyle{amsplain}

\begin{thebibliography}{10}

\bibitem{lagrange}
Y.~Bernard and F.~Finster, \emph{On the structure of minimizers of causal
  variational principles in the non-compact and equivariant settings},
  \href{https://arxiv.org/abs/1205.0403}{arXiv:1205.0403 [math-ph]}, Adv. Calc.
  Var. \textbf{7} (2014), no.~1, 27--57.

\bibitem{DDRZ}
C.~Dappiaggi, N.~Drago, P.~Rinaldi, and L.~Zambotti, \emph{A microlocal
  approach to renormalization in stochastic {PDE}s},
  \href{https://arxiv.org/abs/2009.07640}{arXiv:2009.07640 [math-ph]}, Commun.
  Contemp. Math. \textbf{24} (2022), no.~7, Paper No. 2150075, 74.

\bibitem{linhyp}
C.~Dappiaggi and F.~Finster, \emph{Linearized fields for causal variational
  principles: {E}xistence theory and causal structure},
  \href{https://arxiv.org/abs/1811.10587}{arXiv:1811.10587 [math-ph]}, Methods
  Appl. Anal. \textbf{27} (2020), no.~1, 1--56.

\bibitem{qftlimit}
C.~Dappiaggi, F.~Finster, N.~Kamran, and M.~Reintjes, \emph{The quantum field
  theory limit of causal fermion systems}, in preparation.

\bibitem{endlich}
F.~Finster, \emph{Derivation of field equations from the principle of the
  fermionic projector},
  \href{https://arxiv.org/abs/gr-qc/9606040}{arXiv:gr-qc/9606040} (unpublished
  preprint in German) (1996).

\bibitem{sea}
\bysame, \emph{Definition of the {D}irac sea in the presence of external
  fields}, \href{https://arxiv.org/abs/hep-th/9705006}{arXiv:hep-th/9705006},
  Adv. Theor. Math. Phys. \textbf{2} (1998), no.~5, 963--985.

\bibitem{pfp}
\bysame, \emph{The {P}rinciple of the {F}ermionic {P}rojector},
  \href{https://arxiv.org/abs/hep-th/0001048}{hep-th/0001048},
  \href{https://arxiv.org/abs/hep-th/0202059}{hep-th/0202059},
  \href{https://arxiv.org/abs/hep-th/0210121}{hep-th/0210121}, AMS/IP Studies
  in Advanced Mathematics, vol.~35, American Mathematical Society, Providence,
  RI, 2006.

\bibitem{continuum}
\bysame, \emph{Causal variational principles on measure spaces},
  \href{https://arxiv.org/abs/0811.2666}{arXiv:0811.2666 [math-ph]}, J. Reine
  Angew. Math. \textbf{646} (2010), 141--194.

\bibitem{entangle}
\bysame, \emph{Entanglement and second quantization in the framework of the
  fermionic projector}, \href{https://arxiv.org/abs/0911.0076}{arXiv:0911.0076
  [math-ph]}, J. Phys. A: Math. Theor. \textbf{43} (2010), 395302.

\bibitem{qft}
\bysame, \emph{Perturbative quantum field theory in the framework of the
  fermionic projector}, \href{https://arxiv.org/abs/1310.4121}{arXiv:1310.4121
  [math-ph]}, J. Math. Phys. \textbf{55} (2014), no.~4, 042301.

\bibitem{cfs}
\bysame, \emph{The {C}ontinuum {L}imit of {C}ausal {F}ermion {S}ystems},
  \href{https://arxiv.org/abs/1605.04742}{arXiv:1605.04742 [math-ph]},
  Fundamental Theories of Physics, vol. 186, Springer, Cham, 2016.

\bibitem{nrstg}
\bysame, \emph{Causal fermion systems: {A} primer for {L}orentzian geometers},
  \href{https://arxiv.org/abs/1709.04781}{arXiv:1709.04781 [math-ph]}, J.
  Phys.: Conf. Ser. \textbf{968} (2018), 012004.

\bibitem{positive}
\bysame, \emph{Positive functionals induced by minimizers of causal variational
  principles}, \href{https://arxiv.org/abs/1708.07817}{arXiv:1708.07817
  [math-ph]}, Vietnam J. Math. \textbf{47} (2019), 23--37.

\bibitem{action}
\bysame, \emph{The causal action in {M}inkowski space and surface layer
  integrals}, \href{https://arxiv.org/abs/1711.07058}{arXiv:1711.07058
  [math-ph]}, SIGMA Symmetry Integrability Geom. Methods Appl. \textbf{16}
  (2020), no.~091, 83pp.

\bibitem{perturb}
\bysame, \emph{Perturbation theory for critical points of causal variational
  principles}, \href{https://arxiv.org/abs/1703.05059}{arXiv:1703.05059
  [math-ph]}, Adv. Theor. Math. Phys. \textbf{24} (2020), no.~3, 563--619.

\bibitem{nonlocal}
\bysame, \emph{Solving the linearized field equations of the causal action
  principle in {M}inkowski space},
  \href{https://arxiv.org/abs/2304.00965}{arXiv:2304.00965 [math-ph]}, Adv.
  Theor. Math. Phys. \textbf{27} (2023), no.~7, 2087--2217.

\bibitem{grotz}
F.~Finster and A.~Grotz, \emph{The causal perturbation expansion revisited:
  {R}escaling the interacting {D}irac sea},
  \href{https://arxiv.org/abs/0901.0334}{arXiv:0901.0334 [math-ph]}, J. Math.
  Phys. \textbf{51} (2010), no.~7, 072301.

\bibitem{lqg}
\bysame, \emph{A {L}orentzian quantum geometry},
  \href{https://arxiv.org/abs/1107.2026}{arXiv:1107.2026 [math-ph]}, Adv.
  Theor. Math. Phys. \textbf{16} (2012), no.~4, 1197--1290.

\bibitem{baryogenesis}
F.~Finster, M.~Jokel, and C.F. Paganini, \emph{A mechanism of baryogenesis for
  causal fermion systems},
  \href{https://arxiv.org/abs/2111.05556}{arXiv:2111.05556 [gr-qc]}, Class.
  Quant. Gravity \textbf{39} (2022), no.~16, 165005, 50.

\bibitem{fockbosonic}
F.~Finster and N.~Kamran, \emph{Complex structures on jet spaces and bosonic
  {F}ock space dynamics for causal variational principles},
  \href{https://arxiv.org/abs/1808.03177}{arXiv:1808.03177 [math-ph]}, Pure
  Appl. Math. Q. \textbf{17} (2021), no.~1, 55--140.

\bibitem{fockfermionic}
\bysame, \emph{Fermionic {F}ock spaces and quantum states for causal fermion
  systems}, \href{https://arxiv.org/abs/2101.10793}{arXiv:2101.10793
  [math-ph]}, Ann. Henri Poincar\'{e} \textbf{23} (2022), no.~4, 1359--1398.

\bibitem{dirac}
F.~Finster, N.~Kamran, and M.~Oppio, \emph{The linear dynamics of wave
  functions in causal fermion systems},
  \href{https://arxiv.org/abs/2101.08673}{arXiv:2101.08673 [math-ph]}, J.
  Differential Equations \textbf{293} (2021), 115--187.

\bibitem{fockentangle}
F.~Finster, N.~Kamran, and M.~Reintjes, \emph{Entangled quantum states of
  causal fermion systems and unitary group integrals},
  \href{https://arxiv.org/abs/2207.13157}{arXiv:2207.13157 [math-ph]}, Adv.
  Theor. Math. Phys. \textbf{27} (2023), no.~5, 1463--1589.

\bibitem{gaugefix}
F.~Finster and S.~Kindermann, \emph{A gauge fixing procedure for causal fermion
  systems}, \href{https://arxiv.org/abs/1908.08445}{arXiv:1908.08445
  [math-ph]}, J. Math. Phys. \textbf{61} (2020), no.~8, 082301.

\bibitem{intro}
F.~Finster, S.~Kindermann, and J.-H. Treude, \emph{{C}ausal {F}ermion
  {S}ystems: {A}n {I}ntroduction to {F}undamental {S}tructures, {M}ethods and
  {A}pplications}, \href{https://arxiv.org/abs/2411.06450}{arXiv:2411.06450
  [math-ph]}, to appear in Cambridge Monographs on Mathematical Physics,
  Cambridge University Press, 2025.

\bibitem{noether}
F.~Finster and J.~Kleiner, \emph{Noether-like theorems for causal variational
  principles}, \href{https://arxiv.org/abs/1506.09076}{arXiv:1506.09076
  [math-ph]}, Calc. Var. Partial Differential Equations \textbf{55:35} (2016),
  no.~2, 41.

\bibitem{jet}
\bysame, \emph{A {H}amiltonian formulation of causal variational principles},
  \href{https://arxiv.org/abs/1612.07192}{arXiv:1612.07192 [math-ph]}, Calc.
  Var. Partial Differential Equations \textbf{56:73} (2017), no.~3, 33.

\bibitem{osi}
\bysame, \emph{A class of conserved surface layer integrals for causal
  variational principles},
  \href{https://arxiv.org/abs/1801.08715}{arXiv:1801.08715 [math-ph]}, Calc.
  Var. Partial Differential Equations \textbf{58:38} (2019), no.~1, 34.

\bibitem{collapse}
F.~Finster, J.~Kleiner, and C.~Paganini, \emph{Causal fermion systems as an
  effective collapse theory},
  \href{https://arxiv.org/abs/2405.19254}{arXiv:2405.19254 [math-ph]}, J. Phys.
  A: Math. Theor. \textbf{57} (2024), no.~39, 395303.

\bibitem{localize}
F.~Finster and M.~Kraus, \emph{Construction of global solutions to the
  linearized field equations for causal variational principles},
  \href{https://arxiv.org/abs/2210.16665}{arXiv:2210.16665 [math-ph]}, Methods
  Appl. Anal. \textbf{30} (2023), no.~2, 77--94.

\bibitem{banach}
F.~Finster and M.~Lottner, \emph{Banach manifold structure and
  infinite-dimensional analysis for causal fermion systems},
  \href{https://arxiv.org/abs/2101.11908}{arXiv:2101.11908 [math-ph]}, Ann.
  Global Anal. Geom. \textbf{60} (2021), no.~2, 313--354.

\bibitem{cauchynonloc}
F.~Finster, S.~Murro, and G.~Schmid, \emph{The {C}auchy problem for symmetric
  hyperbolic systems with nonlocal potentials}, in preparation.

\bibitem{heatcoll}
F.~Finster and C.~Paganini, \emph{A collapse mechanism without heating}, in
  preparation.

\bibitem{loop}
F.~Finster and J.~Tolksdorf, \emph{Bosonic loop diagrams as perturbative
  solutions of the classical field equations in $\phi^4$-theory},
  \href{https://arxiv.org/abs/1201.5497}{arXiv:1201.5497 [math-ph]}, J. Math.
  Phys. \textbf{53} (2012), no.~5, 052305.

\bibitem{norm}
\bysame, \emph{Perturbative description of the fermionic projector:
  {N}ormalization, causality and {F}urry's theorem},
  \href{https://arxiv.org/abs/1401.4353}{arXiv:1401.4353 [math-ph]}, J. Math.
  Phys. \textbf{55} (2014), no.~5, 052301.

\bibitem{glimm+jaffe}
J.~Glimm and A.~Jaffe, \emph{Quantum {P}hysics: A {F}unctional {I}ntegral
  {P}oint of {V}iew}, second ed., Springer-Verlag, New York, 1987.

\bibitem{grigis}
A.~Grigis and J.~Sj\"ostrand, \emph{Microlocal {A}nalysis for {D}ifferential
  {O}perators}, London Mathematical Society Lecture Note Series, vol. 196,
  Cambridge University Press, Cambridge, 1994, An introduction.

\bibitem{Gubinelli}
M.~Gubinelli, P.~Imkeller, and N.~Perkowski, \emph{Paracontrolled distributions
  and singular {PDE}s}, \href{https://arxiv.org/abs/1210.2684}{arXiv:1210.2684
  [math.PR]}, Forum Math. Pi \textbf{3} (2015), e6, 75.

\bibitem{haag-nogo}
R.~Haag, \emph{On quantum field theories}, Danske Vid. Selsk. Mat.-Fys. Medd.
  \textbf{29} (1955), no.~12, 37.

\bibitem{Hairer}
M.~Hairer, \emph{A theory of regularity structures},
  \href{https://arxiv.org/abs/1303.5113}{arXiv:1303.5113 [math.AP]}, Invent.
  Math. \textbf{198} (2014), no.~2, 269--504.

\bibitem{hall-wightman}
D.~Hall and A.S. Wightman, \emph{A theorem on invariant analytic functions with
  applications to relativistic quantum field theory}, Mat.-Fys. Medd. Danske
  Vid. Selsk. \textbf{31} (1957), no.~5, 41.

\bibitem{kleinert}
H.~Kleinert, \emph{Path {I}ntegrals in {Q}uantum {M}echanics, {S}tatistics,
  {P}olymer {P}hysics, and {F}inancial {M}arkets}, fourth ed., World Scientific
  Publishing Co. Pte. Ltd., Hackensack, NJ, 2006.

\bibitem{namiki}
M.~Namiki, \emph{Stochastic {Q}uantization}, Lecture Notes in Physics. New
  Series m: Monographs, vol.~9, Springer-Verlag, Berlin, 1992.

\bibitem{nelson}
E.~Nelson, \emph{Quantum {F}luctuations}, Princeton Series in Physics,
  Princeton University Press, Princeton, NJ, 1985.

\bibitem{ParisiWu}
G.~Parisi and Y.S. Wu, \emph{Perturbation theory without gauge fixing}, Sci.
  Sinica \textbf{24} (1981), no.~4, 483--496.

\bibitem{powers}
R.T. Powers, \emph{Absence of interaction as a consequence of good ultraviolet
  behavior in the case of a local {F}ermi field}, Comm. Math. Phys. \textbf{4}
  (1967), no.~3, 145--156.

\bibitem{surya}
S.~Surya, \emph{The causal set approach to quantum gravity},
  \href{https://arxiv.org/abs/1903.11544}{arXiv:1903.11544 [gr-qc]}, Living
  Rev. Relativ. \textbf{22} (2019), no.~5, 75pp.

\bibitem{thaller}
B.~Thaller, \emph{The {D}irac {E}quation}, Texts and Monographs in Physics,
  Springer-Verlag, Berlin, 1992.

\end{thebibliography}
\providecommand{\bysame}{\leavevmode\hbox to3em{\hrulefill}\thinspace}
\providecommand{\MR}{\relax\ifhmode\unskip\space\fi MR }
\providecommand{\MRhref}[2]{%
  \href{http://www.ams.org/mathscinet-getitem?mr=#1}{#2}
}
\providecommand{\href}[2]{#2}

\end{document}